\documentclass[notitlepage,11pt,pra,superscriptaddress, onecolumn]{revtex4-1}

\usepackage{multirow}    
\usepackage{relsize}

\usepackage{enumerate}
\usepackage{amsmath} 
\usepackage{amssymb} 
\usepackage{graphicx}
\usepackage{color}
\usepackage{hhline}
\usepackage{theorem}
\usepackage{hyperref}

\usepackage{amsfonts}
\usepackage{amstext}
\usepackage[sort&compress]{natbib}
\usepackage{latexsym}
\usepackage{textcomp}
\usepackage{color}
\usepackage{mathtools}

\newcommand{\mc}[1]{\mathcal{#1}}

\newcommand{\tr}{\mathrm{Tr}}

{\theorembodyfont{\normalfont\it}
\theoremheaderfont{\normalfont\bf}
\newtheorem{thm}{ Theorem}
\newtheorem{dfn}[thm]{ Definition}
\newtheorem{lmm}[thm]{ Lemma}

\newtheorem{crl}[thm]{ Corollary}
\newtheorem{asm}[thm]{ Assumption}
\newtheorem{prp}[thm]{ Proposition}
\newtheorem{cjt}[thm]{ Conjecture}}

{\theorembodyfont{\normalfont}
\theoremheaderfont{\normalfont\it}
\newtheorem{prf}{ Proof:}}

{\theorembodyfont{\normalfont}
\theoremheaderfont{\normalfont\it}
}

{\theorembodyfont{\normalfont}
\theoremheaderfont{\normalfont\it}
}

{\theorembodyfont{\normalfont}
\theoremheaderfont{\normalfont\it}
}

{\theorembodyfont{\normalfont}
\theoremheaderfont{\normalfont\it}
}

{\theorembodyfont{\normalfont}
\theoremheaderfont{\normalfont\it}
\newtheorem{rmk}{ Remark.}}

\newcommand{\bra}[1]{\mbox{$\langle#1|$}}

\newcommand{\ket}[1]{\mbox{$|#1\rangle$}}

\newcommand{\inpro}[2]{\mbox{$\left\langle#1|#2\right\rangle$}}
\newcommand{\outpro}[2]{\mbox{$\ket{#1}\!\bra{#2}$}}

\newcommand{\proj}[1]{\mbox{$\ket{#1}\!\bra{#1}$}}

\newcommand{\alg}[1]{\begin{align}#1\end{align}}

\newcommand{\nn}{\nonumber}
\newcommand{\ca}[1]{{\mathcal #1}}
\newcommand{\mbb}[1]{{\mathbb #1}}
\newcommand{\mfk}[1]{{\mathfrak #1}}

\newcommand{\bthm}[1]{\begin{thm}\label{thm:#1}}
\newcommand{\ethm}{\end{thm}}
\newcommand{\rthm}[1]{\ref{thm:#1}}
\newcommand{\rThm}[1]{Theorem \ref{thm:#1}}
\newcommand{\blmm}[1]{\begin{lmm}\label{lmm:#1}}
\newcommand{\elmm}{\end{lmm}}
\newcommand{\rLmm}[1]{Lemma \ref{lmm:#1}}
\newcommand{\rlmm}[1]{\ref{lmm:#1}}
\newcommand{\bdfn}[1]{\begin{dfn}\label{dfn:#1}}
\newcommand{\edfn}{\end{dfn}}

\newcommand{\basm}[1]{\begin{asm}\label{asm:#1}}
\newcommand{\easm}{\end{asm}}
\newcommand{\bprp}[1]{\begin{prp}\label{prp:#1}}
\newcommand{\eprp}{\end{prp}}

\newcommand{\bcrl}[1]{\begin{crl}\label{crl:#1}}
\newcommand{\ecrl}{\end{crl}}

\newcommand{\rCrl}[1]{Corollary \ref{crl:#1}}
\newcommand{\bcjt}[1]{\begin{cjt}\label{cjt:#1}}
\newcommand{\ecjt}{\end{cjt}}

\newcommand{\bprf}{\begin{prf}}
\newcommand{\eprf}{\end{prf}}
\newcommand{\brmk}{\begin{rmk}}
\newcommand{\ermk}{\end{rmk}}
\newcommand{\laeq}[1]{\label{eq:#1}}
\newcommand{\req}[1]{(\ref{eq:#1})}

\newcommand{\QED}{\hfill$\blacksquare$}
\newcommand{\lsec}[1]{\label{sec:#1}}
\newcommand{\rsec}[1]{\ref{sec:#1}}
\newcommand{\rSec}[1]{Section \ref{sec:#1}}
\newcommand{\lapp}[1]{\label{app:#1}}

\newcommand{\rApp}[1]{Appendix \ref{app:#1}}

\newcommand{\bitem}{\begin{itemize}}
\newcommand{\entem}{\end{itemize}}
\newcommand{\benum}{\begin{enumerate}}
\newcommand{\ennum}{\end{enumerate}}

\newcommand{\otm}{\otimes}

\begin{document}

\title{One-Shot Randomized and Nonrandomized Partial Decoupling}

\author{Eyuri Wakakuwa}
\affiliation{Department of Communication Engineering and Informatics, Graduate School of Informatics and Engineering, The University of Electro-Communications, Tokyo 182-8585, Japan}

\author{Yoshifumi Nakata}
\affiliation{Photon Science Center, Graduate School of Engineering, The University of Tokyo, Bunkyo-ku, Tokyo 113-8656, Japan}
\affiliation{Yukawa Institute for Theoretical Physics, Kyoto university, Kitashirakawa Oiwakecho, Sakyo-ku, Kyoto, 606-8502, Japan}
\affiliation{JST, PRESTO, 4-1-8 Honcho, Kawaguchi, Saitama, 332-0012, Japan}

\begin{abstract}
We introduce a task that we call {\it partial decoupling}, in which a bipartite quantum state is transformed by a unitary operation on one of the two subsystems  and then is subject to the action of a quantum channel. 
We assume that the subsystem is decomposed into a direct-sum-product form, which often appears in the context of quantum information theory. The unitary is chosen at random from the set of unitaries having a simple form under the decomposition. 
The goal of the task is to make the final state, for typical choices of the unitary, close to the averaged final state over the unitaries. 
We consider a one-shot scenario, and derive upper and lower bounds on the average distance between the two states.
The bounds are represented simply in terms of smooth conditional entropies of quantum states involving the initial state, the channel and the decomposition.
Thereby we provide generalizations of the one-shot decoupling theorem.
The obtained result would lead to further development of the decoupling approaches in quantum information theory and fundamental physics.
\end{abstract}


\maketitle

\section{Introduction}

{\it Decoupling} refers to the fact that we may destroy correlation between two quantum systems by applying an operation on one of the two subsystems. It has played significant roles in the development of quantum Shannon theory for a decade, particularly in proving the quantum capacity theorem \cite{HHWY2008}, unifying various quantum coding theorems \cite{ADHW2009}, analyzing a multipartite quantum communication task \cite{horo05,HOW07} and in quantifying correlations in quantum states \cite{GPW2005,berta2018conditional}. 
It has also been applied to various fields of physics, such as the black hole information paradox \cite{HP2007}, quantum many-body systems \cite{brandao2015exponential} and quantum thermodynamics \cite{dRARDV2011, dRHRW2014}.  
Dupuis et al.~\cite{DBWR2010} provided one of the most general formulations of decoupling, which is often referred to as the {\it decoupling theorem}.
The decoupling approach simplifies many problems of our interest, mostly due to the fact that
any purification of a mixed quantum state is convertible to another reversibly \cite{uhlmann1976transition}.

All the above studies rely on the notion of {\it random unitary}, i.e., unitaries drawn at random from the set of all unitaries acting on the system, which leads to the full randomization over the whole Hilbert space. 
In various situations, however, the full randomization is a too strong demand. 
In the context of communication theory, for example, the full randomization leads to reliable transmission of {\it quantum} information, while we may be interested in sending {\it classical} information at the same time \cite{devetak2005capacity}, for which the full randomization is more than necessary. 
In the context of quantum many-body physics, the random process caused by the complexity of dynamics is in general restricted by symmetry, and thus no randomization occurs among different values of conserved quantities. 
Hence, in order that the random-unitary-based method fits into broader context in quantum information theory and fundamental physics, it would be desirable to generalize the previous studies using the full-random unitary, to those based on random unitaries that are not fully random but with a proper structure.

As the first step toward this goal, we consider a scenario in which the unitaries take a simple form under the following direct-sum-product (DSP) decomposition of the Hilbert space:
\begin{align}
{\ca H}=\bigoplus_{j=1}^J{\ca H}_j^l\otimes{\ca H}_j^r.
\laeq{decHilb}
\end{align}
Here, the superscripts $l$ and $r$ stand for ``left'' and ``right'', respectively, and $j$ is the index of the diagonal subspaces.
This decomposition often appears in the context of quantum information theory, such as information-preserving structure \cite{robin10,robin08}, the Koashi-Imoto decomposition \cite{koashi02}, data compression of quantum mixed-state source \cite{mixcomp1}, quantum Markov chains \cite{hayden04,wakakuwa2017markovianizing} and  simultaneous transmission of classical and quantum information \cite{devetak2005capacity}. 
Also, quantum systems with symmetry are represented by the Hilbert spaces decomposed into this form (see e.g.~\cite{bartlett07}),  in which case $j$ is the label of irreducible representations of a compact group $G$, ${\ca H}_j^l$ is the representation space and ${\ca H}_j^r$ is the multiplicity space for each $j$.

In this paper, we introduce and analyze a task that we call {\it partial decoupling}. 
We consider a scenario in which a bipartite quantum state $\Psi$ on system $AR$ is subject to a unitary operation $U$ on $A$, followed by the action of a quantum channel (CP map) $\ca{T}:A\rightarrow E$. 
The unitary is assumed to be chosen at random, not from the set of all unitaries on $A$, but from the subset of unitaries that take a simple form under the DSP decomposition. 
Thus, partial decoupling is a generalization of the decoupling theorem \cite{DBWR2010} that incorporates the DSP decomposition.
 Along the similar line as \cite{DBWR2010}, we analyze how close the final state $\ca{T}^{A \rightarrow E}  (U^A \Psi^{AR} U^{\dagger A})$ is, on average over the unitaries, to the averaged final state $\mbb{E}_{U} [ \ca{T}^{A \rightarrow E}(U^A ( \Psi^{AR} ) U^{\dagger A})]$.

The main result in this paper is that we derive  upper and lower bounds on the average distance between the final state and the averaged one. The bounds are represented in terms of the smooth conditional entropies of quantum states involving the initial state, the  channel and the decomposition. 
For a particular case where $J=1$ and $\dim{\ca{H}_j^{A_l}}=1$, the obtained formulae are equivalent to those given by the decoupling theorem \cite{DBWR2010}.

The result in this paper is applicable for generalizing any problems within the scope of the decoupling theorem by incorporating  the DSP structure.
Some of the applications are investigated in our papers \cite{wakakuwa2020randomized,nakata2020one,wakakuwa2020oneshothybrid,nakata2020black}.

In Refs.~\cite{wakakuwa2020randomized,wakakuwa2020oneshothybrid,nakata2020one}, we investigate communication tasks between two parties in which the information to be transmitted has both classical and quantum components.
In this case, the Hilbert space ${\ca H}_j^l$ in ~\req{decHilb} is assumed to be a one-dimensional space $\mbb{C}$, and ${\ca H}_j^l$ to be the spaces with the same dimension for all $j$:
\begin{equation}
{\ca H}=\bigoplus_{j=1}^J {\ca H}_j^r, \ \ \ \ \ \ \dim {\ca H}_j^r = \dim {\ca H}_{j'}^r \ (\forall j, j').
\end{equation}
Here, $j \in [1, J]$ and ${\cal H}_j^r$ correspond to the degrees of freedom related to classical and quantum components of the information to be transmitted, respectively.
We investigate the tasks of channel coding in \cite{wakakuwa2020randomized,nakata2020one} and source coding in \cite{wakakuwa2020oneshothybrid} in the one-shot regime.
Based on the result in this paper, we obtain general trade-off relations among the resources of classical communication, quantum communication and entanglement for those tasks.

In Ref.~\cite{nakata2020black}, we apply the result of partial decoupling to investigate the information paradox of quantum black holes with symmetry. 
Our analysis is based on the framework of Hayden-Preskill model~\cite{HP2007}, where a decoupling technique is used under the postulate that the internal dynamics of the system is given by a fully random unitary. 
This postulate should be modified when the system has symmetry since the dynamics cannot be fully random due to a conserved quantity. 
By letting $j$ be the labeling of the conserved quantity, the internal dynamics randomizes only the multiplicity spaces $\{ {\ca H}_j^r\}$ and should be in the form of 
\begin{align}
U=\bigoplus_{j=1}^J I_j^l \otimes U_j^r,
\end{align}
where $I_j^l$ is the identity on ${\ca H}_j^l$ and $U_j^r$ is a random unitary on ${\ca H}_j^r$. 
Hence, this case is also in the scope of partial decoupling with a DSP decomposition given by the symmetry. 
Similarly, all physical phenomena investigated based on decoupling~\cite{HP2007, brandao2015exponential, dRARDV2011, dRHRW2014,DBWR2010} can be lifted up by partial decoupling to the situation with symmetry.
We think that further significant implications on various topics will be obtained beyond these examples.

This paper is organized as follows. 
In \rSec{prelimi}, we introduce notations and definitions. 
In \rSec{mainresults}, we present formulations of the problem and the main results. 
Before we prove our main results, we provide discussions about implementations of our protocols by quantum circuits in \rSec{Discussions}. \rSec{STRPRF} describes the structure of the proofs of the main results, and provides lemmas that will be used in the proofs.
The detailed proofs of the main theorems are provided in Section \ref{sec:prfPD}-\ref{sec:converse}.
Conclusions are given in \rSec{cncl}.
Some technical lemmas and proofs are provided in Appendices.

\section{Preliminaries}
\lsec{prelimi}

We summarize notations and definitions that will be used throughout this paper. See also \rApp{listofnotations} for the list of notations.

\subsection{Notations}
We denote the set of linear operators and that of Hermitian operators on a Hilbert space $\ca{H}$ by $\ca{L}(\ca{H})$ and ${\rm Her}(\ca{H})$, respectively.
For positive semidefinite operators, density operators and sub-normalized density operators, we use the following notations, respectively:
\begin{align}
&
\ca{P}(\ca{H}) = \{\rho \in {\rm Her}(\ca{H}) : \rho \geq 0 \},
\\
&
\ca{S}_=(\ca{H}) = \{\rho \in \ca{P}(\ca{H}) : \tr [\rho]=1 \},
\\
&
\ca{S}_{\leq}(\ca{H}) = \{\rho \in \ca{P}(\ca{H}) : \tr [\rho] \leq 1 \}.
\end{align}
A Hilbert space associated with a quantum system $A$ is denoted by ${\mathcal H}^A$, and its dimension is denoted by $d_A$. A system composed of two subsystems $A$ and $B$ is denoted by $AB$. When $M$  and $N$ are linear operators on ${\mathcal H}^A$ and ${\mathcal H}^B$, respectively, we denote $M\otimes N$ as $M^A\otimes N^B$ for clarity. In the case of pure states, we often abbreviate $|\psi\rangle^A\otimes|\phi\rangle^B$ as $|\psi\rangle^A|\phi\rangle^B$. 
For $\rho^{AB} \in \ca{L}(\ca{H}^{AB})$, $\rho^{A}$ represents ${\rm Tr}_B[\rho^{AB}]$.  
We denote $|\psi\rangle\!\langle\psi|$ simply by $\psi$.
The maximally entangled state between $A$ and $A'$, where $\ca{H}^{A} \cong \ca{H}^{A'}$, is denoted by $\ket{\Phi}^{AA'}$ or $\Phi^{AA'}$.
The identity operator is denoted by $I$. 
We denote $(M^A\otimes I^B)\ket{\psi}^{AB}$ as $M^A\ket{\psi}^{AB}$, and $(M^A\otimes I^B)\rho^{AB}(M^A\otimes I^B)^{\dagger}$ as $M^A\rho^{AB}M^{A\dagger}$. 

When ${\mathcal E}$ is a supermap from $\ca{L}(\ca{H}^{A})$ to $\ca{L}(\ca{H}^{B})$, we denote it by $\ca{E}^{A \rightarrow B}$. When $A = B$, we use $\ca{E}^{A}$ for short.
We also denote $({\mathcal E}^{A \rightarrow B} \otimes{\rm id}^C)(\rho^{AC})$ by ${\mathcal E}^{A \rightarrow B} (\rho^{AC})$.  
The set of  linear completely-positive (CP) supermaps from $A$ to $B$ is denoted by $\ca{C}\ca{P}(A\rightarrow B)$, and the subset of trace non-increasing (resp. trace preserving) ones by $\ca{C}\ca{P}_\leq(A\rightarrow B)$ (resp. $\ca{C}\ca{P}_=(A\rightarrow B)$).
When a supermap is given by a conjugation of a unitary $U^A$ or an isometry $W^{A \rightarrow B}$, we especially denote it by its calligraphic font such as 
\begin{align}
\ca{U}^{A}(X^A):= (U^{A }) X^A (U^{A })^{\dagger},
\quad
\ca{W}^{A \rightarrow B}(X^A):= (W^{A \rightarrow B}) X^A (W^{A \rightarrow B})^{\dagger}.
\end{align}

Let $A$ be a quantum system such that the associated Hilbert space $\ca{H}^A$ is decomposed into the DSP form as
\begin{align}
{\ca H}^A=\bigoplus_{j=1}^J{\ca H}_j^{A_l}\otimes{\ca H}_j^{A_r}.
\laeq{decHilbA}
\end{align} 
For the dimension of each subspace, we introduce the following notation:
\alg{
l_j:=\dim{{\ca H}_j^{A_l}},\quad r_j:=\dim{{\ca H}_j^{A_r}}.
}
We denote by $\Pi_j^A$ the projection onto a subspace ${\ca H}_j^{A_l}\otimes{\ca H}_j^{A_r}\subseteq{\ca H}^A$ for each $j$. For any quantum system $R$ and any $X\in\ca{L}(\ca{H}^A\otm\ca{H}^R)$, we introduce a notation
\alg{
X_{jk}^{AR}:=\Pi_j^AX^{AR}\Pi_k^A,
\laeq{haihai}
}
which leads to 
$X^{AR}=\sum_{j,k=1}^JX_{jk}^{AR}$.

\subsection{Norms and Distances}

For a linear operator $X$, the trace norm is defined as $|\! | X |\! |_1 = \tr[ \sqrt{X^{\dagger}X}]$, and the Hilbert-Schmidt norm as $|\! | X |\! |_2 = \sqrt{\tr[ X^{\dagger}X]}$. 
The trace distance between two unnormalized states $\rho,\rho'\in\ca{P}(\ca{H})$ is defined by $\|\rho-\rho'\|_1$. 
For subnormalized states $\rho,\rho'\in\ca{S}_\leq(\ca{H})$, the generalized fidelity and the purified distance are defined by
\alg{
\bar{F}(\rho,\rho')
:=
\|\sqrt{\rho}\sqrt{\rho'}\|_1
+
\sqrt{(1-{\rm Tr}[\rho])(1-{\rm Tr}[\rho'])},
\quad
P(\rho,\rho')
:=
\sqrt{1-\bar{F}(\rho,\rho')^2},
\laeq{dfnPD}
}
respectively \cite{tomamichel2010duality}.
The epsilon ball of a subnormalized state $\rho\in\ca{S}_\leq(\ca{H})$ is defined by
\begin{align}
\ca{B}^\epsilon(\rho):=\{\rho'\in\ca{S}_\leq(\ca{H})|\:P(\rho,\rho')\leq\epsilon\}.
\label{eq:epsilon}
\end{align}

For a linear superoperator $\ca{E}^{A \rightarrow B}$, we define 
 the {\it DSP norm} by
\begin{align}
\| \ca{E}^{A \rightarrow B} \|_{\rm DSP}:= \sup_{C,\:\xi} \|  \ca{E}^{A \rightarrow B}(\xi^{AC}) \|_1,
\laeq{dfnDSPnorm}
\end{align}
where the supremum is taken over all finite dimensional quantum systems $C$ and all subnormalized states $\xi\in\ca{S}_\leq(\ca{H}^{AC})$ such that the reduced state on $A$ is decomposed in the form of
\begin{align}
\xi^A=\bigoplus_{j=1}^Jq_j \varpi_j^{A_l}\otimes \pi_j^{A_r}.
\end{align}
Here, $\{q_j\}_{j=1}^J$ is a probability distribution, $\{\varpi_j\}_{j=1}^J$ is a set of subnormalized states on $\ca{H}_j^{A_l}$ and $\pi_j^{A_r}$ is the maximally mixed state on $\ca{H}_j^{A_r}$.
The epsilon ball of linear CP maps 
with respect to the DSP norm is defined by
\begin{align}
\ca{B}_{\rm DSP}^\epsilon(\ca{E}):=
\{\ca{E}'\in\ca{C}\ca{P}_\leq(A\rightarrow B)\:|\:\|\ca{E}'-\ca{E}\|_{\rm DSP}\leq\epsilon\}.
\label{eq:epsilonTnorm}
\end{align}

For quantum systems $V$, $W$, a linear operator $X\in\ca{L}(\ca{H}^{VW})$ and a subnormalized state $\varsigma\in\ca{S}_\leq(\ca{H}^W)$, we introduce the following notation:
\alg{
|\!| X^{VW} |\!|_{2,\varsigma^W}:=|\!| (\varsigma^W)^{-1/4} X^{VW} (\varsigma^W)^{-1/4} |\!|_{2}.
\laeq{cond2on}
}
This includes the case where $V$ is a trivial (one-dimensional) system, in which case $X^{VW}=X^W$. We omit the superscript $W$ for $\varsigma$ when there is no fear of confusion.

\subsection{One-shot entropies}

For any subnormalized state $\rho\in\ca{S}_\leq(\ca{H}^{AB})$ and normalized state $\varsigma\in\ca{S}_=(\ca{H}^{B})$, define
\alg{
&
H_{\rm min}(A|B)_{\rho|\varsigma} := \sup \{ \lambda \in \mathbb{R}| 2^{-\lambda} I^A \otimes \varsigma^B \geq \rho^{AB} \}, \label{eq:condmins} \\
&
H_{\rm max}(A|B)_{\rho|\varsigma} := \log{\|\sqrt{\rho^{AB}}\sqrt{I^A\otm\varsigma^B}\|_1^2},
\label{eq:condmaxs}
\\
&
H_2(A|B)_{\rho|\varsigma} :=  - \log \tr \bigl[ \bigl(  (\varsigma^B)^{-1/4} \rho^{AB} (\varsigma^B)^{-1/4} \bigr)^2 \bigr].
\label{eq:cond2vs}
}
The conditional min-, max- and collision entropies (see e.g.~\cite{T16}) are defined by
\begin{align}
H_{\rm min}(A|B)_{\rho}& := \sup_{\varsigma^B \in \ca{S}_=(\ca{H}^B)}H_{\rm min}(A|B)_{\rho|\varsigma},\label{eq:condmin} \\
H_{\rm max}(A|B)_{\rho}& := \sup_{\varsigma^B \in \ca{S}_=(\ca{H}^B)}H_{\rm max}(A|B)_{\rho|\varsigma},\label{eq:condmax} \\
H_2(A|B)_{\rho} &:= \sup_{\varsigma^B \in \ca{S}_=(\ca{H}^B)}H_2(A|B)_{\rho|\varsigma},\label{eq:cond2}
\end{align}
respectively. The smoothed versions are of the key importance when we are interested in the one-shot scenario. We particularly use the smooth conditional min- and max-entropies:
\begin{align}
H_{\rm min}^\epsilon(A|B)_{\rho}& := \sup_{\hat{\rho}^{AB} \in \ca{B}^\epsilon(\rho)}H_{\rm min}(A|B)_{\hat\rho},\label{eq:condminsm} \\
H_{\rm max}^\epsilon(A|B)_{\rho}& := \inf_{\hat{\rho}^{AB} \in \ca{B}^\epsilon(\rho)}H_{\rm max}(A|B)_{\hat\rho}\label{eq:condmaxsm} 
\end{align}
for $\epsilon\geq0$.
Note that Expressions \req{condmins}-\req{cond2} can be generalized to the case where $\rho\in\ca{P}(\ca{H})$.

\subsection{Choi-Jamiolkowski representation}

Let $\ca{T}^{A \rightarrow B}$ be a linear supermap from  $\ca{L}(\ca{H}^A)$ to $\ca{L}(\ca{H}^B)$, and let $\Phi^{AA'}$ be the maximally entangled state between $A$ and $A'$. A linear operator $\mfk{J}(\ca{T}^{A \rightarrow B})\in\ca{L}(\ca{H}^{AB})$ defined by $\mfk{J}(\ca{T}^{A \rightarrow B}) := \ca{T}^{A' \rightarrow B}(\Phi^{AA'})$ is called the {\it Choi-Jamio\l kowski representation of $\ca{T}$} \cite{J1972,C1975}. The representation is an isomorphism. The inverse map is given by, for an operator $X^{AB} \in \ca{L}(\ca{H}^{AB})$,
\alg{
\mfk{J}^{-1}(X^{AB}) (\varsigma^A) = d_A \tr_A \bigl[ (\varsigma^{A^T} \otimes I^B) X^{AB} \bigr],
\laeq{CJinverse}
}
where $A^T$ denotes the transposition of $A$ with respect to the Schmidt basis of $\Phi^{AA'}$.
When $\ca{T}$ is completely positive, then $\mfk{J}(\ca{T}^{A \rightarrow B})$ is an unnormalized state on $AB$ and is called the {\it Choi-Jamio\l kowski state of $\ca{T}$}.

Note that the Choi-Jamio\l kowski representation depends on the choice of the maximally entangled state $\Phi^{AA'}$, i.e., the Schmidt basis thereof. When $\ca{H}^{A}$ is decomposed into the DSP form as \req{decHilbA}, the isomorphic space $\ca{H}^{A'}$ is decomposed into the same form. In the rest of this paper, we fix the maximally entangled state $\Phi^{AA'}$, which is decomposed as
\alg{
\ket{\Phi}^{AA'}=\bigoplus_{j=1}^J\sqrt{\frac{l_jr_j}{d_A}}|\Phi_j^l\rangle^{A_lA_l'}|\Phi_j^r\rangle^{A_rA_r'},
\laeq{dfnmaxentDSP}
}
where $\Phi_j^l$ and $\Phi_j^r$ are fixed maximally entangled states on $\ca{H}^{A_l}_j\otm\ca{H}^{A_l'}_j$ and $\ca{H}^{A_r}_j\otm\ca{H}^{A_r'}_j$, respectively.

\subsection{Random unitaries}

Random unitaries play a crucial role in the analyses of one-shot decoupling. By using them, it can be shown that there exists at least one unitary that achieves the desired task.
In particular, the Haar measure on the unitary group is often used. The Haar measure ${\sf H}$ on the unitary group is the unique unitarily invariant provability measure, often called uniform distribution of the unitary group. When a random unitary $U$ is chosen uniformly at random with respect to the Haar measure, it is referred to as a \emph{Haar random unitary} and is denoted by $U \sim {\sf H}$. 

The most important property of the Haar measure is the left- and right-unitary invariance: for a Haar random unitary $U \sim {\sf H}$ and any unitary $V$, the random unitaries $VU$ and $UV$ are both distributed uniformly with respect to the Haar measure. This property combined with the Schur-Weyl duality enables us to explicitly study the averages of many functions on the unitary group over the Haar measure.
In the following, the average of a function $f(U)$ on the unitary group over the Haar measure is denoted by $\mbb{E}_{U \sim {\sf H}} [f]$.

In this paper, however, we are interested in the case where the Hilbert space is decomposed into the DSP form: ${\ca H}^A=\bigoplus_{j=1}^J{\ca H}_j^{A_l}\otimes {\ca H}_j^{A_r}$, and mainly consider the unitaries that act non-trivially only on $\{ {\ca H}_j^{A_r} \}_{j=1}^{J}$ such as the untiary in the form of $\bigoplus_{j=1}^J I_j^{A_l}\otimes U_j^{A_r}$, where $U_j^{A_r}$ is a unitary on ${\ca H}_j^{A_r}$. In this case, we can naturally introduce a product ${\sf H}_{\times}$ of the Haar measures by
\begin{equation}
{\sf H}_{\times} = {\sf H}_{1} \times \dots \times {\sf H}_{J},
\end{equation}
where ${\sf H}_{j}$ is the Haar measure on the unitary group on ${\ca H}_j^{A_r}$ for any $j$. Hence, when we write $U \sim {\sf H}_{\times}$ below, it means that $U$ is in the form of $\bigoplus_{j=1}^J I_j^{A_l}\otimes U_j^{A_r}$ and $U_j^{A_r} \sim {\sf H}_j$.

\section{Main Results}\lsec{mainresults}

We consider two scenarios in which a bipartite quantum state $\Psi^{AR}$ is transformed by a unitary operation on $A$ and then is subject to the action of a quantum channel (linear CP map) $\ca{T}^{A\rightarrow E}$. 
The unitary is chosen at uniformly random from the set of unitaries that take a simple form under the DSP decomposition \req{decHilb}. 

In the first scenario, which we call {\it non-randomized partial decoupling}, the unitaries are such that they completely randomize the space $\ca{H}_j^{A_r}$ for each $j$, while having no effect on $j$ or the space $\ca{H}_j^{A_l}$. 
This scenario may find applications when complex quantum many-body systems are investigated based on the decoupling approach, in which case the DSP decomposition is, for instance, induced by the symmetry the system has.
In the second scenario, which we refer to as {\it randomized partial decoupling}, we assume that ${\rm dim}\ca{H}_j^{A_l}=1$ and that ${\rm dim}\ca{H}_j^{A_r}$ does not depend on $j$. The unitaries do not only completely randomize the space $\ca{H}^{A_r}$, but also randomly permute $j$. This scenario may fit to the communication problems. For instance, one of the applications may be classical-quantum hybrid communicational tasks, where the division of the classical and quantum information leads to the DSP decomposition.

For both scenarios, our concern is how close the final state is, after the action of the unitary and the quantum channel, to the averaged final state over all unitaries. 
It should be noted that the averaged final state is in the form of a block-wise decoupled state in general. This is in contrast to the decoupling theorem, in which the averaged final state is a fully decoupled state.

\subsection{Non-Randomized Partial Decoupling}
Let us consider the situation where $U$ has the DSP form: $U:=\bigoplus_{j=1}^J I_j^{A_l} \otimes U_j^{A_r}$. For any state $\Psi^{AR}$, the averaged state obtained after the action of the random unitary $U \sim {\sf H}_{\times}$ is given by
\begin{align}
\Psi_{\rm av}^{AR}
:=\mbb{E}_{U \sim {\sf H}_{\times}} [ 
U^A ( \Psi^{AR} ) U^{\dagger A}]
= \bigoplus_{j=1}^J \Psi_{jj}^{A_lR}\otimes \pi_j^{A_r}. \laeq{taisho}
\end{align}
Here, $\pi_j^{A_r}$ is the maximally mixed state on $\ca{H}_j^{A_r}$, and $\Psi_{jj}^{A_lR}$ is an unnormalized state on $\ca{H}_j^{A_l}\otm\ca{H}^R$ defined by
\alg{
\Psi_{jj}^{A_lR}:={\rm Tr}_{A_r}[\Psi_{jj}^{AR}]={\rm Tr}_{A_r}[\Pi_j^A\Psi^{AR}\Pi_j^A].
\laeq{spoon}
}
Our interest is on the average distance between the state $\ca{T}^{A \rightarrow E} (U^A \Psi^{AR} U^{\dagger A}) $ and the averaged state $\ca{T}^{A \rightarrow E}(\Psi_{\rm av}^{AR})$ over all $U \sim {\sf H}_{\times}$. 

For expressing the upper bound on the average distance, we introduce a quantum system $A^*$ represented by a Hilbert space 
\alg{
{\ca H}^{A^*}:=\bigoplus_{j=1}^J{\ca H}_j^{A_r}\otimes{\ca H}_j^{\bar{A}_r},
\laeq{decHAstas}
}
and a linear operator $F^{A\bar{A}\rightarrow A^*}:
\ca{H}^A\otm\ca{H}^{\bar{A}}
\rightarrow
{\ca H}^{A^*}$ defined by 
\begin{align}
&F^{A\bar{A}\rightarrow A^*}:=
\bigoplus_{j=1}^J \sqrt{\frac{d_Al_j}{r_j}} \langle\Phi_j^l|^{A_l\bar{A}_l}(\Pi_j^{A} \otimes \Pi_j^{\bar{A}}),
 \label{Eq:26}
\end{align}
where ${\ca H}_j^{\bar{A}_l}\cong{\ca H}_j^{A_l}$, ${\ca H}_j^{\bar{A}_r}\cong{\ca H}_j^{A_r}$ and $\ca{H}^{\bar{A}}\cong\ca{H}^{A}$.

The following is our first main theorem about the upper bound:

\bthm{SmoothMarkov}[{\bf Main result 1: One-shot non-randomized partial decoupling}]
For any $\epsilon,\mu\geq0$, any subnormalized state $\Psi^{AR} \in \ca{S}_\leq(\ca{H}^{AR})$ and any linear CP map $\ca{T}^{A \rightarrow E}$,
it holds that
\begin{align}
\mbb{E}_{U \sim {\sf H}_{\times}} 
\left[ \left\|
\ca{T}^{A \rightarrow E} \circ \ca{U}^A ( \Psi^{AR} ) 
- \ca{T}^{A \rightarrow E}(\Psi_{\rm av}^{AR})
\right\|_1 \right]
\leq  2^{-\frac{1}{2} H_{\rm min}^{\epsilon,\mu}(A^*|RE)_{{\Lambda}(\Psi,\ca{T})}}+2(\epsilon\|\ca{T}\|_{\rm DSP}+\mu+\epsilon\mu).\label{eq:smthMark}
\end{align}
Here, $H_{\rm min}^{\epsilon,\mu}(A^*|RE)_{{\Lambda}(\Psi,\ca{T})}$ is the smooth conditional min-entropy for an unnormalized state ${\Lambda}(\Psi,\ca{T})$, defined by $F(\Psi^{AR}\otimes\tau^{\bar{A}E})F^\dagger$ with $\tau^{AE} = \mfk{J}(\ca{T}^{A \rightarrow E})$ being the Choi-Jamio\l kowski representation of $\ca{T}^{A \rightarrow E}$. It is explicitly given by
\begin{align}
H_{\rm min}^{\epsilon,\mu}(A^*|RE)_{{\Lambda}(\Psi,\ca{T})}
:=\sup_{\Psi'\in\ca{B}^\epsilon(\Psi)}\sup_{\ca{T}'\in\ca{B}_{\rm DSP}^{\mu}(\ca{T})}
H_{\rm min}(A^*|RE)_{{\Lambda}(\Psi',\ca{T}')},\label{eq:dfnhhat2}
\end{align}
where $\ca{B}_{\rm DSP}^{\mu}(\ca{T})$ is the set of $\mu$-neighbourhoods of $\ca{T}$, defined by \req{epsilonTnorm}.
\ethm

In the literature of chaotic quantum many-body systems, it is often assumed that the dynamics is approximated well by a random unitary channel, which is sometimes called scrambling \cite{HP2007, SS2008, LSHOH2013}. Despite the fact that a number of novel research topics have been opened based on the idea of scrambling, some of which are using the decoupling approach \cite{HP2007, dRARDV2011, dRHRW2014}, symmetry of the physical systems has rarely been taken into account properly. 
When the system has symmetry, the associated Hilbert space is naturally decomposed into a DSP form as 
\begin{align}
{\ca H}^A=\bigoplus_{j=1}^J{\ca H}_j^{A_l}\otimes{\ca H}_j^{A_r},
\end{align}
where $j$ is the label of irreducible representations of a compact group of the symmetry, ${\ca H}_j^{A_l}$ is the irreducible representation and ${\ca H}_j^{A_r}$ corresponds to the multiplicity for each $j$. Due to the conservation law, the scrambling dynamics in the system should be compatible with this decomposition and should be in the form of $U^A=\bigoplus_{j=1}^J I_j^{A_l}\otimes U_j^{A_r}$.
Hence, Theorem \ref{thm:SmoothMarkov} is applicable to the study of complex physics in chaotic quantum many-body systems with symmetry.

Theorem \ref{thm:SmoothMarkov} reduces to a simpler form when the symmetry is abelian. In this case, all the irreducible representation one-dimensional, i.e., $\dim {\ca H}_j^{A_l} =1$. 
The averaged output state is explicitly calculated to be
\begin{equation}
\ca{T}^{A \rightarrow E}(\Psi_{\rm av}^{AR})=
\bigoplus_{j=1}^J \frac{d_A}{r_j} (\tr_A[ \Pi_j^{A} \Psi^{AR} ]) \otimes (\tr_{\bar{A}}[\Pi_j^{\bar{A}} \tau^{\bar{A}E}]).
\end{equation}
The operator $F^{A \bar{A} \rightarrow A^*}$ in~\eqref{Eq:26} reduces to a direct sum of operators that are proportional to projectors, and the operator $\Lambda(\Psi, {\cal T}) \in {\cal S}_{\leq}({\cal H}^{A^*RE})$ in Theorem \ref{thm:SmoothMarkov} reduces to 
\begin{equation}
\Lambda(\Psi, {\cal T})=
\bigoplus_{j, j'=1}^J \frac{d_A}{\sqrt{r_j r_{j'}}} (\Pi_j^{A} \Psi^{AR} \Pi_{j'}^{A}) \otimes (\Pi_j^{\bar{A}} \tau^{\bar{A}E} \Pi_{j'}^{\bar{A}}).
\end{equation}
Theorem \ref{thm:SmoothMarkov} implies that, if the smooth conditional min-entropy of the unnormalized state $\Lambda(\Psi, {\cal T})$ is sufficiently large, the final state $\ca{T}^{A \rightarrow E} \circ \ca{U}^A ( \Psi^{AR} )$ is close to $\ca{T}^{A \rightarrow E}(\Psi_{\rm av}^{AR})$.

\subsection{Randomized Partial Decoupling}

Next we assume that
\alg{
\dim\ca{H}_j^l=1,\quad\dim\ca{H}_j^r=r \quad(j=1,\cdots, J).
\laeq{condPDandCO}
} 
The Hilbert space $\ca{H}^A=\oplus_{j=1}^J{\mathcal H}_j^{A_r}$ is then isomorphic to a tensor product Hilbert space ${\ca H}^{A_c} \otimes{\ca H}^{A_r}$, i.e., $A\cong A_cA_r$.
Here, ${\mathcal H}^{A_c}$ is a $J$-dimensional Hilbert space with a fixed orthonormal basis $\{|j\rangle\}_{j=1}^J$, and ${\mathcal H}^{A_r}$ is an $r$-dimensional Hilbert space. 
We consider a random unitary $U$ on system $A$ of the form
\begin{align}
U:=\sum_{j=1}^J\outpro{j}{j}^{A_c}  \otimes U_j^{A_r},
\label{eq:RUrpd}
\end{align}
which we also denote by $U \sim {\sf H}_{\times}$.
In addition, let $\mbb{P}$ be the permutation group on $[1,\cdots,J]$, and ${\sf P}$ be the uniform distribution on $\mbb{P}$. We define a unitary $G_\sigma$ for any $\sigma\in\mbb{P}$ by
\begin{equation}
G_\sigma:=\sum_{j=1}^J\outpro{\sigma(j)}{j}^{A_c}  \otimes I^{A_r}.
\label{eq:RPrpd}
\end{equation}
We denote the supermap given by conjugation of $G_\sigma$ by the calligraphic font as $\ca{G}_\sigma(\cdot)=G_\sigma(\cdot)G_\sigma^\dagger$.
For the initial state, we use the notion of \emph{classically coherent} states, defined as follows:

\begin{dfn}[classically coherent states \cite{dupuis2014decoupling}]
Let $K_1$ and $K_2$ be $d$-dimensional quantum systems with fixed orthonormal bases $\{|k_1\rangle\}_{k_1=1}^d$ and $\{|k_2\rangle\}_{k_2=1}^d$, respectively, and let $W$ be a quantum system. An unnormalized state $\varrho\in{\ca P}(\ca{H}^{K_1K_2W})$ is said to be classically coherent in $K_1K_2$ if it satisfies $\varrho\ket{k}^{K_1}\ket{k'}^{K_2}=0$ for any $k\neq k'$, or equivalently, if $\varrho$ is in the form of
\begin{align}
\varrho^{K_1K_2W}=\sum_{k,k'=1}^d\outpro{k}{k'}^{K_1}\otimes\outpro{k}{k'}^{K_2}\otm \varrho_{kk'}^W,
\end{align}
where $\varrho_{kk'}\in\ca{L}(\ca{H}^{W})$ for each $k$ and $k'$.
\end{dfn}

We now provide our second main result:

\bthm{SmoothExMarkov}[{\bf Main result 2: One-shot randomized partial decoupling}]
Let $\epsilon,\mu\geq0$, $\Psi^{AR}$ be a subnormalized state that is classically coherent in $A_cR_c$, and $\ca{T}^{A \rightarrow E}$ be a linear CP map such that the Choi-Jamio\l kowski representation $\tau^{AE} = \mfk{J}(\ca{T}^{A\rightarrow E})$ satisfies ${\rm Tr}[\tau]\leq1$. It holds that
\begin{multline}
\mbb{E}_{U \sim {\sf H}_{\times}, \sigma\sim{\sf P} } \left[ \left\|
\ca{T}^{A \rightarrow E} \circ \ca{G}_\sigma^A  \circ \ca{U}^A ( \Psi^{AR} ) -\ca{T}^{A \rightarrow E} \circ \ca{G}_\sigma^A ( \Psi_{\rm av}^{AR} )
\right\|_1 \right]\\
\!\leq  
\sqrt{\alpha(J)}\cdot2^{-\frac{1}{2}H_I}
+\beta(A_r)\cdot2^{-\frac{1}{2}H_{I\!I}}
+4(\epsilon+\mu+\epsilon\mu), 
\laeq{SmExMama}
\end{multline}
where 
$\Psi_{\rm av}^{AR}:=\mbb{E}_{U \sim {\sf H}_{\times}} [ \ca{U}^A ( \Psi^{AR} )  ]$.
The function $\alpha(J)$ is $0$ for $J=1$ and $\frac{1}{J-1}$ for $J\geq2$, and
$\beta(A_r)$ is $0$ for ${\rm dim}\ca{H}^{A_r}=1$ and $1$ for ${\rm dim}\ca{H}^{A_r}\geq2$.
The exponents $H_I$ and $H_{I\!I}$ are given by
\alg{
H_I=
H_{\rm min}^\epsilon(A|R)_{\Psi}-H_{\rm max}^\mu(A|B)_{\ca{C}(\tau)},
\quad
H_{I\!I}=
H_{\rm min}^\epsilon(A|R)_{\ca{C}(\Psi)}-H_{\rm max}^\mu(A_r|BA_c)_{\ca{C}(\tau)}.
}
Here, $\ca{C}$ is the completely dephasing channel on $A_c$ with respect to the basis $\{|j\rangle\}_{j=1}^J$, and $\tau^{AB} = \mfk{J}(\ca{T}^{A\rightarrow B})$ is the Choi-Jamiolkowski representation of the complementary channel $\ca{T}^{A\rightarrow B}$ of $\ca{T}^{A\rightarrow E}$.
\ethm

\noindent
Note that, since the subnormalized state $\Psi^{AR}$ is classically coherent in $A_cR_c$, the averaged state $\Psi_{\rm av}^{AR}$ is explicitly given by
\begin{align}
\Psi_{\rm av}^{AR}=\sum_{j=1}^J\outpro{j}{j}^{A_c}\otimes\pi^{A_r}\otm\Psi_{jj}^{R_r}\otimes\outpro{j}{j}^{R_c}.\laeq{romanoffofo}
\end{align}

Small error for one-shot randomized partial decoupling implies that the third party having the purifying system of the final state may recover both classical and quantum parts of correlation in $\Psi^{AR}$. Thus, it will be applicable, e.g., for analyzing simultaneous transmission of classical and quantum information in the presence of quantum side information. In this context, $H_I$ in the above expression quantifies how well the total correlation in $\Psi^{AR}$ can be transmitted by the channel $\ca{T}^{A\rightarrow B}$, whereas $H_{I\!I}$ for only quantum part thereof (see  \cite{wakakuwa2020randomized,wakakuwa2020oneshothybrid,nakata2020one}).

\subsection{A Converse Bound}

So far, we have presented achievabilities of non-randomized and randomized partial decoupling. At this point, we do not know whether the obtained bounds are ``sufficiently tight''. To address this question, we prove a converse bound for partial decoupling. 
We assume the following two conditions for the converse:
\begin{itemize}
\setlength{\leftskip}{3.5cm}
\item[{\bf Converse Condition 1}] $\dim\ca{H}_j^l=1,\quad\dim\ca{H}_j^r=r \quad(j=1,\cdots, J)$,
\item[{\bf Converse Condition 2}] the initial (normalized) state $\Psi^{AR}$ is classically coherent in $A_cR_c$.
\end{itemize}
Throughout the paper, we refer to the conditions as {\bf CC1} and {\bf CC2}, respectively. 
The two conditions are always satisfied in the case of randomized partial decoupling, but not necessarily satisfied in the case of non-randomize one.
Consequently, the converse bound we prove below is directly applicable to randomized partial decoupling, but is not applicable to non-randomized partial decoupling in general.

The converse bound is stated by the following theorem.

\bthm{converse}[{\bf Main result 3: Converse for partial decoupling}] \label{Thm:converse}
Suppose that {\bf CC1} and {\bf CC2} are satisfied. Let $|\Psi\rangle^{ARD}$ be a purification of a normalized state $\Psi^{AR}\in\ca{S}_=(\ca{H}^{AR})$, which is classically coherent in $A_cR_c$ due to  {\bf CC2}, and $\ca{T}^{A \rightarrow E}$ be a trace preserving CP map with the complementary channel $\ca{T}^{A \rightarrow B}$.
Suppose that, for $\delta>0$, there exists a normalized state $\Omega^{ER}:=\sum_{j=1}^J\varsigma_j^E\otm\Psi_{jj}^{R_r}\otm\proj{j}^{R_c}$, where $\{\varsigma_j\}_{j=1}^J$ are normalized states on $E$, such that
\alg{
\left\|
\ca{T}^{A \rightarrow E} ( \Psi^{AR} ) -\Omega^{ER}
\right\|_1
\leq
\delta.
}
Then, for any $\upsilon\in[0,1/2)$ and $\iota\in(0,1]$, it holds that
\alg{
&
\!\!
H_{\rm min}^{\lambda}(A|R)_\Psi
-H_{\rm min}^{\upsilon}(RD|B)_{\ca{T}\circ\ca{C}(\Psi)}+\log{J}
\geq
\log{\iota},
\laeq{tokiyotomare}
\\
&
\!\!
H_{\rm min}^{\lambda'}(A|R)_{\ca{C}(\Psi)}
-H_{\rm min}^{\upsilon}(R_rD|BR_c)_{\ca{T}\circ\ca{C}(\Psi)}
\geq
\log{\iota}+\log{(1-2\upsilon)},
\laeq{tokiyougoke}
}
where $\ca{C}$ is the completely dephasing channel on $A_c$, and the smoothing parameters $\lambda$ and $\lambda'$ are defined by
\alg{
&
\lambda:=
2\sqrt{\iota+4\sqrt{20\upsilon+2\delta}}
+\sqrt{2\sqrt{20\upsilon+2\delta}}+2\sqrt{2\delta}
+2\sqrt{20\upsilon+2\delta}
+3\upsilon,
\laeq{dfnsmtlambda}
\\
&
\lambda':=
\upsilon+\sqrt{4\sqrt{\iota+2x}+2\sqrt{x}+(4\sqrt{\iota+8}+24) x
}
}
and $x:=\sqrt{2}\sqrt[4]{24\upsilon+2\delta}$.
\ethm

\noindent
Note that, when a quantum channel $\ca{T}^{A\rightarrow E}$ achieves partial decoupling for a state $\Psi^{AR}$ within a small error, it follows from the decomposition of $\Psi_{\rm av}$ (see \req{romanoffofo}) that
\alg{
\ca{T}^{A\rightarrow E}(\Psi^{AR})\approx\ca{T}^{A\rightarrow E}(\Psi_{\rm av}^{AR})=\sum_{j=1}^J\hat{\tau}_j^E\otm\Psi_{jj}^{R_r}\otm\proj{j}^{R_c},
} 
where $\hat{\tau}_j^E:=\ca{T}^{A\rightarrow E}(\proj{j}^{A_c}\otm\pi^{A_r})=(\ca{H}^E)$. This is in the same form as the assumption of \rThm{converse}.

Let us compare the direct part of randomized partial decoupling (\rThm{SmoothExMarkov}) and the converse bound presented above. The first term in the R.H.S. of the achievability bound \req{SmExMama} is calculated to be
\alg{
-2\log{\left(\sqrt{\alpha(J)}\cdot2^{-\frac{1}{2}H_I}\right)}
=
H_{\rm min}^\epsilon(A|R)_{\Psi}-H_{\rm max}^\mu(A|B)_{\ca{C}(\tau)}+\log{(J-1)}.
}
On the other hand,  the converse bound \req{tokiyotomare} yields
\alg{
H_{\rm min}^{\lambda}(A|R)_\Psi
-H_{\rm min}^\mu(A|B)_{\ca{C}(\psi)}+\log{J}
\geq
\log{\iota},
}
where $\psi^{AB}:=\ca{T}^{A'\rightarrow B}(\Psi_p^{AA'})$, with $|\Psi_p\rangle^{AA'}$ being a purification of $\Psi^A$ and $\ca{H}^A\cong\ca{H}^{A'}$.
Note that there exists a linear isometry from $A'$ to $RD$ that maps $|\Psi_p\rangle$ to $|\Psi\rangle$ \cite{uhlmann1976transition},
and that the conditional max entropy is invariant under local isometry (see \rLmm{invCEiso} below).
A similar argument also applies to the second term in \req{SmExMama} and \req{tokiyougoke}. Thus, when $\Psi^A$ is the maximally mixed state, in which case $|\Psi_p\rangle^{AA'}=|\Phi\rangle^{AA'}$ and thus $\psi=\tau$, the gap between the two bounds is only due to the difference in values of smoothing parameters and types of conditional entropies. By the fully quantum asymptotic equipartition property \cite{tomamichel2009fully}, this gap vanishes in the limit of infinitely many copies. From this viewpoint, we conclude that the achievability bound of randomized partial decoupling and the converse bound are sufficiently tight.

\subsection{Reduction to The Existing Results}

We briefly show that the existing results on one-shot decoupling \cite{DBWR2010} and dequantization \cite{dupuis2014decoupling} are obtained from Theorems \rthm{SmoothMarkov}, \rthm{SmoothExMarkov} and \ref{Thm:converse} as corollaries, up to changes in smoothing parameters. Thus, our results are indeed generalizations of these two tasks.

First, by letting $J=1$ in \rThm{SmoothExMarkov}, we obtain the achievability of one-shot decoupling:

\bcrl{OSdecach}[{\bf Achievability for one-shot decoupling (Theorem 3.1 in \cite{DBWR2010})}]
Let $\epsilon,\mu\geq0$, $\Psi^{AR}$ be a subnormalized state, and $\ca{T}^{A \rightarrow E}$ be a linear CP map such that the Choi-Jamio\l kowski representation $\tau^{AE} = \mfk{J}(\ca{T}^{A\rightarrow E})$ satisfies ${\rm Tr}[\tau]\leq1$. Let $U\sim {\sf H}$ be the Haar random unitary on $\ca{H}^A$. Then, it holds that
\begin{align}
\mbb{E}_{U \sim {\sf H}} \left[ \left\|
\ca{T}^{A \rightarrow E}\circ \ca{U}^A ( \Psi^{AR} ) -\tau^E\otm  \Psi^{R}
\right\|_1 \right]
\leq  
2^{-\frac{1}{2}[H_{\rm min}^\epsilon(A|R)_{\Psi}+H_{\rm min}^\mu(A|E)_{\tau}]}
+4(\epsilon+\mu+\epsilon\mu). 
\end{align}
\ecrl

\noindent
Note that the duality of the conditional min and max entropies (\cite{tomamichel2010duality}: see also \rLmm{duality} in \rSec{propCE}) implies $H_{\rm min}^\mu(A|E)_{\tau}=-H_{\rm max}^\mu(A|B)_{\tau}$, with $\tau^{AB} = \mfk{J}(\ca{T}^{A\rightarrow B})$ being the Choi-Jamiolkowski representation of the complementary channel $\ca{T}^{A\rightarrow B}$ of $\ca{T}^{A\rightarrow E}$.
A similar bound is also obtained by letting $J=1$ and ${\rm dim}\ca{H}_j^{A_l}=1$ in \rThm{SmoothMarkov}.
A converse bound for one-shot decoupling is obtained by letting $J=1$ in Theorem \ref{Thm:converse}, and by using the duality of the conditional entropies, as follows:

\bcrl{OSdecconverse}[{\bf Converse for one-shot decoupling (Theorem 4.1 in \cite{DBWR2010})}]
Consider a normalized state $\Psi^{AR}\in\ca{S}_=(\ca{H}^{AR})$ and a trace preserving CP map $\ca{T}^{A \rightarrow E}$.
Suppose that, for $\delta>0$, there exists a normalized state $\varsigma\in\ca{S}_=(\ca{H}^E)$, such that
$
\|
\ca{T}^{A \rightarrow E} ( \Psi^{AR} ) -\varsigma^E\otm\Psi^R
\|_1
\leq
\delta
$.
Then, for any $\upsilon\in[0,1/2)$ and $\iota\in(0,1]$, it holds that
\alg{
&
\!\!
H_{\rm min}^{\lambda}(A|R)_\Psi
+H_{\rm max}^{\upsilon}(A|E)_{\ca{T}^{A'\rightarrow E}(\Psi_p^{AA'})}
\geq
\log{\iota},
}
where $|\Psi_p\rangle^{AA'}$ is a purification of $\Psi^A$, $\ca{H}^A\cong\ca{H}^{A'}$, and
the smoothing parameter $\lambda$ is defined by \req{dfnsmtlambda}.
\ecrl

Next, we consider the opposite extreme for Theorem \rthm{SmoothExMarkov}, i.e., we consider the case where ${\rm dim}\ca{H}^{A_r}=1$. This case yields the {\it dequantizing theorem}: 

\bcrl{OSdeqach}[{\bf Achievability for dequantization (Theorem 3.1 in \cite{dupuis2014decoupling})}]
Let $A$ be a quantum system with a fixed basis $\{|j\rangle\}_{j=1}^{d_A}$, $\ca{H}^R\cong\ca{H}^A$ and 
$\epsilon,\mu\geq0$. Consider a subnormalized state $\Psi^{AR}$ that is classically coherent in $AR$, and a linear CP map $\ca{T}^{A \rightarrow E}$ such that the Choi-Jamio\l kowski representation $\tau^{AE} = \mfk{J}(\ca{T}^{A\rightarrow E})$ satisfies ${\rm Tr}[\tau]\leq1$. 
Let $\sigma$ be the random permutation on $[1,\cdots,d_A]$ with the associated unitary $G_\sigma:=\sum_{j=1}^{d_A}\outpro{\sigma(j)}{j}$.
Then, it holds that
\begin{multline}
\mbb{E}_{\sigma\sim{\sf P} } \left[ \left\|
\ca{T}^{A \rightarrow E} \circ \ca{G}_\sigma^A ( \Psi^{AR} ) -\ca{T}^{A \rightarrow E} \circ \ca{G}_\sigma^A \circ  \ca{C}^A( \Psi^{AR} )
\right\|_1 \right]\\
\leq  
\frac{1}{\sqrt{d_A-1}}\cdot2^{-\frac{1}{2}[H_{\rm min}^\epsilon(A|R)_{\Psi}-H_{\rm max}^\mu(A|B)_{\ca{C}(\tau)}]}
+4(\epsilon+\mu+\epsilon\mu), 
\end{multline}
where $\ca{C}$ is the completely dephasing channel on $A$ with respect to the basis $\{|j\rangle\}_{j=1}^J$,
 and $\tau^{AB} = \mfk{J}(\ca{T}^{A\rightarrow B})$ is the Choi-Jamiolkowski representation of the complementary channel $\ca{T}^{A\rightarrow B}$ of $\ca{T}^{A\rightarrow E}$.
\ecrl

\noindent
In the same extreme, Theorem \ref{Thm:converse} provides a converse bound for dequantization, which has not been known so far:

\bcrl{OSdeqconverse}[{\bf Converse for dequantization}]
Consider the same setting as in \rCrl{OSdeqach}, and assume that $\Psi^{AR}$ is normalized, and that $\ca{T}^{A \rightarrow E}$ is trace preserving. Let $|\Psi\rangle^{ARD}$ be a purification of $\Psi^{AR}$.
Suppose that, for $\delta>0$, there exists a normalized state $\Omega^{ER}:=\sum_{j=1}^Jp_j\varsigma_j^E\otm\proj{j}^{R}$, where $\{p_j,\varsigma_j\}_{j=1}^J$ is an ensemble of normalized states on $E$, such that
$
\|
\ca{T}^{A \rightarrow E} ( \Psi^{AR} ) -\Omega^{ER}
\|_1
\leq
\delta
$.
Then, for any $\upsilon\in[0,1/2)$ and $\iota\in(0,1]$, it holds that
\alg{
&
\!\!
H_{\rm min}^{\lambda}(A|R)_\Psi
-H_{\rm min}^{\upsilon}(RD|B)_{\ca{T}\circ\ca{C}(\Psi)}+\log{J}
\geq
\log{\iota},
}
where the smoothing parameter $\lambda$ is defined by \req{dfnsmtlambda}.
\ecrl

\section{Implementing the random unitary with the DSP form} \lsec{Discussions}

Before we proceed to the proofs, we here briefly discuss how the random unitaries $U \sim {\sf H}_{\times}$ that respect the DSP form can be implemented by quantum circuits. Since Haar random unitaries are in general hard to implement, unitary $t$-designs, mimicking the $t$-th statistical moments of the Haar measure on average~\cite{DLT2002,DCEL2009,GAE2007}, have been exploited in many cases. Since the decoupling method makes use of the second statistical moments of the Haar measure, we could use the unitary $2$-designs instead of the Haar measure for our tasks. Although a number of efficient implementations of unitary $2$-designs have been discovered~\cite{DLT2002,BWV2008a,WBV2008,GAE2007,DCEL2009,HL2009,DJ2011,CLLW2015,NHMW2015-1}, 
and it is also shown that decoupling can be achieved using unitaries less random than unitary 2-designs~\cite{BF2013, NHMW2015-2}, we here need unitary designs in a given DSP form, which we refer to as the \emph{DSP unitary designs}. Thus, we cannot directly use the existing constructions, posing a new problem about efficient implementations of DSP unitary designs. Although this problem is out of the scope in this paper, we will briefly discuss possible directions toward the solution.

One possible way is to simply modify the constructions of unitary designs known so far. This could be done by regarding each Hilbert space $\mc{H}^{A_r}_j$, on which each random unitary $U_j^{A_r} \sim {\sf H}_j$ acts, as the Hilbert space of ``virtual'' qubits. The complexity of the implementation, i.e. the number of quantum gates, is then determined by how complicated the unitary is that transforms the basis in each $\mc{H}^{A_r}_j$ into the standard basis of the virtual qubits.
Another way is to use the implementation of designs on one \emph{qudit}~\cite{NHKW2017}, where it was shown that alternate applications of random diagonal unitaries in two complementary bases achieves unitary designs. This implementation would be suited in quantum many-body systems because we can choose two natural bases, position and momentum bases, and just repeat switching random potentials in those bases under the condition that the potentials satisfy the DSP form. 
Finally, when the symmetry-induced DSP form is our concern, unitary designs with symmetry may possibly be implementable by applying random quantum gates that respects the symmetry. 

In any case, the implementations of DSP unitary designs, or the symmetric unitary designs, and their efficiency are left fully open. Further analyses are desired.

\begin{figure}[tb!]
\centering
\includegraphics[bb={0 0 1180 582},scale=0.31]{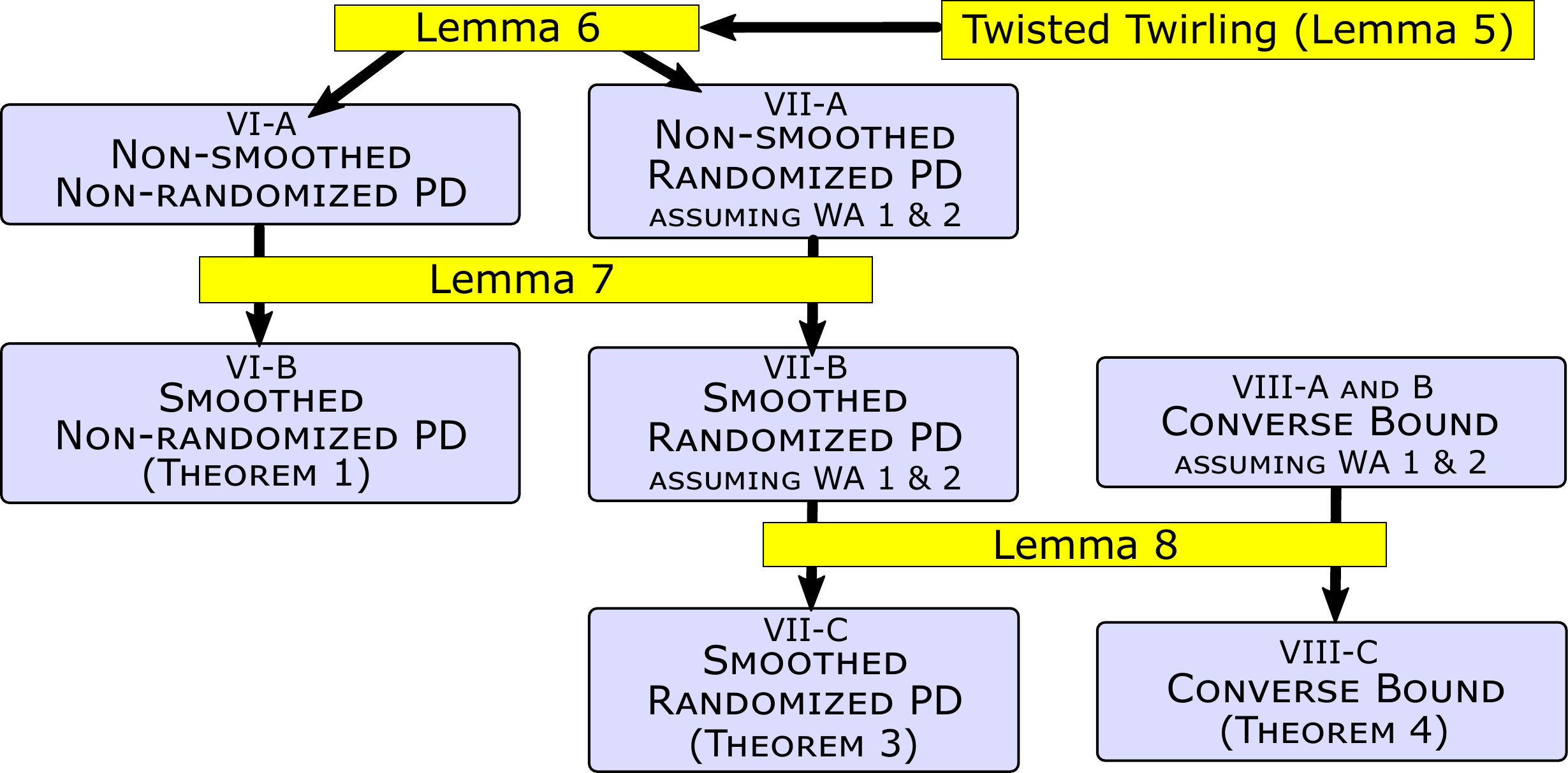}
\caption{Outline of our proofs. The PD stands for the partial decoupling. For the smoothed randomized partial decoupling and the converse bound, we first assume two conditions, {\bf WA 1} and {\bf WA 2}, but will remove them later to complete the proof. The details of the conditions are given in the main text.}
\label{Fig:outline}
\end{figure}

\section{Structure of the Proof}
\lsec{STRPRF}
In the rest of the paper, we prove the three main theorems, Theorem \ref{thm:SmoothMarkov}, \rThm{SmoothExMarkov} and \rThm{converse} in \rSec{prfPD}, \rSec{prfRPD} and \rSec{converse}, respectively. 
For the sake of clarity, we sketch the outline of the proofs in Subsection~\ref{SS:KLSP} (see also Fig.~\ref{Fig:outline}). We then list useful lemmas in Subsection~\ref{SS:LUL}. See also \rApp{listofnotations} for the list of notations used in the proofs.

\subsection{Key lemmas and the structure of the proofs} \label{SS:KLSP}

For the achievability statements (Theorem \ref{thm:SmoothMarkov} and \rThm{SmoothExMarkov}), the key technical lemma is the \emph{twisted twirling}, which can be seen as a generalization of the twirling method often used in quantum information science. See \rApp{TT} for the proof. 

\begin{lmm}[Twisted Twirling]\label{Lemma:av}
Let $\ca{H}_j^{A_r}$ be a $r_j$-dimensional subspace of $\ca{H}^{A_r}$, and $\Pi_j^{A_r}$ be the projector onto $\ca{H}_j^{A_r}\subset\ca{H}^{A_r}$ for each of $j=1,\cdots,J$.
Let $\mbb{I}^{A_rA_r'}$ be $I^{A_r} \otimes I^{A_r'}$, and $\mbb{F}^{A_rA_r'} \in \ca{L}(\ca{H}^{A_rA_r'})$ be the swap operator defined by $\sum_{a,b} |a\rangle \langle b|^{A_r} \otimes |b\rangle \langle a|^{A_r'}$ for any orthonormal basis $\{ \ket{a} \}$ in $\mathcal{H}^{A_r}$ and $\mathcal{H}^{A_r'}$.
In addition, let $\mbb{I}_{jk}^{A_rA_r'}$ and $\mbb{F}_{jk}^{A_rA_r'}$ be $\Pi_j^{A_r} \otimes \Pi_k^{A_r'}$ and $( \Pi_j^{A_r} \otimes \Pi_k^{A_r'})\mbb{F}^{A_rA_r'}$, respectively.
For any $M^{A_rA_r'BB'}\in\ca{L}(\ca{H}^{A_rA_r'BB'})$, define 
\alg{
M^{BB'}_{\mbb{I},jk}:=\tr_{A_rA_r'}[\mbb{I}_{jk}^{A_rA_r'}M^{A_rA_r'BB'}],
\quad
M^{BB'}_{\mbb{F},kj}:=\tr_{A_rA_r'}[\mbb{F}_{kj}^{A_rA_r'}M^{A_rA_r'BB'}].
}
Then, it holds that, for $j \neq k$,
\begin{align}
&\mbb{E}_{U_j \sim {\sf H}_j,U_k \sim {\sf H}_k} \bigl[ (U_j^{A_r} \otimes U_k^{A_r'}) M^{A_rA_r'BB'} (U_j^{A_r} \otimes  U_k^{A_r'})^{\dagger} \bigr]
= \frac{\mbb{I}_{jk}^{A_rA_r'}}{r_jr_k} \otimes M_{\mbb{I},jk}^{BB'},\\
&\mbb{E}_{U_j \sim {\sf H}_j,U_k \sim {\sf H}_k} \bigl[ (U_j^{A_r} \otimes U_k^{A_r'}) M^{A_rA_r'BB'} (U_k^{A_r} \otimes  U_j^{A_r'})^{\dagger} \bigr]
= \frac{\mbb{F}_{jk}^{A_rA_r'}}{r_jr_k} \otimes  M^{BB'}_{\mbb{F},kj}.
\end{align}
Moreover, 
\begin{multline}
\mbb{E}_{U_j \sim {\sf H}_j} \bigl[ (U_j^{A_r} \otimes U_j^{A_r'}) M^{A_rA_r'BB'} (U_j^{A_r} \otimes  U_j^{A_r'})^{\dagger} \bigr]\\
=\frac{1}{r_j (r_j^2-1)}
\left[
(r_j\mbb{I}_{jj}^{A_rA_r'}- \mbb{F}_{jj}^{A_rA_r'})\otimes M_{\mbb{I},jj}^{BB'} 
+
(r_j\mbb{F}_{jj}^{A_rA_r'} -\mbb{I}_{jj}^{A_rA_r'})\otimes M^{BB'}_{\mbb{F},jj} 
\right].
\end{multline}
Otherwise,
$\mbb{E}_{U_j,U_k,U_m,U_n} \bigl[ (U_j^{A_r} \otimes U_k^{A_r'}) M^{A_rA_r'BB'} (U_m^{A_r} \otimes  U_n^{A_r'})^{\dagger} \bigr]=0.$
\end{lmm}

The twisted twirling enables us to show the following lemma (see \rApp{mother}).

\begin{lmm}
\label{lmm:nenchalu}
For any $\varsigma^{ER} \in \ca{S}_=(\ca{H}^{ER})$ and any $X\in{\rm Her}(\ca{H}^{AR})$ such that
$X_{jj}^{A_lR}=0$, the following inequality holds for any possible permutation $\sigma \in \mbb{P}$:
\begin{align}
\mbb{E}_{U \sim {\sf H}_{\times}} \left[\left\|\ca{T}^{A \rightarrow E} \circ\ca{G}_{\sigma^{-1}}^A  \circ \ca{U}^A ( X^{AR} )\right\|_{2,\varsigma^{ER}}^2\right]
\leq 
\sum_{j,k=1}^J \frac{d_A^2}{r_j r_k}
\left\|   \tr_{A_l}\left[X_{\sigma(j)\sigma(k)}^{A_l^T A_r R}\tau^{A_l \bar{A}_r E}_{jk} \right]  \right\|^2_{2, \varsigma^{ER}} 
.
\end{align}
Here, $A_l^T$ denotes the transposition of $A_l$ with respect to the Schmidt basis of the maximally entangled state $|\Phi_j^l\rangle^{A_lA_l'}$ in \req{dfnmaxentDSP}, and the norm in the R.H.S. is defined by \req{cond2on}.
\end{lmm}

\noindent
Based on this lemma, we can prove the non-smoothed versions of Theorem \ref{thm:SmoothMarkov} and \rThm{SmoothExMarkov} in Subsections~\ref{sec:prfNSPD} and~\ref{SS:prNSRPD}, respectively.

To complete the proofs of Theorem \ref{thm:SmoothMarkov} and \rThm{SmoothExMarkov}, smoothing the statements is needed, which is done in Subsections~\ref{SS:prSNRPD} and~\ref{SS:SRPD} based on the following lemma proven in \rApp{trianglemother}.

\blmm{trianglemother}
Consider arbitrary unnormalized states $\Psi^{AR},\hat{\Psi}^{AR}\in\ca{P}(\ca{H}^{AR})$ and arbitrary CP maps $\ca{T},\hat{\ca{T}}:A\rightarrow E$.  
Let $\ca{D}_+^{A \rightarrow E}$ and $\ca{D}_-^{A \rightarrow E}$ be arbitrary CP maps such that $\ca{T}-\hat{\ca{T}}=\ca{D}_+-\ca{D}_-$.
Let $\delta_+^{AR}$ and $\delta_-^{AR}$ be linear operators on $\ca{H}^A\otimes\ca{H}^{R}$, such that
\alg{
\delta_+^{AR}\geq0,\quad\delta_-^{AR}\geq0,
\quad
{\rm supp}[\delta_+^{AR}]\perp{\rm supp}[\delta_-^{AR}]
}
and that
\alg{
\hat{\Psi}^{AR} -\Psi^{AR}=\delta_+^{AR}-\delta_-^{AR}.
}
The following inequality holds for any possible permutation $\sigma \in \mbb{P}$ and for both ${\Psi}_*={\Psi}_{\rm av}$ and ${\Psi}_*=\ca{C}^A(\Psi)$:
\begin{multline}
\mbb{E}_{U \sim {\sf H}_{\times}}\left[
\left\|{\ca{T}}^{A \rightarrow E}  \circ \ca{G}_\sigma^A \circ \ca{U}^A ( {\Psi}^{AR}  - {\Psi}_*^{AR} )\right\|_1
\right]\\
\leq
\mbb{E}_{U \sim {\sf H}_{\times}}\left[
\left\|\hat{\ca{T}}^{A \rightarrow E} \circ \ca{G}_\sigma^A \circ \ca{U}^A ( \hat{\Psi}^{AR}  -  \hat{\Psi}_*^{AR} )\right\|_1
\right]+2 \:
\tr[(\ca{D}_+^{A \rightarrow E}+ \ca{D}_-^{A \rightarrow E}) \circ \ca{G}_\sigma^A   ( \Psi_{\rm av}^{AR} )]
\\
+2 \:\mbb{E}_{U \sim {\sf H}_{\times}}\tr[\hat{\ca{T}}^{A \rightarrow E}  \circ \ca{G}_\sigma^A \circ\ca{U}^A (\delta_+^{AR}+\delta_-^{AR})].
\laeq{pracap}
\end{multline}
Here, $\hat{\Psi}_*=\mbb{E}_{U\sim{\sf H}_\times}[\ca{U}^A(\hat{\Psi}^{AR})]$ for ${\Psi}_*={\Psi}_{\rm av}$ and $\hat{\Psi}_*=\ca{C}^A(\hat{\Psi})$ for ${\Psi}_*=\ca{C}^A(\Psi)$.
\elmm

The converse statements are proved independently in Section~\rsec{converse}.\\

When we prove the one-shot randomized decoupling theorem (Theorem \rthm{SmoothExMarkov}) and the converse (Theorem \rthm{converse}), we first put the following two working assumptions:
\begin{description}
\item[{\bf WA 1}]  $E\cong E_cE_r$, where $E_c$ is a quantum system of dimension $J$. \vspace{-3mm}
\item[{\bf WA 2}] The CP map $\ca{T}^{A \rightarrow E}$ is decomposed into
\begin{align}
\ca{T}^{A \rightarrow E}(X)=\sum_{j,k=1}^J\outpro{j}{k}^{E_c}\otimes\ca{T}_{jk}^{A_r \rightarrow E_r}(X_{jk}),
\end{align}
in which $\ca{T}_{jk}$ is a linear supermap from $\ca{L}(\ca{H}^{A_r})$ to $\ca{L}(\ca{H}^{E_r})$ defined by $\ca{T}_{jk}(\zeta)=\ca{T}(\outpro{j}{k}\otimes\zeta)$ for each $j,k$. 
\end{description}
These assumptions are finally dropped in Subsections~\ref{SS:DA1} and \ref{SS:DA2} using the following lemma (see \rApp{prflmmC} for a proof).

\blmm{dropassump}
Let $\ca{T}^{A\rightarrow E}$ be a linear CP map that does not necessarily satisfies {\bf WA 1} and {\bf WA 2}.
By introducing a quantum system $E_c$ with dimension $J$, define an isometry $Y^{A_c \rightarrow A_c E_c}:=\sum_{j}\ket{jj}^{A_cE_c}\bra{j}^{A_c}$, and a linear map $\check{\ca{T}}^{A \rightarrow EE_c}$ by $\ca{T}^{A \rightarrow E} \circ \mathcal{Y}^{A_c \rightarrow A_c E_c}$.
Then, $\check{\ca{T}}^{A \rightarrow EE_c}$ is a linear CP map and, for any $\Psi^{AR}$ that is classically coherent in $A_cR_c$, the following equalities hold:
\begin{align}
&\left\|\check{\ca{T}}^{A \rightarrow EE_c}  ( {\Psi}^{AR}  -  \Psi_{\rm av}^{AR} )\right\|_1
=
\left\|\ca{T}^{A \rightarrow E}  ( {\Psi}^{AR}  -  \Psi_{\rm av}^{AR} )\right\|_1,\\
&\left\|\check{\ca{T}}^{A \rightarrow EE_c}  \circ \ca{G}_\sigma^A \circ \ca{U}^A ( {\Psi}^{AR}  -  \Psi_{\rm av}^{AR} )\right\|_1
=
\left\|\ca{T}^{A \rightarrow E}  \circ \ca{G}_\sigma^A \circ \ca{U}^A ( {\Psi}^{AR}  -  \Psi_{\rm av}^{AR} )\right\|_1.
\end{align}
\elmm

\subsection{List of useful lemmas} \label{SS:LUL}

We here provide several useful lemmas, some of which are in common with those in the proof of the one-shot decoupling theorem~\cite{DBWR2010}. Proofs of Lemmas \ref{lmm:propPD2}--\rlmm{projHSN} and \ref{lmm:condmaxCQCQ}--\ref{lmm:inprothree} will be provided in \rApp{prflmma}.

\subsubsection{Properties of Norms and Distances}

\begin{lmm}[Lemma 3.6 in  \cite{DBWR2010}] \label{Lemma:purity}
For any $\xi^{AB} \in {\rm Her}(\ca{H}^{AB})$,
$|\!|\xi^{AB}|\!|_2 \leq \sqrt{d_A}|\!|\xi^B|\!|_2$.
\end{lmm}

\begin{lmm}[Lemma 3.7 in  \cite{DBWR2010}]
\label{lmm:holder}
For any $X\in{\rm Her}(\ca{H})$ and $\gamma\in\ca{P}(\ca{H})$, it holds that
\begin{align}
\left\|X\right\|_1
\leq \sqrt{{\rm Tr}[\gamma]} \left\|X\right\|_{2, \gamma}
= \sqrt{{\rm Tr}[\gamma]\cdot{\rm Tr}[(\gamma^{-1/4}X\gamma^{-1/4})^2]}.
\label{eq:holdertype}
\end{align}
\end{lmm}

\begin{lmm}[Sec.~I\!I in  \cite{tomamichel2010duality}]
\label{lmm:propPD}
The purified distance defined by \req{dfnPD} satisfies the following properties:
\benum

\item{\it triangle inequality:}
For any $\rho,\varsigma,\tau\in\ca{S}_\leq(\ca{H})$, it holds that $P(\rho,\varsigma)\leq P(\rho,\tau)+P(\tau,\varsigma)$. \vspace{-3mm}
\item{\it monotonicity:} 
For any $\rho,\varsigma\in\ca{S}_\leq(\ca{H})$ and trace-nonincreasing CP map $\ca{E}$, it holds that $P(\rho,\varsigma)\geq P(\ca{E}(\rho),\ca{E}(\varsigma))$.\vspace{-3mm}
\item{\it Uhlmann's theorem:}
For any $\rho,\varsigma\in\ca{S}_\leq(\ca{H})$ and any purification $|\varphi_\rho\rangle\in\ca{H}\otm\ca{H}'$ of $\rho$, where $\ca{H}'\cong\ca{H}$, there exists a purification $|\varphi_\varsigma\rangle\in\ca{H}\otm\ca{H}'$ of $\varsigma$ such that $P(\rho,\varsigma)=P(\varphi_\rho,\varphi_\varsigma)$.
\ennum
\elmm

\begin{lmm}
\label{lmm:propPD2}
The purified distance defined by \req{dfnPD} satisfies the following properties:
\benum

\item{\it pure states:}
For any subnormalized pure state $|\psi\rangle\in\ca{H}$ and any normalized pure state $|\phi\rangle\in\ca{H}$, $P(\psi,\phi)=\sqrt{1-|\langle\psi|\phi\rangle|^2}$.\vspace{-3mm}

\item{\it relation to the trace distance:}
For any $\rho,\varsigma\in\ca{S}_\leq(\ca{H})$, $
\frac{1}{2}\|\rho-\varsigma\|_1
\leq
P(\rho,\varsigma)
\leq
\sqrt{2\|\rho-\varsigma\|_1}$.\vspace{-3mm}

\item{\it Inequality for subnormalized pure states:}
For any subnormalized pure states $\ket{\psi},\ket{\phi}\in\ca{H}$, 
$P(\psi,\phi)
\leq
\sqrt{1-|\inpro{\psi}{\phi}|^2}
+
\sqrt{1-\inpro{\phi}{\phi}}$.
\ennum
\elmm

\blmm{kazuhito}
Let $\{p_k\}_k$ be a normalized probability distribution, $\{\rho_k\}_k$ be a set of normalized states on $AB$, and $\{\hat{\rho}_k\}_k$ be that of subnormalized ones. For $\rho^{ABK}:=\sum_kp_k\rho_k^{AB}\otm\proj{k}^K$ and $\hat{\rho}^{ABK}:=\sum_kp_k\hat{\rho}_k^{AB}\otm\proj{k}^K$, the purified distance satisfies
\alg{
P(\rho^{ABK},\hat{\rho}^{ABK})\leq\sqrt{2\sum_kp_kP(\rho_k^{AB},\hat{\rho}_k^{AB})}.
}
\elmm

\blmm{kazufusa}
Let $\{p_k\}_k$ and $\{q_k\}_k$ be subnormalized probability distributions, and $\{\rho_k\}_k$ and $\{\varsigma_k\}_k$ be sets of normalized states on $A$. For $\rho^{AK}:=\sum_kp_k\rho_k^{A}\otm\proj{k}^K$ and $\varsigma^{AK}:=\sum_kq_k\varsigma_k^{A}\otm\proj{k}^K$, it holds that
\alg{
\left|
\sum_kp_k\left\|\rho_k-\varsigma_k\right\|_1
-
\left\|\rho^{AK}-\varsigma^{AK}\right\|_1
\right|
\leq
\sum_k|p_k-q_k|
\leq
\left\|\rho^{AK}-\varsigma^{AK}\right\|_1.
\laeq{somodey}
}
\elmm

\begin{lmm}
\label{lmm:triDSP}
The DSP norm defined by \req{dfnDSPnorm} satisfies the triangle inequality, i.e., for any superoperators $\ca{E}$ and $\ca{F}$ from $\ca{L}(\ca{H}^A)$ to $\ca{L}(\ca{H}^B)$,
$\|\ca{E}+\ca{F}\|_{\rm DSP}
\leq
\|\ca{E}\|_{\rm DSP}
+
\|\ca{F}\|_{\rm DSP}$.
\elmm

\blmm{projHSN}
Let $\{\Pi_j\}_j$ be a set of orthogonal projectors on $\ca{H}$ such that $\sum_j\Pi_j=I$. For any $\varrho\in\ca{P}(\ca{H})$,
$\left\|\varrho\right\|_2^2=\sum_{j,k}\left\|\Pi_j\varrho \Pi_k\right\|_2^2$.
\elmm

\subsubsection{Properties of Conditional Entropies}
\lsec{propCE}

\begin{lmm}[Corollary of Lemma 13 in  \cite{tomamichel2010duality}]\label{lmm:invCEiso}
For any $\epsilon\geq0$, $\rho^{AB} \in \ca{S}_\leq(\ca{H}^{AB})$ and any linear isometry $V:A\rightarrow C$, $H_{\rm min}^\epsilon(A|B)_\rho=H_{\rm min}^\epsilon(C|B)_{\ca{V}(\rho)}$.
\elmm

\begin{lmm}[Corollary of Lemma 15 in  \cite{tomamichel2010duality}]\label{lmm:invCEmaxiso}
For any $\epsilon\geq0$, $\rho^{AB} \in \ca{S}_\leq(\ca{H}^{AB})$ and any linear isometry $W:B\rightarrow D$, $H_{\rm max}^\epsilon(A|B)_\rho=H_{\rm max}^\epsilon(A|D)_{\ca{W}(\rho)}$.
\elmm

\begin{lmm}[Lemma A.1 in  \cite{DBWR2010}]\label{Lemma:Candmin}
For any $\rho^{AB} \in \ca{S}_\leq(\ca{H}^{AB})$ and $\varsigma^{B} \in \ca{S}_=(\ca{H}^{B})$, it holds that
\begin{align}
 H_2(A|B)_{\rho|\varsigma} \geq H_{\rm min}(A|B)_{\rho|\varsigma},\; H_2(A|B)_{\rho} \geq H_{\rm min}(A|B)_{\rho} . 
\end{align}
\end{lmm}

\begin{lmm}
[Definition 14, Equality (6) and Lemma 16 in \cite{tomamichel2010duality}]\label{lmm:duality}
For any subnormalized pure state $|\psi\rangle$ on system $ABC$, and for any $\epsilon>0$, 
$H_{\rm max}^\epsilon(A|B)_\psi=
-
H_{\rm min}^\epsilon(A|C)_\psi$.
\end{lmm}

\begin{lmm}[Lemma B.2 in  \cite{DBWR2010}]\label{lmm:ZYX}
Let $\psi^{ABC}\in\ca{S}_\leq(\ca{H}^{ABC})$ be a subnormalized pure state. For any full-rank state $\varsigma^B\in\ca{S}_=(\ca{H}^B)$, it holds that
$\psi^{ABC}\leq Z^{AB}\otm I^C$,
where
\begin{equation}
Z^{AB}:=
2^{\frac{1}{2}H_{\rm max}(A|B)_{\psi|\varsigma}}\cdot
(\varsigma^B)^{-\frac{1}{2}}
\sqrt{
(\varsigma^B)^{\frac{1}{2}}
\psi^{AB}
(\varsigma^B)^{\frac{1}{2}}
}
(\varsigma^B)^{-\frac{1}{2}}.
\end{equation}
\elmm

\begin{lmm}[Lemma A.5 in  \cite{DBWR2010}]\label{lmm:condminCQ}
For any state $\rho^{ABK} \in \ca{S}_=(\ca{H}^{ABK})$ in the form of
\alg{
\rho^{ABK}=\sum_kp_k\rho_k^{AB}\otm\proj{k}^K,
}
where $\rho_k \in \ca{S}_=(\ca{H}^{AB})$, $\inpro{k}{k'}=\delta_{k,k'}$ and $\{p_k\}_k$ is a normalized probability distribution, it holds that 
\alg{
&
H_{\rm min}(A|BK)_\rho=-\log\left(\sum_kp_k\cdot2^{-H_{\rm min}(A|B)_{\rho_k}}\right),
\\
&
H_{\rm max}(A|BK)_\rho=\log\left(\sum_kp_k\cdot2^{H_{\rm max}(A|B)_{\rho_k}}\right).
}
(It is straightforward to show that the above equalities also hold for $\rho^{ABK} \in \ca{S}_\leq(\ca{H}^{ABK})$ and $\rho_k \in \ca{S}_\leq(\ca{H}^{AB})$, by noting that $H_{\rm min}(A|BK)_\rho=H_{\rm min}(A|BK)_{\rho/{\rm Tr}[\rho]}-\log{{\rm Tr}[\rho]}$ and that $H_{\rm max}(A|BK)_\rho=H_{\rm max}(A|BK)_{\rho/{\rm Tr}[\rho]}+\log{{\rm Tr}[\rho]}$.)
\elmm

\begin{lmm}[Lemma A.7 in  \cite{DBWR2010}]\label{lmm:condminCQCQ}
For any state $\rho^{ABK_1K_2} \in \ca{S}_\leq(\ca{H}^{ABK_1K_2})$ in the form of
\alg{
\rho^{ABK_1K_2}=\sum_kp_k\rho_k^{AB}\otm\proj{k}^{K_1}\otm\proj{k}^{K_2},
\laeq{rhoclk1k2}
}
where the notations are the same as in \rLmm{condminCQ}, and for any $\epsilon\geq0$ it holds that 
\alg{
H_{\rm min}^\epsilon(AK_1|BK_2)_\rho=H_{\rm min}^\epsilon(A|BK_2)_\rho.
}
(Note that, although Lemma A.7 in \cite{DBWR2010} assumes that $\rho^{ABK_1K_2}$ is normalized, the condition is not used in the proof thereof.)
\elmm

\begin{lmm}[Lemma A.1 in  \cite{dupuis2014decoupling}]\label{lmm:clch}
Let $\rho\in\ca{S}_\leq(\ca{H}^{K_1K_2AB})$ be a subnormalized state that is classically coherent in $K_1K_2$. For any $\epsilon\geq0$, there exists $\hat{\rho}\in\ca{B}^\epsilon(\rho)$ that is classically coherent in $K_1K_2$, and $\varsigma\in\ca{S}_=(\ca{H}^{K_2B})$ that is decomposed as $\varsigma=\sum_k\proj{k}^{K_2}\otm\varsigma_k^B$, such that
\alg{
H_{\rm min}^\epsilon(K_1A|K_2B)_\rho
=H_{\rm min}(K_1A|K_2B)_{\hat\rho}
=H_{\rm min}(K_1A|K_2B)_{\hat{\rho}|\varsigma}.
}
\elmm

\begin{lmm}\label{lmm:condmaxCQCQ}
In the same setting as in \rLmm{condminCQCQ}, it holds that 
\alg{
H_{\rm max}^\epsilon(AK_1|BK_2)_\rho=H_{\rm max}^\epsilon(A|BK_2)_\rho.
}
\elmm

\begin{lmm}\label{lmm:clchtwo}
Let $\rho\in\ca{S}_\leq(\ca{H}^{K_1K_2AB})$ be a subnormalized state that is classically coherent in $K_1K_2$. For any $\epsilon\geq0$, there exists $\hat{\rho}\in\ca{B}^\epsilon(\rho)$ that is classically coherent in $K_1K_2$, such that
\alg{
H_{\rm max}^\epsilon(K_1A|K_2B)_\rho
=H_{\rm max}(K_1A|K_2B)_{\hat\rho}.
}
If $\rho$ is also diagonal in $K_1K_2$ (i.e., if $\rho$ is in the form of \req{rhoclk1k2}), there exists $\hat{\rho}$, satisfying the above conditions, that is diagonal in $K_1K_2$.
\elmm

\blmm{minentave}
Consider the same setting as in \rLmm{condminCQ}. For any $\{\epsilon_k\}_k$ such that $\epsilon_k\geq0$, it holds that
\alg{
H_{\rm min}^{\sqrt{2\varepsilon}}(A|BK)_\rho\geq-\log\left(\sum_kp_k\cdot2^{-H_{\rm min}^{\epsilon_k}(A|B)_{\rho_k}}\right),
}
where $\varepsilon:=\sum_kp_k\epsilon_k$.
\elmm

\subsubsection{Other Technical Lemmas}
\label{sec:lemmasC}

\begin{lmm} \label{Lemma:PTtrace}
Consider two linear operators $X,Y:\ca{H}^A\rightarrow\ca{H}^B$ and assume that $A\cong A'$, $B\cong B'$.  Let $\ket{\Phi}^{AA'}$ and $\ket{\Phi}^{BB'}$ be maximally entangled states between $A$ and $A'$, and $B$ and $B'$, respectively. Then,
${\rm Tr}[X^TY]=\sqrt{d_Ad_B}\bra{\Phi}^{BB'}(X\otimes Y)\ket{\Phi}^{AA'}$, 
where $d_A:=\dim\ca{H}^A$, $d_B:=\dim\ca{H}^B$ and the transposition is taken with respect to the Schmidt bases of $\ket{\Phi}^{AA'}$ and $\ket{\Phi}^{BB'}$.
\end{lmm}

\blmm{clchsq}
If $\varrho^2$ is classically coherent in $XY$
for a positive semidefinite operator $\varrho\in\ca{P}(\ca{H}^{AXY})$, so is $\varrho$.
\elmm

\blmm{OPvar}
Let $\pi$ be the maximally mixed state on system $A$, and
let $\ca{C}$ be the completely dephasing operation on $A$ with respect to a fixed basis $\{|i\rangle\}_{i=1}^{d_A}$. 
For any $\rho\in\ca{P}(\ca{H}^{AB})$, it holds that
\alg{
&
\left\|\rho^{AR}-\pi^A\otm\rho^{R}\right\|_2^2
\leq
\left\|\rho^{AR}\right\|_2^2,
\laeq{deasons}\\
&
\left\|\rho^{AR}-\ca{C}^A(\rho^{AR})\right\|_2^2
\leq
\left\|\rho^{AR}\right\|_2^2.
\laeq{deasons2}
}
\elmm

\blmm{inprothree}
For subnormalized pure states $\ket{\psi},\ket{\phi}\in\ca{H}$ and a real number $c>0$,
suppose that there exists a normalized pure state $\ket{e}\in\ca{H}$ that satisfies
$\inpro{e}{\psi}\geq c$ and $\inpro{e}{\phi}\geq c$.
Then,
$|\inpro{\psi}{\phi}|\geq 2c^2-1$.
\elmm

\begin{lmm}\label{lmm:yoshiki}
(Lemma 35 in \cite{wakakuwa2017coding})
Let $c\in(0,\infty)$ be a constant, $f:[0,c]\rightarrow{\mathbb R}$ be a monotonically nondecreasing function that  satisfies $f(c)<\infty$, and $\{p_k\}_{k\in{\mathbb K}}$ be a probability distribution on a countable set ${\mathbb K}$.  Suppose $\epsilon_k\:(k\in{\mathbb K})$ satisfies $\epsilon_k\in[0,c]$, and $\sum_{k\in{\mathbb K}}p_k\epsilon_k\leq\epsilon$ for a given $\epsilon\in(0,c^2]$. Then we have
\begin{eqnarray}
\sum_{k\in{\mathbb K}}p_kf(\epsilon_k)\leq f(\sqrt{\epsilon})+f( c)\cdot\sqrt{\epsilon}.\label{eq:detectivewall}
\end{eqnarray}
\end{lmm}

\section{Proof of The Non-Randomized Partial Decoupling (Theorem~\ref{thm:SmoothMarkov})} 
\label{sec:prfPD}

We now prove the non-randomized partial decoupling (Theorem~\ref{thm:SmoothMarkov}). As sketched in Subsection~\ref{SS:KLSP}, we proceed the proof in two steps: showing the non-smoothed version in Subsection~\ref{sec:prfNSPD}, and then smoothing it in Subsection~\ref{SS:prSNRPD}.

\subsection{Proof of The Non-Smoothed Non-randomized Partial Decoupling}
\label{sec:prfNSPD}

The non-smoothed version of Theorem \ref{thm:SmoothMarkov} is given by
\begin{align}
\mbb{E}_{U \sim {\sf H}_{\times}} 
\left[ \left\|
\ca{T}^{A \rightarrow E} \circ \ca{U}^A ( \Psi^{AR} ) 
- \ca{T}^{A \rightarrow E}(\Psi_{\rm av}^{AR})
\right\|_1 \right]
\leq  2^{-\frac{1}{2} H_{\rm min}(A^*|RE)_{\Lambda(\Psi,\ca{T})}},\label{eq:smthMark2}
\end{align}
where $\Psi_{\rm av}^{AR}
= \bigoplus_{j=1}^J \Psi_{jj}^{A_lR}\otimes \pi_j^{A_r}$.
Note that, due to the definition of the conditional collision entropy \req{cond2vs}, \req{cond2} and its relation to the conditional min-entropy (see Lemma \ref{Lemma:Candmin}), we have
\begin{align}
\left\| 
{\Lambda}(\Psi,\ca{T}) 
  \right\|^2_{2, \varsigma^{ER}}
= 2^{-\frac{1}{2}H_2(A^*|RE)_{{\Lambda}(\Psi,\ca{T})}}
\leq 2^{-\frac{1}{2}H_{\rm min}(A^*|RE)_{{\Lambda}(\Psi,\ca{T})}} \label{34revspkm}
\end{align}
for a proper choice of $\varsigma^{ER}\in\ca{S}_=(\ca{H}^{ER})$. In addition, it holds that
\alg{
\left\| 
{\Lambda}(\Psi,\ca{T}) 
  \right\|^2_{2, \varsigma^{ER}}
  =
  \sum_{j,k=1}^J \frac{d_A^2}{r_j r_k}
\left\|   \tr_{A_l} \left[ \Psi_{jk}^{A_l^T A_r R} \tau^{A_l \bar{A}_r E}_{jk} \right]  \right\|^2_{2, \varsigma}.
\laeq{kazegusuri}
}
We first show this relation.

Let $\Pi_j^{A^*}$ be the projection onto a subspace ${\ca H}_j^{A_r}\otimes{\ca H}_j^{\bar{A}_r}\subset{\ca H}^{A^*}$ for each $j$. Due to the definition of $F^{A\bar{A}\rightarrow A^*}$ given by (\ref{Eq:26}), it holds that
\alg{
\Pi_j^{A^*}F^{A\bar{A}\rightarrow A^*}
=\sqrt{\frac{d_Al_j}{r_j}} \langle\Phi_j^l|^{A_l\bar{A}_l}(\Pi_j^{A} \otimes \Pi_j^{\bar{A}}).
\laeq{PiF}
}
Using the property of the Hilbert-Schmidt norm (\rLmm{projHSN}), we have
\alg{
\left\|{\Lambda}(\Psi,\ca{T})\right\|_{2,\varsigma}^2
&
=
\left\|(\varsigma^{ER})^{-1/4}{\Lambda}(\Psi,\ca{T})(\varsigma^{ER})^{-1/4}\right\|_2^2
\nn\\
&
=
\sum_{j,k=1}^J
\left\|\left(\Pi_j^{A^*}\otm(\varsigma^{ER})^{-1/4}\right)
{\Lambda}_\varsigma(\Psi,\ca{T})
\left(\Pi_k^{A^*}\otm(\varsigma^{ER})^{-1/4}\right)\right\|_2^2
\nn\\
&
=
\sum_{j,k=1}^J
\left\|\Pi_j^{A^*}
{\Lambda}(\Psi,\ca{T})
\Pi_k^{A^*}\right\|_{2,\varsigma}^2.
\laeq{kimitake}
}
Using Eq.~\req{PiF} and the explicit form of ${\Lambda}(\Psi,\ca{T})$, i.e. ${\Lambda}(\Psi,\ca{T}):=F(\Psi^{AR}\otimes\tau^{\bar{A}E})F^\dagger$, each term in the summand is given by
\alg{
\Pi_j^{A^*}
{\Lambda}(\Psi,\ca{T})
\Pi_k^{A^*}
&
=
(\Pi_j^{A^*}\!F^{A\bar{A}\rightarrow A^*})
(\Psi^{AR}\otimes\tau^{\bar{A}E})
(\Pi_k^{A^*}\!F^{A\bar{A}\rightarrow A^*})^\dagger
\nn\\
&=
\frac{d_A}{\sqrt{r_jr_k}}\cdot
\sqrt{l_jl_k} \langle\Phi_j^l|^{A_l\bar{A}_l}(\Pi_j^{A}\Psi^{AR}\Pi_k^{A} \otimes \Pi_j^{\bar{A}}\tau^{\bar{A}E}\Pi_k^{\bar{A}})|\Phi_k^l\rangle^{A_l\bar{A}_l}
\nn\\
&=
\frac{d_A}{\sqrt{r_jr_k}}\cdot
\sqrt{l_jl_k} \langle\Phi_j^l|^{A_l\bar{A}_l}(\Psi_{jk}^{A_lA_rR} \otimes \tau_{jk}^{\bar{A}_l\bar{A}_rE})|\Phi_k^l\rangle^{A_l\bar{A}_l}
\nn\\
&=
\frac{d_A}{\sqrt{r_jr_k}}
{\rm Tr}_{A_l}\left[\Psi_{jk}^{A_l^TA_rR} \tau_{jk}^{{A}_l\bar{A}_rE}\right],
}
where the last line follows from Lemma \ref{Lemma:PTtrace}. Thus, we obtain \req{kazegusuri}.

From Eqs.~\eqref{34revspkm} and \req{kazegusuri}, it suffices to prove that
\begin{align}
\mbb{E}_{U \sim {\sf H}_{\times}} 
\left[ \left\|
\ca{T}^{A \rightarrow E} \circ \ca{U}^A ( \Psi^{AR} ) 
- \ca{T}^{A \rightarrow E}(\Psi_{\rm av}^{AR})
\right\|_1 \right]
\leq  \sum_{j,k=1}^J \frac{d_A^2}{r_j r_k}
\left\|   \tr_{A_l} \left[ \Psi_{jk}^{A_l^T A_r R} \tau^{A_l \bar{A}_r E}_{jk} \right]  \right\|^2_{2, \varsigma}\label{eq:smthMark23}
\end{align}
for any $\varsigma^{ER}\in\ca{S}_=(\ca{H}^{ER})$. In the following, we denote the L.H.S. of Ineq.~\req{smthMark23} by $\kappa$.
Due to Lemma \ref{lmm:holder}, for any $\varsigma\in\ca{S}_=(\ca{H}^{ER})$, we have
\begin{equation}
\left\|\ca{T}^{A \rightarrow E} \circ\ca{U}^A (  \Psi^{AR}-\Psi_{\rm av}^{AR} )\right\|_1
\leq\left\|\ca{T}^{A \rightarrow E} \circ\ca{U}^A (  \Psi^{AR}-\Psi_{\rm av}^{AR} )\right\|_{2,\varsigma^{ER}}.
\end{equation}
Using this and Jensen's inequality, we obtain
\begin{equation}
\kappa^2
\leq\mbb{E}_{U \sim {\sf H}_{\times}}\left[|\! | \ca{T}^{A \rightarrow E} \circ \ca{U}^A ( \Psi^{AR} ) - \ca{T}^{A \rightarrow E} (\Psi_{\rm av}^{AR} ) |\! |_{2, \varsigma}^2\right].
\laeq{sitasita}
\end{equation}
Noting that
$\Psi_{jj}^{A_lR}
=
{\rm Tr}_{A_r}[\Psi_{jj}^{A_lA_rR}]
=
{\rm Tr}_{A_r}[\Psi_{{\rm av}, jj}^{A_lR}\otm\pi_j^{A_r}]=\Psi_{{\rm av}, jj}^{A_lR}$,
we can apply Lemma~\ref{lmm:nenchalu} for $X^{AR}=\Psi^{AR}-\Psi_{\rm av}^{AR}$ and $\sigma={\rm id}$. 
This yields
\begin{align}
\kappa^2
&\leq
\sum_{i,j =1}^J \frac{d_A^2}{r_j r_k}
\left\|   \tr_{A_l} \left[ \left( \Psi_{jk}^{A_l^T A_r R} - \Psi_{{\rm av}, jk}^{A_l^T A_r R} \right )\tau^{A_l \bar{A}_r E}_{jk} \right]  \right\|^2_{2, \varsigma}
\nn\\
&=
\sum_{j =1}^J \frac{d_A^2}{r_j^2}
\left\|   \tr_{A_l} \left[ \left (\Psi_{jj}^{A_l^T A_r R} - \Psi_{jj}^{A_l^T  R}\otimes \pi_{jj}^{A_r}\right)\tau^{A_l \bar{A}_r E}_{jj} \right]  \right\|^2_{2, \varsigma}
+
\sum_{j\neq k} \frac{d_A^2}{r_j r_k}
\left\|   \tr_{A_l} \left[ \Psi_{jk}^{A_l^T A_r R} \tau^{A_l \bar{A}_r E}_{jk} \right]  \right\|^2_{2, \varsigma},
\laeq{sitasita69}
\end{align}
where the second line follows from the fact that $\Psi_{{\rm av},jk}^{A_lA_rR}=0$ for $j\neq k$. 
To calculate the first term in \req{sitasita69}, note that
\alg{
\tr_{A_l} [ \Psi_{jj}^{A_l^T A_r R} \tau^{A_l \bar{A}_r E}_{jj} ]
\in
\ca{P}(\ca{H}^{A_r\bar{A}_rRE})
}
and that
\alg{
\tr_{A_l} \left[ \left( \Psi_{jj}^{A_l^T  R}\otimes \pi_{jj}^{A_r}\right)\tau^{A_l \bar{A}_r E}_{jj} \right] 
=
\tr_{A_lA_r} [ \Psi_{jj}^{A_l^T A_r R} \tau^{A_l \bar{A}_r E}_{jj} ]
\otimes \pi_{jj}^{A_r}.
}
Thus, we simply apply \rLmm{OPvar} to obtain
\alg{
\left\|   \tr_{A_l} \left[ \left (\Psi_{jj}^{A_l^T A_r R} - \Psi_{jj}^{A_l^T  R}\otimes \pi_{jj}^{A_r}\right)\tau^{A_l \bar{A}_r E}_{jj} \right]  \right\|^2_{2, \varsigma}
\leq
\left\|   \tr_{A_l} \left[ \Psi_{jj}^{A_l^T A_r R} \tau^{A_l \bar{A}_r E}_{jj} \right]  \right\|^2_{2, \varsigma}
}
for each $j$. Substituting this to \req{sitasita69}, we arrive at Ineq.~\req{smthMark23}. \hfill$\blacksquare$

\subsection{Proof of The Smoothed Non-Randomized Partial Decoupling} \label{SS:prSNRPD}

We now smoothen the conditional min-entropy to complete the proof of Theorem~\ref{thm:SmoothMarkov}. To this end, fix $\hat{\Psi}\in\ca{B}^\epsilon(\Psi)$ and $\hat{\ca{T}}\in\ca{B}_{\rm DSP}^\mu(\ca{T})$ so that 
\begin{align}
H_{\rm min}^{\epsilon,\mu}(A^*|RE)_{\Lambda(\Psi,\ca{T})}=H_{\rm min}(A^*|RE)_{\Lambda(\hat{\Psi},\hat{\ca{T}})}.\laeq{yakuhin}
\end{align}
Let $|\Psi_{p,{\rm av}}\rangle^{AA'}$ be a purification of $\Psi_{\rm av}^A$.
Noting that $\Psi_{\rm av}$ is decomposed in the form of \req{taisho}, by properly choosing a DSP decomposition for $A'$, it holds that
\begin{align}
(\Pi_j^A\otm\Pi_k^{A'})|\Psi_{p,{\rm av}}\rangle^{AA'}=
\delta_{jk}\sqrt{q_j}|\varpi_j\rangle^{A_lA_l'}|\Phi_j^r\rangle^{A_rA_r'},\label{eq:itsw}
\end{align}
where $q_j:=\tr{\Psi_{jj}}$ and $\varpi_j$ is a purification of $\Psi_{jj}^{A_l}/q_j$ for each $j$. 
Let $\Delta_+^{A'E}$ and $\Delta_-^{A'E}$ be linear operators on $\ca{H}^E\otimes\ca{H}^{A'}$ such that 
$\Delta_+^{A'E}\geq0,\; \Delta_-^{A'E}\geq0,
\;
{\rm supp}[\Delta_+^{A'E}]\perp{\rm supp}[\Delta_-^{A'E}]$
and that
\begin{align}
\ca{T}^{A \rightarrow E}(\Psi_{p,{\rm av}}^{AA'}) -\hat{\ca{T}}^{A \rightarrow E}(\Psi_{p,{\rm av}}^{AA'})=\Delta_{+}^{A'E}-\Delta_{-}^{A'E}.
\label{eq:plmn}
\end{align} 
In addition, let $\ca{D}_+^{A\rightarrow E}$ and $\ca{D}_-^{A\rightarrow E}$ be superoperators such that
\alg{
\ca{D}_{+}^{A\rightarrow E}(\Psi_{p,{\rm av}}^{AA'})=\Delta_{+}^{A'E},
\quad
\ca{D}_{-}^{A\rightarrow E}(\Psi_{p,{\rm av}}^{AA'})=\Delta_{-}^{A'E},
\laeq{dfnDeltapm}
}
which yields $\ca{T}-\hat{\ca{T}}=\ca{D}_{+}-\ca{D}_{-}$.
Note that, in general, it does not necessarily imply that $\ca{D}_{+}=\ca{T}$ and $\ca{D}_{-}=\hat{\ca{T}}$.

We now apply Lemma \ref{lmm:trianglemother} for the case where $\sigma={\rm id}$. To obtain the explicit forms, we compute 
\begin{align}
\tr[(\ca{D}_+^{A \rightarrow E}+ \ca{D}_-^{A \rightarrow E})   ( \Psi_{\rm av}^{AR} )]
&=
\tr[(\ca{D}_+^{A \rightarrow E}+ \ca{D}_-^{A \rightarrow E})   ( \Psi_{\rm av}^{A} )]
\nn\\
&
=\tr[(\ca{D}_+^{A \rightarrow E}+ \ca{D}_-^{A \rightarrow E}) ( \Psi_{p,{\rm av}}^{AA'} )]
\nonumber\\
&
=\tr[\Delta_{+}^{A'E}+\Delta_{-}^{A'E}]
\nonumber\\
&
=\left\|\Delta_{+}^{A'E}-\Delta_{-}^{A'E}\right\|_1
\nonumber\\
&=\left\|\ca{T}^{A \rightarrow E}(\Psi_{p,{\rm av}}^{AA'}) -\hat{\ca{T}}^{A \rightarrow E}(\Psi_{p,{\rm av}}^{AA'})\right\|_1
\nn\\
&
\leq 
\left\|\ca{T}^{A \rightarrow E}-\hat{\ca{T}}^{A \rightarrow E}\right\|_{\rm DSP}\leq\mu,\laeq{grapefrp}
\end{align}
where we have used the properties of $\Psi_{p,{\rm av}}^{AA'}$, $\Delta_{\pm}^{A'E}$, and $\mathcal{D}_{\pm}^{A \rightarrow E}$ described above. The last line follows from the definition of the DSP norm.
Furthermore, introducing a notation $\bar{\ca{U}}(\cdot):=\mbb{E}_{U \sim {\sf H}_{\times}}[\:\ca{U}(\cdot)]$, we also have (see Lemma \ref{lmm:trianglemother} for the definition and properties of $\delta_{\pm}^{AR}$)
\begin{align}
&\tr[\hat{\ca{T}}^{A \rightarrow E}  \circ \bar{\ca{U}}^A (\delta_+^{AR}+\delta_-^{AR})]\nonumber\\
&=\left\|\hat{\ca{T}}^{A \rightarrow E} \circ \bar{\ca{U}}^A (\delta_+^{AR})\right\|_1+\left\|\hat{\ca{T}}^{A \rightarrow E} \circ \bar{\ca{U}}^A (\delta_-^{AR}) \right\|_1\nonumber\\
&=\tr{[\delta_+^{AR}]}\cdot\left\|\hat{\ca{T}}^{A \rightarrow E} \circ {\bar{\ca{U}}}^A (\delta_+^{AR}/\tr{[\delta_+^{AR}]})\right\|_1
+\tr{[\delta_-^{AR}]}\cdot\left\|\hat{\ca{T}}^{A \rightarrow E} \circ \bar{\ca{U}}^A (\delta_-^{AR}/\tr{[\delta_-^{AR}]})\right\|_1\nonumber\\
&\leq(\tr{[\delta_+^{AR}]}+\tr{[\delta_-^{AR}]})\cdot\left\|\hat{\ca{T}}^{A \rightarrow E}\right\|_{\rm DSP}\nonumber\\
&=\left\|\delta_+^{AR} -\delta_-^{AR}\right\|_1\cdot\left\|\hat{\ca{T}}^{A \rightarrow E}\right\|_{\rm DSP}
\nonumber\\
&=\left\|\hat{\Psi}^{AR} -\Psi^{AR}\right\|_1\cdot\left\|\hat{\ca{T}}^{A \rightarrow E}\right\|_{\rm DSP}
\nonumber\\
&\leq \left\|\hat{\Psi}^{AR} \!-\!\Psi^{AR}\right\|_1\cdot\left(\left\|{\ca{T}}^{A \rightarrow E}\right\|_{\rm DSP}\!+\!\left\|\hat{\ca{T}}^{A \rightarrow E}\!-\!{\ca{T}}^{A \rightarrow E}\right\|_{\rm DSP}\right)\nonumber\\
&\leq\epsilon\left\|{\ca{T}}^{A \rightarrow E}\right\|_{\rm DSP}+\epsilon\mu,\laeq{grapefrp2}
\end{align}
where the fourth line follows from the definition of the DSP norm \req{dfnDSPnorm}, and the seventh line from the triangle inequality for the DSP norm (Lemma \ref{lmm:triDSP}).
Applying the non-smoothed version of the non-randomized partial decoupling (Ineq. \req{smthMark2}) to a state $\hat{\Psi}$ and a CP map $\hat{\ca{T}}$, we have
\begin{align}
\mbb{E}_{U \sim {\sf H}_{\times}} \bigl[ \bigl|\! \bigl|
\hat{\ca{T}}^{A \rightarrow E} \circ \ca{U}^A ( \hat{\Psi}^{AR} ) - {\ca{T}}^{A \rightarrow E}(\hat{\Psi}_{\rm av}^{AR})
\bigr|\! \bigr|_1 \bigr]
\leq 2^{-\frac{1}{2} H_{\rm min}(A_cA_r\bar{A}_r|RE)_{\Lambda(\hat{\Psi},\hat{\ca{T}})}}.\label{eq:boundthird}
\end{align}

All together, Ineq.~\req{pracap} in \rLmm{trianglemother} leads to
\begin{multline}
\mbb{E}_{U \sim {\sf H}_{\times}}\left[
\left\|{\ca{T}}^{A \rightarrow E}  \circ \ca{U}^A ( {\Psi}^{AR} ) - {\ca{T}}^{A \rightarrow E}    ( \Psi_{\rm av}^{AR} )\right\|_1
\right] \\
\leq
2^{-\frac{1}{2} H_{\rm min}(A_cA_r\bar{A}_r|RE)_{\Lambda(\hat{\Psi},\hat{\ca{T}})}}
+2 \left(
\mu
+\epsilon\left\|{\ca{T}}^{A \rightarrow E}\right\|_{\rm DSP}+\epsilon\mu
\right),
\label{eq:boundhathat2}
\end{multline}
which, together with \req{yakuhin}, concludes the proof of \rThm{SmoothMarkov}.
\hfill$\blacksquare$

\section{Proof of The Randomized Partial Decoupling (Theorem~\ref{thm:SmoothExMarkov})} 
\lsec{prfRPD}

We here show Theorem~\ref{thm:SmoothExMarkov}. We first put the following two assumptions, which simplify the proof:
\begin{description}
\item[{\bf WA 1}]  $E\cong E_cE_r$, where $E_c$ is a quantum system of dimension $J$ \vspace{-3mm}
\item[{\bf WA 2}] The CP map $\ca{T}^{A \rightarrow E}$ is decomposed into
\begin{align}
\ca{T}^{A \rightarrow E}(X)=\sum_{j,k=1}^J\outpro{j}{k}^{E_c}\otimes\ca{T}_{jk}^{A_r \rightarrow E_r}(X_{jk}),
\end{align}
in which $\ca{T}_{jk}$ is a linear supermap from $\ca{L}(\ca{H}^{A_r})$ to $\ca{L}(\ca{H}^{E_r})$ defined by $\ca{T}_{jk}(\zeta)=\ca{T}(\outpro{j}{k}\otimes\zeta)$ for each $j,k$.
\end{description}
We show the non-smoothed version in Subsection~\ref{SS:prNSRPD} and the smoothed version in Subsection~\ref{SS:SRPD}. The above assumptions are then dropped in Subsection~\ref{SS:DA1}.

\subsection{Proof of The Non-Smoothed Randomized Partial Decoupling under {\bf WA 1} and {\bf WA 2}} \label{SS:prNSRPD}

Under the assumptions {\bf WA 1} and {\bf WA 2}, the non-smoothed version of the randomized partial decoupling is given by
\begin{multline}
\mbb{E}_{U \sim {\sf H}_{\times}, \sigma\sim{\sf P}} \left[ \left\|
\ca{T}^{A \rightarrow E} \circ \ca{G}_\sigma^A  \circ \ca{U}^A ( \Psi^{AR} ) -\ca{T}^{A \rightarrow E} \circ \ca{G}_\sigma^A ( \Psi_{\rm av}^{AR} )
\right\|_1 \right]\\
\leq  
\sqrt{\alpha(J)}\cdot2^{-\frac{1}{2}H_{\rm min}(A|R)_{\Psi}-\frac{1}{2}H_{\rm min}(A|E)_{\tau}}
+\beta(A_r)\cdot2^{-\frac{1}{2}H_{\rm min}(A|R)_{\ca{C}(\Psi)}-\frac{1}{2}H_{\rm min}(A|E)_{\ca{C}(\tau)}}.
\laeq{SmExMaNS}
\end{multline}
Note that, as we will describe in Subsection \ref{SS:DA1} for general cases, the min entropies $H_{\rm min}(A|E)_{\tau}$ and $H_{\rm min}(A|E)_{\ca{C}(\tau)}$ are equal to the max entropies $-H_{\rm max}(A|B)_{\ca{C}(\tau)}$ and $-H_{\rm max}(A_r|BA_c)_{\ca{C}(\tau)}$, respectively, due to the duality of the conditional entropies for pure states (\rLmm{duality}). 
The proof of this inequality will be divided into three steps.

\subsubsection{Upper bound on the average trace norm}

To prove Ineq. \req{SmExMaNS}, we first introduce the following lemma that relates the average trace norm of an operator $\ca{T}^{A \rightarrow E} \circ \ca{G}_\sigma^A  \circ \ca{U}^A ( X^{AR})$ to the average Hilbert-Schmidt norm. 

\begin{lmm}
\label{lmm:aidaga}
Let $X^{AR}$ be an arbitrary Hermitian operator such that $X^{AR}=\sum_{j,k=1}^J\outpro{j}{k}^{A_c}\otimes X_{jk}^{A_rR_r}\otimes\outpro{j}{k}^{R_c}$,
and let  $\zeta\in\ca{S}_=(\ca{H}^{E})$ and $\xi\in\ca{S}_=(\ca{H}^{R})$ be arbitrary states that are decomposed as
$\zeta^{E}\!=\!\sum_j\outpro{j}{j}^{E_c}\!\otimes\zeta_j^{E_r}$, $\xi^{R}\!=\!\sum_j\outpro{j}{j}^{R_c}\!\otimes\xi_j^{R_r}$,
respectively.
Then it holds that
\begin{align}
\mbb{E}_{\sigma,U }  \bigl[ \bigl|\! \bigl|
\ca{T}^{A \rightarrow E} \circ \ca{G}_\sigma^A  \circ \ca{U}^A ( X^{AR}) \bigr|\! \bigr|_1 \bigr]
\leq
\frac{1}{\sqrt{J}}\!\cdot\!\sqrt{\mbb{E}_{\sigma,U } \left\|\ca{T}^{A \rightarrow E} \circ\ca{G}_\sigma^A  \circ \ca{U}^A (  X^{AR} )\right\|_{2,\:\zeta^E\otimes\xi^R}^2}\:,\!\laeq{iiwake}
\end{align}
where the norm in the R.H.S. is defined by \req{cond2on}.
\end{lmm}
It should be noted that Lemma~\ref{lmm:aidaga} provides a stronger inequality than that obtained simply using Lemma~\ref{lmm:holder}.

\begin{prf}

We exploit techniques developed in  \cite{dupuis2014decoupling}. 
Recall that $U$ is in the form of $\sum_{j=1}^J\outpro{j}{j}^{A_c}  \otimes U_j^{A_r}$, and $G_\sigma$ is defined by $G_\sigma:=\sum_{j=1}^J\outpro{\sigma(j)}{j}^{A_c}  \otimes I^{A_r}$ for any $\sigma\in\mbb{P}$.

We define a subnormalized state $\gamma_\sigma\in\ca{S}_\leq(\ca{H}^{ER})$ for each $\sigma$ by
$\gamma_\sigma^{ER}:=\sum_{j=1}^J\outpro{\sigma(j)}{\sigma(j)}^{E_c}\otimes\zeta_{\sigma(j)}^{E_r}\otimes\xi_j^{R_r}\otm\outpro{j}{j}^{R_c}$.
Further, by letting $P$ be a quantum system with an orthonormal basis $\{|\sigma\rangle\}_{\sigma\in\mbb{P}}$, we define a subnromalized state $\gamma\in\ca{S}_\leq(\ca{H}^{PER})$ by
\begin{align}
\gamma^{PER}:=\frac{1}{|{\mbb P}|}\sum_{\sigma\in{\mbb P}}\outpro{\sigma}{\sigma}^P\otimes\gamma_\sigma^{ER}.
\end{align}
Using Lemma~\ref{lmm:holder} and Jensen's inequality, we obtain
\begin{align}
\mbb{E}_{\sigma} \bigl[ \bigl|\! \bigl|
\ca{T}^{A \rightarrow E} \circ \ca{G}_\sigma^A  \circ \ca{U}^A ( X^{AR}) \bigr|\! \bigr|_1 \bigr]
&=\left\|\mbb{E}_{\sigma} \bigl[ \outpro{\sigma}{\sigma}^P\otimes
\ca{T}^{A \rightarrow E} \circ \ca{G}_\sigma^A  \circ \ca{U}^A (X^{AR}) \bigr]\right\|_1 \nonumber \\
&\leq \sqrt{{\rm Tr}[\gamma]}\cdot
\left\|\mbb{E}_{\sigma} \bigl[ \outpro{\sigma}{\sigma}^P\otimes \ca{T}^{A \rightarrow E} \circ\ca{G}_\sigma^A  \circ \ca{U}^A (X^{AR} ) \bigr] \right\|_{2,\gamma^{PER}}\nn\\
&= \sqrt{{\rm Tr}[\gamma]}\cdot \mbb{E}_{\sigma} 
\left\| \ca{T}^{A \rightarrow E} \circ\ca{G}_\sigma^A  \circ \ca{U}^A (X^{AR} )  \right\|_{2,\gamma^{ER}_{\sigma}}\nn\\
&= \sqrt{{\rm Tr}[\gamma]}\cdot \mbb{E}_{\sigma} 
\left\| \ca{T}^{A \rightarrow E} \circ\ca{G}_\sigma^A  \circ \ca{U}^A (X^{AR} )  \right\|_{2,\zeta^E \otimes \xi^R}. 
\laeq{ripopo}
\end{align}
In the last line, we used the following relation:
\begin{multline}
(\gamma^{ER}_{\sigma})^{-1/4} \bigl[ \ca{T}^{A \rightarrow E} \circ\ca{G}_\sigma^A  \circ \ca{U}^A (X^{AR} ) \bigr]  (\gamma^{ER}_{\sigma})^{-1/4}\\
= 
(\zeta^E \otimes \xi^R)^{-1/4} \bigl[ \ca{T}^{A \rightarrow E} \circ\ca{G}_\sigma^A  \circ \ca{U}^A (X^{AR} ) \bigr]  (\zeta^E \otimes \xi^R)^{-1/4},
\end{multline}
which can be observed from the fact that, due to the decomposition of $\ca{T}^{A\rightarrow E}$ from {\bf WA 2}, 
\begin{align}
\ca{T}^{A \rightarrow E} \circ\ca{G}_\sigma^A  \circ \ca{U}^A (  X^{AR} )
=\sum_{j,k=1}^J\outpro{\sigma(j)}{\sigma(k)}^{E_c}\otimes \ca{T}_{\sigma(j)\sigma(k)}^{A_r \rightarrow E_r} ( U_j^{A_r}X_{jk}^{A_rR_r}U_k^{\dagger A_r}) \otimes\outpro{j}{k}^{R_c}.
\end{align} 
Due to the fact that
\alg{
\frac{1}{|{\mbb P}|}\sum_{\sigma\in{\mbb P}}{\rm Tr}[\zeta_{\sigma(j)}^{E_r}]=\frac{1}{J}\sum_{j'=1}^J{\rm Tr}[\zeta_{j'}^{E_r}]
}
for all $j$, we obtain
\begin{align}
{\rm Tr}[\gamma]
&=\frac{1}{|{\mbb P}|}\sum_{\sigma\in{\mbb P}}\sum_{j=1}^J{\rm Tr}[\zeta_{\sigma(j)}^{E_r}]{\rm Tr}[\xi_j^{R_r}]
\nn\\
&
=\sum_{j=1}^J\left(\frac{1}{|{\mbb P}|}\sum_{\sigma\in{\mbb P}}{\rm Tr}[\zeta_{\sigma(j)}^{E_r}]\right){\rm Tr}[\xi_j^{R_r}]
\nonumber\\
&=\frac{1}{J}\sum_{j'=1}^J{\rm Tr}[\zeta_{j'}^{E_r}]\cdot \sum_{j=1}^J{\rm Tr}[\xi_{j}^{R_r}]
\nonumber\\
&=\frac{1}{J}{\rm Tr}[\zeta^E]\cdot{\rm Tr}[\xi^R]
=\frac{1}{J}.
\end{align}
Substituting this to \req{ripopo}, and by using Jensen's inequality, we arrive at the desired result.
\QED
\end{prf}

\subsubsection{Generalization of the dequantizing theorem}

Our second step to prove the non-smoothed randomized partial decoupling is to generalize the non-smoothed version of the \emph{dequantizing} theorem (Proposition 3.5 in  \cite{dupuis2014decoupling}).

\begin{lmm}
\label{lmm:extdequant}
In the same setting as in \rThm{SmoothExMarkov}, it holds that
\begin{align}
\mbb{E}_{\sigma,U } \bigl[ \bigl|\! \bigl|
\ca{T}^{A \rightarrow E} \circ \ca{G}_\sigma^A  \circ \ca{U}^A ( \Psi^{AR}  -\Psi_{\rm dp}^{AR} )
\bigr|\! \bigr|_1 \bigr]
\leq \sqrt{\alpha(J)}\cdot2^{-\frac{1}{2}H_{\rm min}(A|R)_{\Psi}-\frac{1}{2}H_{\rm min}(A|E)_{\tau}},
\laeq{ggga}
\end{align}
where we have defined $\Psi_{\rm dp}^{AR}:=\ca{C}^A(\Psi^{AR})=\sum_{j=1}^J\proj{j}^{A_c}\otm\Psi_{jj}^{A_rR}$.
\end{lmm}

Note that $\alpha(J)$ is $0$ for $J=1$ and $\frac{1}{J-1}$ for $J\geq2$.

\begin{prf}
Since $\Psi^{AR}$ and $\Psi_{\rm av}^{AR}$ are classically coherent in $A_cR_c$ by assumption, we can apply 
Lemma \ref{lmm:aidaga}  for $X^{AR}=\Psi^{AR}-\Psi_{\rm dp}^{AR}$ to obtain
\begin{equation}
\mbb{E}_{\sigma,U}  \left[ \left\|
\ca{T}^{A \rightarrow E} \circ \ca{G}_\sigma^A  \circ \ca{U}^A ( \Psi^{AR}-\Psi_{\rm dp}^{AR}) \right\|_1 \right]
\leq
\frac{1}{\sqrt{J}}\cdot\sqrt{\mbb{E}_{\sigma,U} \left\|\ca{T}^{A \rightarrow E} \circ\ca{G}_\sigma^A  \circ \ca{U}^A (  \Psi^{AR}-\Psi_{\rm dp}^{AR} )\right\|_{2,\:\zeta^E\otimes\xi^R}^2}\:.\label{ano2}
\end{equation}
Noting that $\Psi_{jj}^{AR}-\Psi_{{\rm dp},jj}^{AR}=0$, we can also apply Lemma \ref{lmm:nenchalu} under the assumption that $A_l$ is a one-dimensional system, $r_j=r$ and $\varsigma^{ER}=\zeta^E\otimes \xi^R$. Then, we obtain, for any $\sigma\in \mbb{P}$, 
\begin{align}
\mbb{E}_{U} \left[\left\|\ca{T}^{A \rightarrow E} \circ\ca{G}_{\sigma^{-1}}^A  \circ \ca{U}^A (  \Psi^{AR}-\Psi_{\rm dp}^{AR} )\right\|_{2,\:\zeta^E\otimes\xi^R}^2\right]
&
\leq 
\frac{d_A^2}{r^2}
\sum_{j\neq k} 
\bigl| \! \bigl|   \Psi_{\sigma(j)\sigma(k)}^{ A_r R} \otimes  \tau^{\bar{A}_r E}_{jk} \cdot\bigr| \! \bigr|^2_{2,\:\zeta^E\otimes\xi^R}
\nn\\
&
=
J^2
\sum_{j\neq k} 
\bigl| \! \bigl|   \Psi_{\sigma(j)\sigma(k)}^{ A_r R}\bigr| \! \bigr|^2_{2,\:\xi^R}\cdot  \bigl| \! \bigl| \tau^{\bar{A}_r E}_{jk} \bigr| \! \bigr|^2_{2,\:\zeta^E}
,
\end{align}
where we have used $d_A=rJ$ in the last line. 
Taking the case of $J=1$ into account, 
and noting that $\mbb{E}_{\sigma}[g(\sigma)]=\mbb{E}_{\sigma}[g(\sigma^{-1})]$ for any function $g$,
it follows that
\begin{align}
&\mbb{E}_{\sigma,U }  \left[\left\|\ca{T}^{A \rightarrow E} \circ\ca{G}_\sigma^A  \circ \ca{U}^A (  \Psi^{AR}-\Psi_{\rm dp}^{AR} )\right\|_{2,\:\zeta^E\otimes\xi^R}^2\right]\nonumber\\
&=\mbb{E}_{\sigma,U }  \left[\left\|\ca{T}^{A \rightarrow E} \circ\ca{G}_{\sigma^{-1}}^A  \circ \ca{U}^A (  \Psi^{AR}-\Psi_{\rm dp}^{AR} )\right\|_{2,\:\zeta^E\otimes\xi^R}^2\right]\nonumber\\
&\leq
J^2
\sum_{ j\neq k} 
\mbb{E}_{\sigma}\left[\bigl| \! \bigl|\Psi_{\sigma(j)\sigma(k)}^{ A_r R}  \bigr| \! \bigr|^2_{2,\xi^R}\right]
\cdot  \bigl| \! \bigl|   \tau^{A_r E}_{jk} \bigr| \! \bigr|^2_{2,\zeta^E}\nonumber
\\
&=J\alpha(J)
 \sum_{j'\neq k'}\bigl| \! \bigl|\Psi_{j'k'}^{ A_r R}  \bigr| \! \bigr|^2_{2,\xi^R}
 \cdot
 \sum_{j\neq k} 
\bigl| \! \bigl|   \tau^{A_r E}_{jk} \bigr| \! \bigr|^2_{2,\zeta^E}\nonumber\\
&=J\alpha(J)
\left\|\sum_{j'\neq k'}\outpro{j'}{k'}^{A_c}\otimes\Psi^{ A_r R}_{j'k'}
\right\|_{2,\xi^R}^2
\cdot
\left\|\sum_{j\neq k}\outpro{j}{k}^{A_c}\otimes
 \tau^{ A_r E}_{jk}\right\|_{2,\zeta^E}^2 \nonumber\\
&= J\alpha(J)
\left\|\Psi^{AR}-\Psi_{\rm dp}^{AR}\right\|_{2,\xi^R}^2
\cdot
\left\|\tau^{AE}-\tau_{\rm dp}^{AE}\right\|_{2,\zeta^E}^2
\nonumber\\
&\leq J\alpha(J)
\left\|\Psi^{AR}\right\|_{2,\xi^R}^2
\cdot
\left\|\tau^{AE}\right\|_{2,\zeta^E}^2
\nonumber\\
&= J\alpha(J)
\cdot2^{-H_2(A|R)_{\Psi|\xi}-H_2(A|E)_{\tau|\zeta}}.\label{ano}
\end{align}
Here, we have used the definitions $\Psi_{\rm dp}^{AR}:=\ca{C}^A(\Psi^{AR})$ and $\tau_{\rm dp}^{AE}:=\ca{C}^A(\tau^{AE})$ in the sixth line, and \rLmm{OPvar} in the seventh line.
Due the relation between the conditional collision entropy and the conditional min-entropy (Lemma \ref{Lemma:Candmin}), it is further bounded from above by $2^{-H_{\rm min}(A|R)_{\Psi|\xi}-H_{\rm min}(A|E)_{\tau|\zeta}}$.

Finally, we use the property of the the conditional min-entropy (\rLmm{clch}). There exist normalized states $\xi$ and $\zeta$ in the form of 
\begin{align}
\xi^{R} =\!\sum_j\outpro{j}{j}^{R_c}\!\otimes\xi_j^{R_r}, \quad
\zeta^{E} &=\!\sum_j\outpro{j}{j}^{E_c}\!\otimes\zeta_j^{E_r},
\end{align}
such that $H_{\rm min}(A|R)_{\Psi|\xi}=H_{\rm min}(A|R)_{\Psi}$ and $H_{\rm min}(A|E)_{\tau|\zeta}=H_{\rm min}(A|E)_{\tau}$.
Thus, we obtain
\begin{align}
&\mbb{E}_{\sigma,U }  \left[\left\|\ca{T}^{A \rightarrow E} \circ\ca{G}_\sigma^A  \circ \ca{U}^A (  \Psi^{AR}-\Psi_{\rm dp}^{AR} )\right\|_{2,\:\zeta^E\otimes\xi^R}^2\right]
\leq
J\alpha(J)
\cdot2^{-H_{\rm min}(A|R)_{\Psi}-H_{\rm min}(A|E)_{\tau}},
\end{align}
which, together with Ineq.~(\ref{ano2}), complete the proof of \rLmm{extdequant}.
\hfill$\blacksquare$
\end{prf}

\subsubsection{Proof of The Non-Smoothed Randomized Partial Decoupling}

We now prove the non-smoothed randomized partial decoupling, i.e.,
\begin{multline}
\mbb{E}_{U \sim {\sf H}_{\times}, \sigma\sim{\sf P}} \left[ \left\|
\ca{T}^{A \rightarrow E} \circ \ca{G}_\sigma^A  \circ \ca{U}^A ( \Psi^{AR} ) -\ca{T}^{A \rightarrow E} \circ \ca{G}_\sigma^A ( \Psi_{\rm av}^{AR} )
\right\|_1 \right]\\
\leq  
\sqrt{\alpha(J)}\cdot2^{-\frac{1}{2}H_{\rm min}(A|R)_{\Psi}-\frac{1}{2}H_{\rm min}(A|E)_{\tau}}
+\beta(A_r)\cdot2^{-\frac{1}{2}H_{\rm min}(A|R)_{\ca{C}(\Psi)}-\frac{1}{2}H_{\rm min}(A|E)_{\ca{C}(\tau)}}, \label{Eq;areo4f}
\end{multline}
under the assumptions {\bf WA 1} and {\bf WA 2}.
Note that $\beta(A_r)$ is $0$ for ${\rm dim}\ca{H}^{A_r}=1$ and $1$ for ${\rm dim}\ca{H}^{A_r}\geq2$.
By the triangle inequality, we have
\begin{multline}
\mbb{E}_{\sigma,U } \bigl[ \bigl|\! \bigl|
\ca{T}^{A \rightarrow E} \circ \ca{G}_\sigma^A  \circ \ca{U}^A ( \Psi^{AR} ) -\ca{T}^{A \rightarrow E}  \circ \ca{G}_\sigma^A  ( \Psi_{\rm av}^{AR} )
\bigr|\! \bigr|_1 \bigr]\\
\leq
\mbb{E}_{\sigma,U } \bigl[ \bigl|\! \bigl|
\ca{T}^{A \rightarrow E} \circ \ca{G}_\sigma^A  \circ \ca{U}^A ( \Psi^{AR}  - \Psi_{\rm dp}^{AR} )
\bigr|\! \bigr|_1 \bigr]
+
\mbb{E}_{\sigma,U } \bigl[ \bigl|\! \bigl|
\ca{T}^{A \rightarrow E} \circ \ca{G}_\sigma^A  \circ \ca{U}^A ( \Psi_{\rm dp}^{AR}  - \Psi_{\rm av}^{AR} )
\bigr|\! \bigr|_1 \bigr],
\laeq{101112}
\end{multline}
where we have used the fact that the unitary invariance of the Haar measure implies $\ca{U}^A(\Psi_{\rm av}^{AR})=\Psi_{\rm av}^{AR}$ for any unitary $U$.
The first term is bounded by simply using Lemma \ref{lmm:extdequant}.

To bound the second term in \req{101112}, we use \rLmm{aidaga}, leading to
\begin{multline}
\mbb{E}_{\sigma,U }  \bigl[ \bigl|\! \bigl|
\ca{T}^{A \rightarrow E} \circ \ca{G}_\sigma^A  \circ \ca{U}^A ( \Psi^{AR}_{\rm dp}-\Psi_{\rm av}^{AR}) \bigr|\! \bigr|_1 \bigr]\\
\leq
\frac{1}{\sqrt{J}}\cdot\sqrt{\mbb{E}_{\sigma,U } \left\|\ca{T}^{A \rightarrow E}\! \circ\!\ca{G}_\sigma^A \! \circ\! \ca{U}^A (  \Psi^{AR}_{\rm dp}\!-\!\Psi_{\rm av}^{AR} )\right\|_{2,\:\zeta^E\otimes\xi^R}^2}\:.\label{ano3}
\end{multline}
Since $\Psi_{{\rm dp},jj}^{R}=\Psi_{{\rm av},jj}^{R}$ by definition, we can apply \rLmm{nenchalu} for $X^{AR}=\Psi^{AR}_{\rm dp}-\Psi_{\rm av}^{AR}$. 
Noting that $\Psi_{{\rm dp},jk}^{ A_r R}=\Psi_{{\rm av},jk}^{ A_r R}=0$ for $j\neq k$, this yields
\begin{align}
\mbb{E}_{U} \left[\left\|\ca{T}^{A \rightarrow E}  \circ \ca{G}_\sigma^A \circ \ca{U}^A (  \Psi_{\rm dp}^{AR}-\Psi_{\rm av}^{AR} )\right\|_{2,\zeta^E\otimes\xi^R}^2\right]
&\quad
\leq 
\frac{d_A^2}{r^2}
\sum_{j =1}^J 
\left\|    \Psi_{\sigma(j)\sigma(j)}^{ A_r R} \otimes \tau^{ \bar{A}_r E}_{jj}  \right\|^2_{2, \zeta^E\otimes\xi^R}
\nonumber\\
&\quad=J^2
\sum_{j =1}^J 
\bigl| \! \bigl|\Psi_{\sigma(j)\sigma(j)}^{ A_r R}  \bigr| \! \bigr|^2_{2,\xi^R}
\cdot \bigl| \! \bigl|   \tau^{A_r E}_{jj} \bigr| \! \bigr|^2_{2,\zeta^E}.
\end{align}
Thus, similarly to the derivation around Eq.~\eqref{ano}, we obtain
\begin{equation}
\mbb{E}_{\sigma,U }  \left[\left\|\ca{T}^{A \rightarrow E} \circ\ca{G}_\sigma^A  \circ \ca{U}^A (  \Psi^{AR}_{\rm dp}-\Psi_{\rm av}^{AR} )\right\|_{2,\:\zeta^E\otimes\xi^R}^2\right]
\leq J\cdot
2^{-H_{\rm min}(A|R)_{\Psi_{\rm dp}}-H_{\rm min}(A|E)_{\tau_{\rm dp}}}.
\end{equation}
Substituting this into Ineq.~\eqref{ano3}, and noting that $\Psi_{\rm dp}^{AR}  -  \Psi_{\rm av}^{AR}=0$ if ${\rm dim}\ca{H}^{A_r}=1$, we obtain an upper bound on the second term of the R.H.S. in Ineq.~\req{101112}. 

All together, we obtain Ineq.~\eqref{Eq;areo4f} as desired.
\hfill$\blacksquare$

\subsection{Proof of The Randomized Partial Decoupling under The Conditions {\bf WA 1} and {\bf WA 2}} 
\lsec{prfSEM}
\label{SS:SRPD}

We now show, under the conditions {\bf WA 1} and {\bf WA 2}, the randomized partial decoupling:
\begin{multline}
\mbb{E}_{U \sim {\sf H}_{\times}, \sigma\sim{\sf P} } \left[ \left\|
\ca{T}^{A \rightarrow E} \circ \ca{G}_\sigma^A  \circ \ca{U}^A ( \Psi^{AR} ) -\ca{T}^{A \rightarrow E} \circ \ca{G}_\sigma^A ( \Psi_{\rm av}^{AR} )
\right\|_1 \right]\\
\!\leq  
\sqrt{\alpha(J)}\cdot2^{-\frac{1}{2}\tilde{H}_I}
+\beta(A_r)\cdot2^{-\frac{1}{2}\tilde{H}_{I\!I}}
+4(\epsilon\cdot{\rm Tr}[\tau]+\mu+\epsilon\mu), \laeq{SmExMa}
\end{multline}
where 
$\Psi_{\rm av}^{AR}:=\mbb{E}_{U \sim {\sf H}_{\times}} [ \ca{U}^A ( \Psi^{AR} )  ]$.
The function $\alpha(J)$ is $0$ for $J=1$ and $\frac{1}{J-1}$ for $J\geq2$, and $\beta(A_r)$ is $0$ for ${\rm dim}\ca{H}^{A_r}=1$ and $1$ for ${\rm dim}\ca{H}^{A_r}\geq2$.
The exponents $\tilde{H}_I$ and $\tilde{H}_{I\!I}$ are given by
\alg{
\tilde{H}_I=
H_{\rm min}^\epsilon(A|R)_{\Psi} + H_{\rm min}^\mu(A|E)_{\tau},
\quad
\tilde{H}_{I\!I}=
H_{\rm min}^\epsilon(A|R)_{\ca{C}(\Psi)} + H_{\rm min}^\mu(A|E)_{\ca{C}(\tau)}.
}
Note that, the duality of the conditional smooth entropies for pure states (\rLmm{duality}), implies $H_{\rm min}^\mu(A|E)_{\tau}=-H_{\rm max}^\mu(A|B)_{\ca{C}(\tau)}$ and $H_{\rm min}^\mu(A|E)_{\ca{C}(\tau)}=-H_{\rm max}^\mu(A_r|BA_c)_{\ca{C}(\tau)}$ (see Subsection \ref{SS:DA1} for the detail).

To prove the statement, we again start with the triangle inequaltiy:
By the triangle inequality, we have
\begin{multline}
\mbb{E}_{\sigma,U } \bigl[ \bigl|\! \bigl|
\ca{T}^{A \rightarrow E} \circ \ca{G}_\sigma^A  \circ \ca{U}^A ( \Psi^{AR} ) -\ca{T}^{A \rightarrow E}  \circ \ca{G}_\sigma^A  ( \Psi_{\rm av}^{AR} )
\bigr|\! \bigr|_1 \bigr]\\
\leq
\mbb{E}_{\sigma,U } \bigl[ \bigl|\! \bigl|
\ca{T}^{A \rightarrow E} \circ \ca{G}_\sigma^A  \circ \ca{U}^A ( \Psi^{AR}  - \Psi_{\rm dp}^{AR} )
\bigr|\! \bigr|_1 \bigr]
+
\mbb{E}_{\sigma,U } \bigl[ \bigl|\! \bigl|
\ca{T}^{A \rightarrow E} \circ \ca{G}_\sigma^A  \circ \ca{U}^A ( \Psi_{\rm dp}^{AR}  - \Psi_{\rm av}^{AR} )
\bigr|\! \bigr|_1 \bigr]. \label{4mkeroerdds}
\end{multline}
Below, we derive upper bounds on the two terms in the R.H.S. separately.

For an upper bound on the first term, fix $\hat\Psi\in\ca{B}^\epsilon(\Psi)$ and $\hat\tau\in\ca{B}^\mu(\tau)$ so that we have $H_{\rm min}(A|R)_{\hat\Psi}=H_{\rm min}^\epsilon(A|R)_{\Psi}$ and $H_{\rm min}(A|E)_{\hat\tau}=H_{\rm min}^\epsilon(A|E)_{\tau}$.
Let $\Delta_+^{A'E}$ and $\Delta_-^{A'E}$ be linear operators on $\ca{H}^{A'}\otimes\ca{H}^E$ such that 
\alg{
\Delta_+^{A'E}\geq0,\; \Delta_-^{A'E}\geq0,
\;
{\rm supp}[\Delta_+^{A'E}]\perp{\rm supp}[\Delta_-^{A'E}]
\laeq{Deltaorthooo}
}
and that
\begin{align}
\tau^{A'E} -\hat{\tau}^{A'E}=\Delta_{+}^{A'E}-\Delta_{-}^{A'E}.
\laeq{plmnnn}
\end{align} 
Let $\ca{D}_+^{A\rightarrow E}$ and $\ca{D}_-^{A\rightarrow E}$ be superoperators such that
\alg{
\ca{D}_{+}^{A\rightarrow E}(\Phi^{AA'})=\Delta_{+}^{A'E},
\quad\ca{D}_{-}^{A\rightarrow E}(\Phi^{AA'})=\Delta_{-}^{A'E},
} 
which yields $\ca{T}-\hat{\ca{T}}=\ca{D}_+-\ca{D}_-$.
From Lemma~\ref{lmm:trianglemother}, the   CP map $\hat{\ca{T}}^{A \rightarrow E}$ having the Choi-Jamio\l kowski state $\hat{\tau}^{AE}$ satisfies
\begin{multline}
\mbb{E}_{\sigma,U }
\left[
\left\|{\ca{T}}^{A \rightarrow E}  \circ \ca{G}_\sigma^A \circ \ca{U}^A ( {\Psi}^{AR}  -  \Psi_{\rm dp}^{AR} )\right\|_1
\right]\\
\leq
\mbb{E}_{\sigma,U }
\left[
\left\|\hat{\ca{T}}^{A \rightarrow E} \circ \ca{G}_\sigma^A \circ \ca{U}^A ( \hat{\Psi}^{AR}  - 
 \hat{\Psi}_{\rm dp}^{AR} )\right\|_1
\right]
+2\: \mbb{E}_{\sigma} \left[
\tr[(\ca{D}_+^{A \rightarrow E}+ \ca{D}_-^{A \rightarrow E}) \circ \ca{G}_\sigma^A  ( \Psi_{\rm av}^{AR} )]
\right]
\\
+2\:\mbb{E}_{\sigma,U}
\left[
\tr[\hat{\ca{T}}^{A \rightarrow E}  \circ \ca{G}_\sigma^A \circ \ca{U}^A (\delta_+^{AR}+\delta_-^{AR})]
\right].
\end{multline}
Due to Lemma \ref{lmm:extdequant}, the first term in the R.H.S. of the above inequality is bounded as
\begin{align}
\mbb{E}_{\sigma,U }
\left[
\left\|\hat{\ca{T}}^{A \rightarrow E} \circ \ca{G}_\sigma^A \circ \ca{U}^A ( \hat{\Psi}^{AR}  -  \hat{\Psi}_{\rm dp}^{AR} )\right\|_1
\right]
\leq \sqrt{\alpha(J)}\cdot2^{-\frac{1}{2}H_{\rm min}(A|R)_{\hat\Psi}-\frac{1}{2}H_{\rm min}(A|E)_{\hat\tau}}.
\end{align}
Similarly to \req{grapefrp} and \req{grapefrp2}, using \req{Deltaorthooo} and \req{plmnnn}, it turns out that the second and the third terms are bounded from above by
\begin{equation}
\mbb{E}_{\sigma} \left[
\tr[(\ca{D}_+^{A \rightarrow E}+ \ca{D}_-^{A \rightarrow E}) \circ \ca{G}_\sigma^A   ( \Psi_{\rm av}^{AR} )]
\right] \leq\mu
\end{equation}
and
\begin{equation}
\mbb{E}_{\sigma,U}
\left[
\tr[\hat{\ca{T}}^{A \rightarrow E}  \circ \ca{G}_\sigma^A \circ \ca{U}^A (\delta_+^{AR}+\delta_-^{AR})]
\right]
\leq
\epsilon\cdot\tr[\tau]+\epsilon\mu,
\end{equation}
respectively.
Hence, we obtain
\begin{multline}
\mbb{E}_{\sigma,U }
\left[
\left\|{\ca{T}}^{A \rightarrow E}  \circ \ca{G}_\sigma^A \circ \ca{U}^A ( {\Psi}^{AR}  -  \Psi_{\rm dp}^{AR} )\right\|_1
\right] \\
\leq
 \sqrt{\alpha(J)}\cdot2^{-\frac{1}{2}H_{\rm min}^\epsilon(A|R)_{\Psi}-\frac{1}{2}H_{\rm min}^\mu(A|E)_{\tau}}
 +
2( \epsilon\cdot\tr[\tau]+\mu+\epsilon\mu).
\end{multline}
In the same way, we also have
\begin{multline}
\mbb{E}_{\sigma,U}
\left[
\left\|\ca{T}^{A \rightarrow E} \circ \ca{G}_\sigma^A \circ \ca{U}^A ( \Psi_{\rm dp}^{AR}  -  \Psi_{\rm av}^{AR} )\right\|_1
\right]
\\
\leq \beta(A_r)\cdot2^{-\frac{1}{2}H_{\rm min}^\epsilon(A|R)_{\Psi_{\rm dp}}-\frac{1}{2}H_{\rm min}^\mu(A|E)_{\tau_{\rm dp}}}
+ 2(\epsilon\cdot\tr[\tau]+\mu+\epsilon\mu).
\nn
\end{multline}
Substituting these inequalities into Eq.~\eqref{4mkeroerdds}, we obtain the desired result (Ineq.~\req{SmExMa}).\QED

\subsection{Dropping Working Assumptions {\bf WA 1} and {\bf WA 2}}
\lsec{prfSEM2}
\label{SS:DA1}

We now drop the working assumptions {\bf WA 1} and {\bf WA 2}, and show that \rThm{SmoothExMarkov} holds in general. 
To remind the working assumptions, we write them down here again: 
\begin{itemize}
\setlength{\leftskip}{0.7cm}
\item[{\bf WA 1}]  $E\cong E_cE_r$, where $E_c$ is a quantum system of dimension $J$ \vspace{-3mm}
\item[{\bf WA 2}] The CP map $\ca{T}^{A \rightarrow E}$ is decomposed into
\begin{align}
\ca{T}^{A \rightarrow E}(X)=\sum_{j,k=1}^J\outpro{j}{k}^{E_c}\otimes\ca{T}_{jk}^{A_r \rightarrow E_r}(X_{jk}),
\end{align}
in which $\ca{T}_{jk}$ is a linear supermap from $\ca{L}(\ca{H}^{A_r})$ to $\ca{L}(\ca{H}^{E_r})$ defined by $\ca{T}_{jk}(\zeta)=\ca{T}(\outpro{j}{k}\otimes\zeta)$ for each $j,k$,
\end{itemize}

To drop these assumptions, we use \rLmm{dropassump}. Using the linear isometry $Y^{A_c\rightarrow A_cE_c}$, given by $Y=\sum_{j}\ket{jj}^{A_cE_c}\bra{j}^{A_c}$, we define a new   CP map $\check{\ca{T}}^{A \rightarrow EE_c}$ by $\ca{T}^{A \rightarrow E} \circ \mathcal{Y}^{A_c \rightarrow A_c E_c}$.
\rLmm{dropassump} states that 
\begin{align}
\left\|\check{\ca{T}}^{A \rightarrow EE_c}  \circ \ca{G}_\sigma^A \circ \ca{U}^A ( {\Psi}^{AR}  -  \Psi_{\rm av}^{AR} )\right\|_1
&=
\left\|\ca{T}^{A \rightarrow E}  \circ \ca{G}_\sigma^A \circ \ca{U}^A ( {\Psi}^{AR}  -  \Psi_{\rm av}^{AR} )\right\|_1. \label{a}
\end{align}

Let $\check{\tau}^{AEE_c}$ be the Choi-Jamio\l kowski state of ${\check{\ca{T}}}^{A \rightarrow EE_c}$, i.e., $\check{\tau}^{AEE_c}:=\mfk{J}(\check{\ca{T}}^{A \rightarrow EE_c})$.
We denote by $|\tau\rangle^{ABE}$ a purification of $\tau^{AE}$  such that the reduced state $\tau^{AB}$ is equal to $\mfk{J}(\ca{T}^{A\rightarrow B})$, where $\ca{T}^{A\rightarrow B}$ is the complementary map of $\ca{T}^{A\rightarrow E}$. 
Then, it is clear that $\check{\tau}^{AEE_c}=\ca{Y}(\tau^{AE})$, which implies that a purification $|\check{\tau}\rangle^{ABEE_c}$ of $\check{\tau}^{AEE_c}$ is given by $|\check{\tau}\rangle^{ABEE_c}=Y|\tau\rangle^{ABE}$. It is also straightforward to verify that $\check{\tau}^{AB}=\ca{C}(\tau^{AB})$.

The new   CP map $\check{\ca{T}}^{A \rightarrow EE_c}$ clearly satisfies {\bf WA 1} and {\bf WA 2}.
Hence, using Eq.~\eqref{a} and achievability of the randomized partial decoupling under those assumptions (Ineq.~\req{SmExMa}), we obtain
\begin{align}
&\mbb{E}_{\sigma,U }
\left[
\left\|\ca{T}^{A \rightarrow E}  \circ \ca{G}_\sigma^A \circ \ca{U}^A ( {\Psi}^{AR}  -  \Psi_{\rm av}^{AR} )\right\|_1
\right] \nn \\
&=
\mbb{E}_{\sigma,U }
\left[
\left\|\check{\ca{T}}^{A \rightarrow EE_c}  \circ \ca{G}_\sigma^A \circ \ca{U}^A ( {\Psi}^{AR}  -  \Psi_{\rm av}^{AR} )\right\|_1
\right]\nn\\
&\leq
 \sqrt{\alpha(J)}\cdot2^{-\frac{1}{2}H_{\rm min}^\epsilon(A|R)_{\Psi}-\frac{1}{2}H_{\rm min}^{\mu}(A|EE_c)_{\check{\tau}}}
 \nn\\
 &\quad\quad
 +\beta(A_r)\cdot2^{-\frac{1}{2}H_{\rm min}^\epsilon(A|R)_{\ca{C}(\Psi)}-\frac{1}{2}H_{\rm min}^{\mu}(A|EE_c)_{\ca{C}(\check{\tau})}}
+
4( \epsilon\cdot\tr[\check{\tau}]+\mu+\epsilon\mu).
\laeq{aldalk2}
\end{align}
Due to the duality of conditional smooth entropies (\rLmm{duality}),
we have
\alg{
H_{\rm min}^{\mu}(A|EE_c)_{\check{\tau}}
=
-H_{\rm max}^{\mu}(A|B)_{\check{\tau}}
=
-H_{\rm max}^{\mu}(A|B)_{\ca{C}(\tau)}.
\laeq{sannnwa3}
}
Using the property of the conditional smooth entropy for classical-quantum states (\rLmm{condminCQCQ}), and noting that $\check{\tau}^{AEE_c}$ is classically coherent in $A_cE_c$, we also have
\alg{
H_{\rm min}^{\mu}(A|EE_c)_{\ca{C}(\check{\tau})}
=
H_{\rm min}^{\mu}(A_r|EE_c)_{\check{\tau}}
=
-H_{\rm max}^{\mu}(A_r|BA_c)_{\check{\tau}}
=
-H_{\rm max}^{\mu}(A_r|BA_c)_{\ca{C}(\tau)}.
\laeq{sannnwa4}
}
Substituting these into \req{aldalk2}, and noting that $\tr[\check{\tau}]=\tr[\tau]\leq1$ by assumption, we obtain \rThm{SmoothExMarkov}. \hfill$\blacksquare$

\section{Proof of The Converse}
\lsec{converse}

We provide the proof of \rThm{converse} under {\bf Converse Conditions 1} and {\bf 2}, which are 
\begin{description}
\setlength{\itemindent}{5pt}
\item[CC 1] $\dim\ca{H}_j^l=1,\quad\dim\ca{H}_j^r=r \quad(j=1,\cdots, J)$, \vspace{-2mm}
\item[CC 2] the initial (normalized) state $\Psi^{AR}$ is classically coherent in $A_cR_c$.
\end{description}

The proof proceeds along the similar line as the proof of the converse part of the one-shot decoupling theorem (see Section 4 in  \cite{DBWR2010}). Suppose that there exists a normalized state $\Omega^{ER}:=\sum_{j=1}^J\varsigma_j^E\otm\Psi_{jj}^{R_r}\otm\proj{j}^{R_c}$, where $\{\varsigma_j\}_{j=1}^J$ are normalized states on $E$, such that, for $\delta>0$,
\alg{
\left\|
\ca{T}^{A \rightarrow E} ( \Psi^{AR} ) -\Omega^{ER}
\right\|_1
\leq
\delta. \laeq{keshikeshiit}
}
We separately prove that,  in this case,
the following inequalities hold for any $\upsilon\in[0,1/2)$ and $\iota\in(0,1]$:
\alg{
&
H_{\rm min}^{\lambda}(A|R)_\Psi
+
H_{\rm max}^{\upsilon}(RD|E)_{\ca{T}(\Psi)}
+\log{J}
\geq
\log{\iota},
\laeq{atatatata}
\\
&
H_{\rm min}^{\lambda'}(A|R)_{\ca{C}(\Psi)}
+
H_{\rm max}^{\upsilon}(RD|E)_{\ca{T}\circ\ca{C}(\Psi)}
\geq
\log{\iota}+\log{(1-2\upsilon)}.
\laeq{atatatate}
}
Here, $\lambda$ and $\lambda'$ are given by
\alg{
&
\lambda:=
2\sqrt{\iota+4\sqrt{20\upsilon+2\delta}}
+\sqrt{2\sqrt{20\upsilon+2\delta}}+2\sqrt{2\delta}
+2\sqrt{20\upsilon+2\delta}
+3\upsilon,
\laeq{dfndeltadelta2}\\
&
\lambda':=
\upsilon+\sqrt{4\sqrt{\iota+2x}+2\sqrt{x}+(4\sqrt{\iota+8}+24) x}
\laeq{dfndeltadelta489}
}
and $x:=\sqrt{2}\sqrt[4]{24\upsilon+2\delta}$.

First, we prove these relations based on the working assumptions {\bf WA 1} and {\bf WA 2} in Subsection~\ref{SS:aaaaaerfe} and \rsec{prfdpconv}. We complete the proof of \rThm{converse} by dropping these assumptions in Subsection~\ref{SS:DA2}.

\subsection{Proof of Ineq.~\req{atatatata} under {\bf WA 1} and {\bf WA 2}} \label{SS:aaaaaerfe}

To prove Ineq.~\req{atatatata}, we introduce the following notations:
\begin{itemize}

\item $\ket{\Psi}^{ARD}$ : A purification of $\Psi^{AR}$. \vspace{-2mm}

\item $V^{A\rightarrow BE}$ : A Stinespring dilation of $\ca{T}^{A\rightarrow E}$.  \vspace{-2mm}

\item $\ket{\Theta}^{BERD}$ : A pure state on $BERD$ defined by
$\ket{\Theta}:=V\ket{\Psi}$.
\vspace{-2mm}

\item $\ket{\theta}^{BERD}$ : A subnormalized pure state on $BERD$ such that
\alg{
H_{\rm max}(RD|E)_\theta=H_{\rm max}^{\upsilon}(RD|E)_{\Theta},
\quad
P(\theta^{BERD},\Theta^{BERD})\leq\upsilon,
\laeq{sigmatildeup}
}
which is classically coherent in $E_cR_c$.

\end{itemize}

\noindent
Note that the existence of $\ket{\theta}$ satisfying the above condition follows from \rLmm{clchtwo} about the property of the conditional max-entropy for classically coherent states.
From the definition of the conditional max-entropy, and from the definitions of $\theta$ and $\Theta$, we have
\alg{
H_{\rm max}(RD|E)_{\theta|\theta}
\leq
H_{\rm max}(RD|E)_{\theta}
=
H_{\rm max}^{\upsilon}(RD|E)_{\Theta}
=
H_{\rm max}^{\upsilon}(RD|E)_{\ca{T}(\Psi)}.
\laeq{rosier}
}

The proof of Ineq.~\req{atatatata} proceeds as follows. First, we prove that for any $X\in\ca{P}(\ca{H}^{ER})$, we can construct a subnormalized pure state $\ket{\theta_X}^{BERD}$ from $\theta$ and $X$ such that
\alg{
\theta^{BER}_X
\leq
\frac{
2^{H_{\rm max}(RD|E)_{\theta|\theta}}
}{\iota}
\cdot 
I^B
\otm
X^{ER}.
\laeq{saayon}
}
Second, we prove that if $X^{ER}$ satisfies certain conditions, the $\theta_X$ satisfies
\alg{
H_{\rm min}(BE|R)_{\theta_X}
\leq
H_{\rm min}^{\lambda}(A|R)_{\Psi}.
 \laeq{gravity}
}
Third, we prove that for a proper choice of $X^{ER}$ satisfying the conditions for \req{gravity}, Ineq.~\req{saayon} implies
\alg{
H_{\rm min}(BE|R)_{\theta_X}
+H_{\rm max}(RD|E)_{\theta|\theta}
+\log{J}
\geq
\log{\iota}.
\laeq{CEUO}
}
Combining \req{rosier}, \req{gravity} and \req{CEUO}, we arrive at \req{atatatata}.

Before we start, we remark that the partial decoupling condition \req{keshikeshiit} is used in the proof of \req{gravity}, particularly when we evaluate the smoothing parameter $\lambda$.

\subsubsection{Proof of Ineq.~\req{saayon}}

Define
$
Y^{ERD}:=
2^{-\frac{1}{2}H_{\rm max}(RD|E)_{\theta|\theta}}
\cdot
(\theta^E)^{-\frac{1}{2}}
\sqrt{
(\theta^E)^{\frac{1}{2}}
\theta^{ERD}(\theta^E)^{\frac{1}{2}}
}
(\theta^E)^{-\frac{1}{2}}.
\laeq{dfnY}
$
Due to \rLmm{ZYX}, it holds that
$
\theta^{BERD}
\leq
2^{H_{\rm max}(RD|E)_{\theta|\theta}}
\cdot
I^B
\otm
Y^{ERD}
$
and thus
\alg{
\theta^{BER}
\leq
2^{H_{\rm max}(RD|E)_{\theta|\theta}}
\cdot
I^B
\otm
Y^{ER}.
\laeq{inhi}
}
Let  $X\in\ca{P}(\ca{H}^{ER})$ be an arbitrary positive semidefinite operator, and
define
\alg{
\Gamma_X^{ER}&:=
\sqrt{1-\iota}\cdot (X^{ER})^{\frac{1}{2}}((1-\iota)\cdot X^{ER}+\iota\cdot Y^{ER})^{-\frac{1}{2}}
\laeq{kajikai}
}
and
$\ket{\theta_X}^{BERD}:=\Gamma_X^{ER}\ket{\theta}^{BERD}$.
From \req{inhi}, $X\geq0$ and the assumption that $\iota\leq1$,
it follows that
\alg{
\theta^{BER}
\leq
\frac{
2^{H_{\rm max}(RD|E)_{\theta|\theta}}
}{\iota}
\cdot
I^B
\otm
((1-\iota)\cdot X^{ER}+\iota\cdot Y^{ER}),
}
and consequently,
\alg{
\theta^{BER}_X
=
\Gamma_X^{ER}\theta^{BER}\Gamma_X^{\dagger ER}
\leq
\frac{
(1-\iota)\cdot
2^{H_{\rm max}(RD|E)_{\theta|\theta}}
}{\iota}
\cdot 
I^B
\otm
X^{ER}
\leq
\frac{
2^{H_{\rm max}(RD|E)_{\theta|\theta}}
}{\iota}
\cdot 
I^B
\otm
X^{ER}.
}

\subsubsection{Proof of Ineq.~\req{gravity}}

Define a subnormalized probability distribution $\bigl\{q_k :=\|\bra{k}^{R_c}\ket{\theta}\|_1^2 \bigr\}_{k=1}^J$,
and normalized pure states $\ket{\theta_k}^{E_rR_r}$ by
$\ket{\theta_k}^{E_rR_r}:=q_k^{-1/2}\bra{k}^{E_c}\bra{k}^{R_c}\ket{\theta}$
for $k$ such that $q_k>0$.
Let $\omega\in\ca{S}_\leq(\ca{H}^{ER})$ be a subnormalized state defined by
\alg{
\omega^{ER}&:=\sum_{k:q_k>0}q_k\proj{k}^{E_c}\otm\theta_k^{E_r}\otm\theta_k^{R_r}\otm\proj{k}^{R_c},
\laeq{dfnbas}
}
where $\theta_k^{E_r}$ and $\theta_k^{R_r}$ are reduced states of $\ket{\theta_k}$ on $E_r$ and $R_r$, respectively.
Consider an arbitrary $X\in\ca{P}(\ca{H}^{ER})$ so that
\alg{
[(X^{ER})^{-\frac{1}{2}},\omega^{ER}]=0
\laeq{mewotoji}
}
and
\alg{
(\theta^E)^{-\frac{1}{2}}(X^{ER})^{-\frac{1}{2}}\omega^{ER}(X^{ER})^{-\frac{1}{2}}(\theta^E)^{-\frac{1}{2}}
=\sum_{k:q_k>0}\proj{k}^{E_c}\otm I_k^{E_r}\otm I_k^{R_r}\otm\proj{k}^{R_c}.
\laeq{moisttt}
}
As we prove in Appendix \rsec{Evdelta}, for any such $X$, the state $\ket{\theta_X}$ is a subnormalized pure state, and the partial decoupling condition \req{keshikeshiit} implies
\alg{
P(\theta^{BER}_X,\Theta^{BER})
\leq
\lambda,
\laeq{laaa}
}
where $\lambda$ is defined by \req{dfndeltadelta2}.
Due to the definition of $\Theta$ and the invariance of min-entropy under local isometry (\rLmm{invCEiso}), we obtain
\alg{
H_{\rm min}(BE|R)_{\theta_X}
\leq
H_{\rm min}^{\lambda}(BE|R)_{\Theta}
=H_{\rm min}^{\lambda}(A|R)_{\Psi}.
}

\subsubsection{Proof of Ineq.~\req{CEUO}}

We choose a proper $X^{ER}$ satisfying Conditions \req{mewotoji} and \req{moisttt}, and
prove Ineq. \req{CEUO} from \req{saayon}.
Define a normalized state
\alg{
\hat{\theta}^R:=\frac{1}{J'}\sum_{k:q_k>0}\proj{k}^{R_c}\otm\theta_k^{R_r}
}
where $J':=|\{k|1\leq k\leq J,\:q_k>0\}|$, and
$X^{ER}:=J'\cdot I^E\otm\hat{\theta}^R$.
Noting that $\theta$ is classically coherent in $E_cR_c$, it is straightforward to verify that
\alg{
(X^{ER})^{-\frac{1}{2}}
=
\sum_{k:q_k>0}I^E\otm\proj{k}^{R_c}\otm(\theta_k^{R_r})^{-\frac{1}{2}},
\quad
(\theta^E)^{-\frac{1}{2}}
=
\sum_{k:q_k>0}
q_k^{-\frac{1}{2}}
\proj{k}^{E_c}\otm(\theta_k^{E_r})^{-\frac{1}{2}}.
}
Consequently, $X^{ER}$ satisfies Conditions \req{mewotoji} and \req{moisttt}.

Using Ineq. \req{saayon}, we have
\alg{
\theta^{BER}_X
\leq
\frac{
J'\cdot 2^{H_{\rm max}(RD|E)_{\theta|\theta}}
}{\iota}
I^{BE}
\otm
\hat{\theta}^R,
\laeq{kisakisa}
}
which implies, together from the definition of the conditional min-entropy and $J'\leq J$, that
\alg{
H_{\rm min}(BE|R)_{\theta_X}
+H_{\rm max}(RD|E)_{\theta|\theta}
+\log{J}
\geq
\log{\iota}.
}

\subsection{Proof of Ineq.~\req{atatatate} under {\bf WA 1} and {\bf WA 2}}
\lsec{prfdpconv}

We prove \req{atatatate}, that is,
\begin{equation}
H_{\rm min}^{\lambda'}(A|R)_{\ca{C}(\Psi)}
+
H_{\rm max}^{\upsilon}(RD|E)_{\ca{T}\circ\ca{C}(\Psi)}
\geq
\log{\iota}+\log{(1-2\upsilon)}, \label{aaargev}
\end{equation}
under the assumptions {\bf WA 1} and {\bf WA 2}.
To show this, we introduce the following notations:
\begin{itemize}

\item $\ket{\Psi}^{ARD}$ : A purification of $\Psi^{AR}$, in the same way as in the  previous subsection. \vspace{-2mm}

\item $\ca{T}_{\ca{C}}^{A\rightarrow E}$ : A trace preserving CP map defined by $\ca{T}_{\ca{C}}^{A\rightarrow E}:=\ca{T}^{A\rightarrow E}\circ\ca{C}^A$. \vspace{-2mm}

\item $\Theta_{\ca{C}}^{ERD}$ : A normalized state on $ERD$ defined by $\Theta_{\ca{C}}^{ERD}:=\ca{T}^{A\rightarrow E}\circ\ca{C}^A(\Psi^{ARD})$. \vspace{-2mm}

\item $\theta_{\ca{C}}^{ERD}$ : A subnormalized state on $ERD$ such that $H_{\rm max}(RD|E)_{\theta_{\ca{C}}}=H_{\rm max}^{\upsilon}(RD|E)_{\Theta_{\ca{C}}}$ and $P(\theta_{\ca{C}},\Theta_{\ca{C}})\leq\upsilon$, which is classically coherent and diagonal in $E_cR_c$. 

\item $\hat{\theta}_{\ca{C}}^{ERD}$ : A normalized state on $ERD$ defined by $\hat{\theta}_{\ca{C}}^{ERD}:=\theta_{\ca{C}}^{ERD}/{\rm Tr}[{\theta_{\ca{C}}}]$.

\end{itemize}

\noindent
The assumptions {\bf WA 1} and {\bf WA 2} imply that  $\Theta_{\ca{C}}^{ERD}$ is classically coherent and diagonal in $E_cR_c$. Thus,
the existence of $\theta_{\ca{C}}$ satisfying the above condition follows from \rLmm{clchtwo}.
By definition, we have
\alg{
H_{\rm max}^{\upsilon}(RD|E)_{\ca{T}\circ\ca{C}(\Psi)}
=
H_{\rm max}(RD|E)_{\theta_{\ca{C}}}
=
H_{\rm max}(RD|E)_{\hat{\theta}_{\ca{C}}}
+
\log{\rm Tr}[\theta_{\ca{C}}].
 \laeq{graviton2}
 }

The proof of Ineq.~(\ref{aaargev}) proceeds as follows. First, we introduce a  quantum state $\hat{\Psi}^{ARD}$ and a quantum channel $\hat{\ca{T}}_{\ca{C}}^{A\rightarrow E}$, such that $\hat{\ca{T}}_{\ca{C}}^{A\rightarrow E}(\hat{\Psi}^{ARD})$ is close to the  state $\ca{T}_{\ca{C}}^{A\rightarrow E}(\Psi^{ARD})$.
Second, we apply the converse inequality \req{atatatata} to the channel $\hat{\ca{T}}_{\ca{C},k}^{A_r\rightarrow E}$ and the state $\hat{\Psi}_{k}^{A_rR_r}$, which are obtained by restricting $\hat{\ca{T}}_{\ca{C}}^{A\rightarrow E}$ and $\hat{\Psi}^{ARD}$ to the $k$-th subspace. 
The obtained inequalities are then averaged over all $k$. Finally, by using the properties of the smooth entropies, we obtain Ineq.~(\ref{aaargev}).

To explicitly define $\hat{\Psi}^{ARD}$ and $\hat{\ca{T}}_{\ca{C}}^{A\rightarrow E}$, 
observe that, since $\Theta_{\ca{C}}$ is a normalized state, we have
\alg{
P(\Theta_{\ca{C}}^{RD},\hat{\theta}_{\ca{C}}^{RD})
\leq
P(\Theta_{\ca{C}}^{ERD},\hat{\theta}_{\ca{C}}^{ERD})
\leq
P(\Theta_{\ca{C}}^{ERD},\theta_{\ca{C}}^{ERD})
\leq
\upsilon.
\laeq{tsukawanai}
}
Thus, due to Uhlmann's theorem, and noting that $\Theta_{\ca{C}}^{RD}=\Psi^{RD}$, there exists a normalized pure state $|\hat{\Psi}\rangle^{ARD}$ such that $P(\Psi^{ARD},\hat{\Psi}^{ARD})\leq\upsilon$ and $\hat{\Psi}^{RD}=\hat{\theta}_{\ca{C}}^{RD}$. It follows from the latter equality that there exists a trace preserving CP map $\hat{\ca{T}}_{\ca{C}}^{A\rightarrow E}$ satisfying $\hat{\theta}_{\ca{C}}^{ERD}=\hat{\ca{T}}_{\ca{C}}^{A\rightarrow E}(\hat{\Psi}^{ARD})$.

\subsubsection{Block-wise application of the converse inequality \req{atatatata}}

Define a normalized probability distribution 
$\{r_k:=\|\bra{k}^{R_c}\ket{\hat{\Psi}}\|_1^2\}_{k=1}^J$,
and let
$
|\hat{\Psi}_{k}\rangle^{A_rR_rD}:=r_k^{-1/2}\bra{k}^{E_c}\bra{k}^{R_c}\ket{\hat{\Psi}}
$
for $k$ such that $r_k>0$. Since $\hat{\Psi}$ is classically coherent in $E_cR_c$, the $\hat{\Psi}_{k}$ are normalized states. Define also a CP map $\hat{\ca{T}}_{\ca{C},k}^{A_r\rightarrow E}$ by 
\alg{
\hat{\ca{T}}_{\ca{C},k}^{A_r\rightarrow E}(\tau)
=
\proj{k}^{E_c}\hat{\ca{T}}_{\ca{C}}^{A\rightarrow E}(\proj{k}^{A_c}\otm\tau^{A_r})\proj{k}^{E_c},
}
which is trace preserving due to the assumptions {\bf WA1} and {\bf WA2}.
We apply the converse inequality \req{atatatata} for $\hat{\Psi}_{k}$ and $\hat{\ca{T}}_{\ca{C},k}^{A_r\rightarrow E}$ for each $k$, by letting $J=1$. We particularly choose $\upsilon=0$, in which case Ineq.~\req{atatatata} leads to
\alg{
H_{\rm min}^{\lambda_k}(A_r|R_r)_{\hat{\Psi}_{k}}
+
H_{\rm max}(R_rD|E_r)_{\hat{\ca{T}}_{\ca{C},k}(\hat{\Psi}_{k})}
\geq
\log{\iota}.
}
The smoothing parameter $\lambda_k$ is given by
\alg{
\lambda_k:=2\sqrt{\iota+4\sqrt{2\delta_k}}
+\sqrt{2\sqrt{2\delta_k}}+4\sqrt{2\delta_k},
\quad
\delta_k:=
\left\|
\hat{\ca{T}}_{\ca{C},k}^{A_r\rightarrow E}(\hat{\Psi}_{k}^{A_rR_r})
-
\varsigma_k^E\otm\hat{\Psi}_{k}^{R_r}
\right\|_1.
}
A simple calculation yields
\alg{
-
\log{
\left(
\sum_kr_k\cdot
2^{-H_{\rm min}^{\lambda_k}(A_r|R_r)_{\hat{\Psi}_{k}}}
\right)
}
&
\geq
-
\log{
\left(
\sum_kr_k\cdot
2^{
H_{\rm max}(R_rD|E_r)_{\hat{\ca{T}}_{\ca{C},k}(\hat{\Psi}_{k})}
}
\right)}
+
\log{\iota}.
\laeq{tiluanin2}
}

\subsubsection{Calculation of Averaged Entropies}

Using the fact that $\hat{\theta}_{\ca{C}}$ is classically coherent and diagonal in $E_cR_c$, it is straightforward to verify that
$
\hat{\theta}_{\ca{C}}^{ERD}=\hat{\ca{T}}_{\ca{C}}^{A\rightarrow E}(\hat{\Psi}^{ARD})=\sum_kr_k\hat{\ca{T}}_{\ca{C},k}^{A_r\rightarrow E}(\hat{\Psi}_{k}^{A_rR_rD})\otm\proj{k}^{R_c}
$.
Thus, by using the property of the smooth conditional entropies (Lemmas \rlmm{condminCQ} and \rlmm{minentave}) and $P(\Psi,\hat{\Psi})\leq\upsilon$, both sides of Ineq.~\req{tiluanin2} are calculated to be
\alg{
&
-
\log{
\left(
\sum_kr_k\cdot
2^{
H_{\rm max}(R_rD|E_r)_{\hat{\ca{T}}_{\ca{C},k}(\hat{\Psi}_{k})}
}
\right)}
=
-H_{\rm max}(RD|E)_{\hat{\theta}_{\ca{C}}},
\\
&
-
\log{
\left(
\sum_kr_k\cdot
2^{-H_{\rm min}^{\lambda_k}(A_r|R_r)_{\hat{\Psi}_{k}}}
\right)}
\leq
H_{\rm min}^{\sqrt{2\bar{\lambda}}}(A|R)_{\ca{C}(\hat{\Psi})}
\leq
H_{\rm min}^{\upsilon+\sqrt{2\bar{\lambda}}}(A|R)_{\ca{C}(\Psi)},
}
where $\bar{\lambda}:=\sum_kr_k\lambda_k$.
Combining these all together with Eq.~\req{graviton2}, we obtain 
\begin{equation}
H_{\rm min}^{\upsilon+\sqrt{2\bar{\lambda}}}(A|R)_{\ca{C}(\Psi)}
+
H_{\rm max}^{\upsilon}(RD|E)_{\ca{T}\circ\ca{C}(\Psi)}
\geq
\log{\iota}
+
\log{\rm Tr}[\theta_{\ca{C}}].
\laeq{pirori}
\end{equation}
As we prove in \rApp{tsuuwamoprf}, the partial decoupling condition \req{keshikeshiit} implies
\alg{
\bar{\lambda}\leq
\lambda(\iota,\sqrt{2}\sqrt[4]{24\upsilon+2\delta})+\lambda(\iota,4)\cdot\sqrt{2}\sqrt[4]{24\upsilon+2\delta},
\laeq{tsuuwamo}
}
where $\lambda(\iota,x):=2\sqrt{\iota+2x}+\sqrt{x}+2x$.
A simple calculation then yields
\begin{equation}
\upsilon+\sqrt{2\bar{\lambda}} \leq \upsilon+\sqrt{4\sqrt{\iota+2x}+2\sqrt{x}+(4\sqrt{\iota+8}+24) x},
\end{equation}
whose right-hand side is exactly $\lambda'$ given in \req{dfndeltadelta489}.
In addition, noting that $\Theta_{\ca{C}}$ is normalized, and by using the relation between the purified distance and the trace distance (Property 2 in \rLmm{propPD2}), the last term in the R.H.S. of \req{pirori} is calculated to be
\alg{
{\rm Tr}[\theta_{\ca{C}}]\geq\|\Theta_{\ca{C}}\|_1-\left\|\theta_{\ca{C}}-\Theta_{\ca{C}}\right\|_1
\geq1-2P(\theta_{\ca{C}},\Theta_{\ca{C}})
\geq1-2\upsilon.
}
Combining these all together, we arrive at
\begin{equation}
H_{\rm min}^{\lambda'}(A|R)_{\ca{C}(\Psi)}
+
H_{\rm max}^{\upsilon}(RD|E)_{\ca{T}\circ\ca{C}(\Psi)}
\geq
\log{\iota}
+
\log{(1-2\upsilon)}.
\end{equation}

\subsection{Dropping the working assumptions {\bf WA 1} and {\bf WA 2}}
\label{SS:DA2}

We here show that the working assumptions {\bf WA 1} and {\bf WA 2} can be dropped. The proof is based on \rLmm{dropassump}. 
Since the   CP map $\check{\mathcal{T}}^{A \rightarrow E E_c}$, defined in \rLmm{dropassump}, satisfy both conditions, it satisfies Ineq. \req{atatatata}, which is
\alg{
&
H_{\rm min}^{\lambda}(A|R)_\Psi
+
H_{\rm max}^{\upsilon}(RD|EE_c)_{\check{\ca{T}}(\Psi)}
+\log{J}
\geq
\log{\iota}.
\laeq{batatatata}
}
Let $V^{A\rightarrow BE}$ be a Stinespring dilation of $\ca{T}^{A\rightarrow E}$, and let $Z^{R_c\rightarrow R_cE_c}$ be a linear isometry defined by 
$Z:=\sum_{j}\ket{jj}^{R_cE_c}\bra{j}^{R_c}$.
A purification $|\vartheta\rangle^{BRDEE_c}$ of $\check{\ca{T}}^{A\rightarrow EE_c}(\Psi^{ARD})$ is given by $|\vartheta\rangle^{BRDEE_c}=(V^{A\rightarrow BE}\otimes Z^{R_c\rightarrow R_cE_c})|\Psi\rangle^{ARD}$, and satisfies $\vartheta^{BRD}=\ca{T}^{A\rightarrow B}\circ\ca{C}^A(\Psi^{ARD})$. 
Hence, due to the duality for the conditional smooth entropy (\rLmm{duality}), it holds that
\alg{
H_{\rm max}^{\upsilon}(RD|EE_c)_{\check{\ca{T}}(\Psi)}
=
H_{\rm max}^{\upsilon}(RD|EE_c)_{\vartheta}
=
-H_{\rm min}^{\upsilon}(RD|B)_{\vartheta}
=
-H_{\rm min}^{\upsilon}(RD|B)_{\ca{T}\circ\ca{C}(\Psi)}.
\laeq{lightsatou}
}
Combining this with \req{batatatata}, we conclude
\begin{equation}
H_{\rm min}^{\lambda}(A|R)_\Psi
-H_{\rm min}^{\upsilon}(RD|B)_{\ca{T}\circ\ca{C}(\Psi)}+\log{J}
\geq
\log{\iota}.
\end{equation}

The map $\check{\mathcal{T}}^{A \rightarrow E E_c}$ also satisfies Ineq. \req{atatatate}:
\alg{
H_{\rm min}^{\lambda'}(A|R)_{\ca{C}(\Psi)}
+
H_{\rm max}^{\upsilon}(RD|EE_c)_{\check{\ca{T}}\circ\ca{C}(\Psi)}
\geq
\log{\iota}+\log{(1-2\upsilon)}.
\laeq{fragile}
}
Similarly to \req{lightsatou} and \req{sannnwa4}, by using the property of the conditional max entropy for classical-quantum states (\rLmm{condmaxCQCQ}), we have
\alg{
&
H_{\rm max}^{\upsilon}(RD|EE_c)_{\check{\ca{T}}\circ\ca{C}(\Psi)}
=
H_{\rm max}^{\upsilon}(R_rD|EE_c)_{\check{\ca{T}}\circ\ca{C}(\Psi)}
=
H_{\rm max}^{\upsilon}(R_rD|EE_c)_{\check{\ca{T}}(\Psi)}
\nn\\
&
=
H_{\rm max}^{\upsilon}(R_rD|EE_c)_{\vartheta}
=
-H_{\rm min}^{\upsilon}(R_rD|BR_c)_{\vartheta}
=
-H_{\rm min}^{\upsilon}(R_rD|BR_c)_{\ca{T}\circ\ca{C}(\Psi)},
}
which leads to
\alg{
H_{\rm min}^{\lambda'}(A|R)_{\ca{C}(\Psi)}
-H_{\rm min}^{\upsilon}(R_rD|BR_c)_{\ca{T}\circ\ca{C}(\Psi)}
\geq
\log{\iota}+\log{(1-2\upsilon)}.
}
This concludes the proof of Theorem~\ref{Thm:converse} for any trace preserving CP map $\mathcal{T}^{A \rightarrow E}$.
\QED

\section{Conclusion}
\lsec{cncl}

In this paper, we have proposed and analyzed a task that we call partial decoupling.  We have presented two different formulations of partial decoupling, and derived lower and upper bounds on how precisely partial decoupling can be achieved. The bounds are represented in terms of the smooth conditional entropies of quantum states involving the initial state, the channel and the decomposition of the Hilbert space. Thereby we provided a generalization of the decoupling theorem in the version of  \cite{DBWR2010}, by incorporating the direct-sum-product decomposition of the Hilbert space. 
Applications of our result to quantum communication tasks and black hole information paradox are provided in Refs.\cite{wakakuwa2020randomized,wakakuwa2020oneshothybrid,nakata2020one} and \cite{nakata2020black}, respectively.
A future direction is to apply the result to various scenarios that have been analyzed in terms of the decoupling theorem, such as relative thermalization \cite{dRHRW2014} and area laws \cite{brandao2015exponential} in the foundation of statistical mechanics.

\section*{Acknowledgement}
This work was supported by JST CREST, Grant Number JPMJCR1671 as well as by JST, PRESTO Grant Number JPMJPR1865, Japan, and by JSPS KAKENHI, Grant Number 18J01329.

\bibliographystyle{unsrt}

\bibliography{bibbib.bib}

\begin{thebibliography}{10}

\bibitem{HHWY2008}
P.~Hayden, M.~Horodecki, A.~Winter, and J.~Yard.
\newblock A decoupling approach to the quantum capacity.
\newblock {\em Open Syst. Inf. Dyn.}, 15:7, 2008.

\bibitem{ADHW2009}
A.~Abeyesinghe, I.~Devetak, P.~Hayden, and A.~Winter.
\newblock The mother of all protocols : Restructuring quantum information's
  family tree.
\newblock {\em Proc. R. Soc. A}, 465:2537, 2009.

\bibitem{horo05}
M.~Horodecki, J.~Oppenheim, and A.~Winter.
\newblock Partial quantum information.
\newblock {\em Nature}, 436:673--676, 2005.

\bibitem{HOW07}
M.~Horodecki, J.~Oppenheim, and A.~Winter.
\newblock Quantum state merging and negative information.
\newblock {\em Comm. Math. Phys.}, 269:107--136, 2007.

\bibitem{GPW2005}
B.~Groisman, S.~Popescu, and A.~Winter.
\newblock Quantum, classical, and total amount of correlations in a quantum
  state.
\newblock {\em Phys. Rev. A}, 72(3):032317, 2005.

\bibitem{berta2018conditional}
Mario Berta, Fernando~GSL Brand{\~a}o, Christian Majenz, and Mark~M Wilde.
\newblock Conditional decoupling of quantum information.
\newblock {\em Phys. Rev. Lett.}, 121(4):040504, 2018.

\bibitem{HP2007}
P.~Hayden and J.~Preskill.
\newblock Black holes as mirrors: quantum information in random subsystems.
\newblock {\em J. High Energy Phys.}, 2007(09):120, 2007.

\bibitem{brandao2015exponential}
F.~GSL Brand{\~a}o and M.~Horodecki.
\newblock Exponential decay of correlations implies area law.
\newblock {\em Comm. Math. Phys.}, 333(2):761--798, 2015.

\bibitem{dRARDV2011}
L.~del Rio, J.~Aberg, R.~Renner, O.~Dahlsten, and V.~Vedral.
\newblock The thermodynamic meaning of negative entropy.
\newblock {\em Nature}, 474(7349):61--63, 2011.

\bibitem{dRHRW2014}
L.~del Rio, A.~Hutter, R.~Renner, and S.~Wehner.
\newblock Relative thermalization.
\newblock {\em Phys. Rev. E}, 94(2):022104, 2016.

\bibitem{DBWR2010}
F.~Dupuis, M.~Berta, J.~Wullschleger, and R.~Renner.
\newblock One-shot decoupling.
\newblock {\em Comm. Math. Phys.}, 328:251, 2014.

\bibitem{uhlmann1976transition}
Armin Uhlmann.
\newblock The ``transition probability'' in the state space of a $c^*$-algebra.
\newblock {\em Rep. Math. Phys.}, 9(2):273--279, 1976.

\bibitem{devetak2005capacity}
I.~Devetak and P.~W. Shor.
\newblock The capacity of a quantum channel for simultaneous transmission of
  classical and quantum information.
\newblock {\em Comm. Math. Phys.}, 256(2):287--303, 2005.

\bibitem{robin10}
R.~B.-Kohout, H.~K. Ng, D.~Poulin, and L.~Viola.
\newblock Information-preserving structures: A general framework for quantum
  zero-error information.
\newblock {\em Phys. Rev. A}, 82:062306, 2010.

\bibitem{robin08}
R.~B.-Kohout, H.~K. Ng, D.~Poulin, and L.~Viola.
\newblock Characterizing the structure of preserved information in quantum
  processes.
\newblock {\em Phys. Rev. Lett.}, 100:030501, 2008.

\bibitem{koashi02}
M.~Koashi and N.~Imoto.
\newblock Operations that do not disturb partially known quantum states.
\newblock {\em Phys. Rev. A}, 66:022318, 2002.

\bibitem{mixcomp1}
M.~Koashi and N.~Imoto.
\newblock Compressibility of quantum mixed-state signals.
\newblock {\em Phys. Rev. Lett.}, 87:017902, 2001.

\bibitem{hayden04}
P.~Hayden, R.~Jozsa, D.~Petz, and A.~Winter.
\newblock Structure of states which satisfy strong subadditivity of quantum
  entropy with equality.
\newblock {\em Comm. Math. Phys.}, 246:359--374, 2004.

\bibitem{wakakuwa2017markovianizing}
E.~Wakakuwa, A.~Soeda, and M.~Murao.
\newblock Markovianizing cost of tripartite quantum states.
\newblock {\em IEEE Trans. Inf. Theory}, 63(2):1280--1298, 2017.

\bibitem{bartlett07}
S.~D. Bartlett, T.~Rudolph, and R.~W. Spekkens.
\newblock Reference frames, superselection rules, and quantum information.
\newblock {\em Rev. Mod. Phys.}, 79:555, 2007.

\bibitem{wakakuwa2020randomized}
Eyuri Wakakuwa and Yoshifumi Nakata.
\newblock Randomized partial decoupling unifies one-shot quantum channel
  capacities.
\newblock {\em arXiv:2004.12593}, 2020.

\bibitem{nakata2020one}
Yoshifumi Nakata, Eyuri Wakakuwa, and Hayata Yamasaki.
\newblock One-shot quantum error correction of classical and quantum
  information: towards demonstration of quantum channel coding.
\newblock {\em arXiv:2011.00668}, 2020.

\bibitem{wakakuwa2020oneshothybrid}
Eyuri Wakakuwa, Yoshifumi Nakata, and Min-Hsiu Hsieh.
\newblock One-shot hybrid state redistribution.
\newblock {\em arXiv:2006.12059}, 2020.

\bibitem{nakata2020black}
Yoshifumi Nakata, Eyuri Wakakuwa, and Masato Koashi.
\newblock Black holes as clouded mirrors: the hayden-preskill protocol with
  symmetry.
\newblock {\em arXiv:2007.00895}, 2020.

\bibitem{tomamichel2010duality}
M.~Tomamichel, R.~Colbeck, and R.~Renner.
\newblock Duality between smooth min-and max-entropies.
\newblock {\em IEEE Trans. Inf. Theory}, 56(9):4674--4681, 2010.

\bibitem{T16}
M.~Tomamichel.
\newblock {\em {\it Quantum Information Processing with Finite Resources}}.
\newblock SpringerBriefs in Mathematical Physics, 2016.

\bibitem{J1972}
A.~Jamio\l kowski.
\newblock Linear transformations which preserve trace and positive
  semidefiniteness of operators.
\newblock {\em Rep. Math. Phys.}, 3:275, 1972.

\bibitem{C1975}
M.~D. Choi.
\newblock Completely positive linear maps on complex matrices.
\newblock {\em Linear Algebra Appl.}, 10:285, 1975.

\bibitem{SS2008}
Y.~Sekino and L.~Susskind.
\newblock Fast scramblers.
\newblock {\em J. High Energy Phys.}, 2008(10):065, 2008.

\bibitem{LSHOH2013}
N.~Lashkari, D.~Stanford, M.~Hastings, T.~Osborne, and P.~Hayden.
\newblock Towards the fast scrambling conjecture.
\newblock {\em J. High Energy Phys.}, 2013(4), 2013.

\bibitem{dupuis2014decoupling}
F.~Dupuis, O.~Szehr, and M.~Tomamichel.
\newblock A decoupling approach to classical data transmission over quantum
  channels.
\newblock {\em IEEE Trans. Inf. Theory}, 60(3):1562--1572, 2014.

\bibitem{tomamichel2009fully}
M.~Tomamichel, R.~Colbeck, and R.~Renner.
\newblock A fully quantum asymptotic equipartition property.
\newblock {\em IEEE Trans. Inf. Theory}, 55(12):5840--5847, 2009.

\bibitem{DLT2002}
D.~P. DiVincenzo, D.~W. Leung, and B.~M. Terhal.
\newblock Quantum data hiding.
\newblock {\em IEEE Trans. Inf. Theory}, 48:580, 2002.

\bibitem{DCEL2009}
C.~Dankert, R.~Cleve, J.~Emerson, and E.~Livine.
\newblock Exact and approximate unitary 2-designs and their application to
  fidelity estimation.
\newblock {\em Phys. Rev. A}, 80:012304, 2009.

\bibitem{GAE2007}
D.~Gross, K.~Audenaert, and J.~Eisert.
\newblock Evenly distributed unitaries: {On} the structure of unitary designs.
\newblock {\em J. of Math. Phys.}, 48(5):052104, 2007.

\bibitem{BWV2008a}
W.~G. Brown, Y.~S. Weinstein, and L.~Viola.
\newblock Quantum pseudorandomness from cluster-state quantum computation.
\newblock {\em Phys. Rev. A}, 77(4):040303(R), 2008.

\bibitem{WBV2008}
Y.~S. Weinstein, W.~G. Brown, and L.~Viola.
\newblock Parameters of pseudorandom quantum circuits.
\newblock {\em Phys. Rev. A}, 78(5):052332, 2008.

\bibitem{HL2009}
A.~W. Harrow and R.~A. Low.
\newblock Random quantum circuits are approximate 2-designs.
\newblock {\em Commun. Math. Phys.}, 291:257, 2009.

\bibitem{DJ2011}
I.~T. Diniz and D.~Jonathan.
\newblock {Comment on ``Random quantum circuits are approximate 2-designs"}.
\newblock {\em Commun. Math. Phys.}, 304:281, 2011.

\bibitem{CLLW2015}
R.~Cleve, D.~Leung, L.~Liu, and C.~Wang.
\newblock Near-linear constructions of exact unitary 2-designs.
\newblock {\em Quant. Info. {\&} Comp.}, 16(9 {\&} 10):0721--0756, 2016.

\bibitem{NHMW2015-1}
Y.~Nakata, C.~Hirche, C.~Morgan, and A.~Winter.
\newblock Unitary $2$-designs from random {$X$}- and {$Z$}-diagonal unitaries.
\newblock arXiv:1502.07514, 2015.

\bibitem{BF2013}
W.~Brown and O.~Fawzi.
\newblock Decoupling with random quantum circuits.
\newblock {\em Commun. Math. Phys.}, 340:867, 2015.

\bibitem{NHMW2015-2}
Y.~Nakata, C.~Hirche, C.~Morgan, and A.~Winter.
\newblock Decoupling with random diagonal unitaries.
\newblock arXiv:1509.05155, 2015.

\bibitem{NHKW2017}
Y.~Nakata, C.~Hirche, M.~Koashi, and A.~Winter.
\newblock Efficient {Quantum} {Pseudorandomness} with {Nearly}
  {Time}-{Independent} {Hamiltonian} {Dynamics}.
\newblock {\em Phys. Rev. X}, 7(2):021006, 2017.

\bibitem{wakakuwa2017coding}
Eyuri Wakakuwa, Akihito Soeda, and Mio Murao.
\newblock A coding theorem for bipartite unitaries in distributed quantum
  computation.
\newblock {\em IEEE Trans. Inf. Theory}, 63(8):5372--5403, 2017.

\bibitem{GR1999}
R.~Goodman and N.~R. Wallach.
\newblock {\em {\it Representations and Invariants of the Classical Groups}}.
\newblock Cambridge University Press, Cambridge, UK, 1999.

\bibitem{hiai2010matrix}
H.~Fumio.
\newblock Matrix analysis: matrix monotone functions, matrix means, and
  majorization.
\newblock {\em Int. Info. Sci.}, 16(2):139--248, 2010.

\bibitem{TomamichelThesis}
M.~Tomamichel.
\newblock {\em A {Framework} for {Non}-{Asymptotic} {Quantum} {Information}
  {Theory}}.
\newblock PhD thesis, ETH Zurich, 2012.
\newblock arXiv:1203.2142.

\end{thebibliography}

\appendix

\section{Proof of the twisted twirling} \lapp{TT}
We here provide the proof of the twisted twirling (Lemma~\ref{Lemma:av}).
The statement is as follows: 
let $\ca{H}_j^{A_r}$ be a subspace of $\ca{H}^{A_r}$ of dimension $r_j$, and $\Pi_j^{A_r}$ be the projector onto $\ca{H}_j^{A_r}\subset\ca{H}^{A_r}$ for $j=1,\cdots,J$.
Let $\mbb{I}^{A_rA_r'}$ be $I^{A_r} \otimes I^{A_r'}$, and $\mbb{F}^{A_rA_r'} \in \ca{L}(\ca{H}^{A_rA_r'})$ be the swap operator defined by $\sum_{a,b} |a\rangle \langle b|^{A_r} \otimes |b\rangle \langle a|^{A_r'}$ for any orthonormal basis $\{ \ket{a} \}$ in $\mathcal{H}^{A_r}$ and $\mathcal{H}^{A_r'}$.
Further, let $\mbb{I}_{jk}^{A_rA_r'}$ and $\mbb{F}_{jk}^{A_rA_r'}$ be $\Pi_j^{A_r} \otimes \Pi_k^{A_r'}$ and $( \Pi_j^{A_r} \otimes \Pi_k^{A_r'})\mbb{F}^{A_rA_r'}$, respectively.
For any $M^{A_rA_r'BB'}\in\ca{L}(\ca{H}^{A_rA_r'BB'})$, define 
\alg{
M^{BB'}_{\mbb{I},jk}:=\tr_{A_rA_r'}[\mbb{I}_{jk}^{A_rA_r'}M^{A_rA_r'BB'}],
\quad
M^{BB'}_{\mbb{F},kj}:=\tr_{A_rA_r'}[\mbb{F}_{kj}^{A_rA_r'}M^{A_rA_r'BB'}].
}
Then, it holds that, for $j \neq k$,
\begin{align}
&\mbb{E}_{U_j \sim {\sf H}_j,U_k \sim {\sf H}_k} \bigl[ (U_j^{A_r} \otimes U_k^{A_r'}) M^{A_rA_r'BB'} (U_j^{A_r} \otimes  U_k^{A_r'})^{\dagger} \bigr]
= \frac{\mbb{I}_{jk}^{A_rA_r'}}{r_jr_k} \otimes M_{\mbb{I},jk}^{BB'},\label{eq:twtw1}\\
&\mbb{E}_{U_j \sim {\sf H}_j,U_k \sim {\sf H}_k} \bigl[ (U_j^{A_r} \otimes U_k^{A_r'}) M^{A_rA_r'BB'} (U_k^{A_r} \otimes  U_j^{A_r'})^{\dagger} \bigr]
= \frac{\mbb{F}_{jk}^{A_rA_r'}}{r_jr_k} \otimes  M^{BB'}_{\mbb{F},kj}.\label{eq:twtw2}
\end{align}
Moreover, \begin{multline}
\mbb{E}_{U_j \sim {\sf H}_j} \bigl[ (U_j^{A_r} \otimes U_j^{A_r'}) M^{A_rA_r'BB'} (U_j^{A_r} \otimes  U_j^{A_r'})^{\dagger} \bigr]\\
=\frac{1}{r_j (r_j^2-1)}
\left[
(r_j\mbb{I}_{jj}^{A_rA_r'}- \mbb{F}_{jj}^{A_rA_r'})\otimes M_{\mbb{I},jj}^{BB'} 
+
(r_j\mbb{F}_{jj}^{A_rA_r'} -\mbb{I}_{jj}^{A_rA_r'})\otimes M^{BB'}_{\mbb{F},jj} 
\right].\label{eq:mindthegap}
\end{multline}
Otherwise, $\mbb{E}_{U_j,U_k,U_m,U_n} \bigl[ (U_j^{A_r} \otimes U_k^{A_r'}) M^{A_rA_r'BB'} (U_m^{A_r} \otimes  U_n^{A_r'})^{\dagger} \bigr]=0$.

\begin{prf}
The equation $\mbb{E}_{U_j,U_k,U_m,U_n} \bigl[ (U_j^{A_r} \otimes U_k^{A_r'}) M^{A_rA_r'BB'} (U_m^{A_r} \otimes  U_n^{A_r'})^{\dagger} \bigr]=0$ for $i \neq j \neq k \neq l$ trivially follows from the fact that the random unitaries $\{ U_j \}_j$ are independent and that
$\mbb{E}_{U_j \sim {\sf H}_j}[U_j]=0$.

Let us consider the case where $j \neq k$ and prove Eqs.~(\ref{eq:twtw1}) and (\ref{eq:twtw2}). 
Note that any $X^{A_rB}\in\ca{L}(\ca{H}^{A_rB})$ is decomposed into
$X^{A_rB}=\sum_{p,q}X_p^{A_r}\otm X_q^B$, where $X_p^{A_r} \in \ca{L}(\ca{H}^{A_r})$ and $X_q^B \in \ca{L}(\ca{H}^B)$.
Using the fact that
\alg{
\mbb{E}_{U_j \sim {\sf H}_j}[U_j^{A_r} X_p^{A_r} U_j^{A \dagger}] = \frac{\tr[\Pi_j^{A_r}X_p^{A_r}]}{r_j} \Pi_j^{A_r}
}
for any $X_p^{A_r}\in \ca{L}(\ca{H}^{A_r})$, which follows from the Schur-Weyl duality~\cite{GR1999}, we have
\begin{align}
\mbb{E}_{U_j \sim {\sf H}_j}[ U_j^{A_r} X^{A_rB} U_j^{A \dagger}]
&=
\sum_{p,q} 
\mbb{E}_{U_j \sim {\sf H}_j}[ U_j^{A_r} X_p^{A_r} U_j^{A \dagger} ] \otimes X_q^B  \nonumber\\
&=
\frac{\Pi_j^{A_r}}{r_j}  \otimes \sum_{p,q} \tr[\Pi_j^{A_r}X_p^{A_r}] X_q^B  \nonumber\\
&=
\frac{\Pi_j^{A_r}}{r_j}  \otimes  \tr_{A_r}[\Pi_j^{A_r}X^{A_rB}].  
\end{align}
Using this equality twice for $j$ and $k$, we obtain Eq.~(\ref{eq:twtw1}). It also leads to Eq.~(\ref{eq:twtw2}) as follows:
\begin{align}
&\mbb{E}_{U_j,U_k } \bigl[ (U_j^{A_r} \otimes U_k^{A_r'}) M^{A_rA_r'BB'} (U_k^{A_r} \otimes  U_j^{A_r'})^{\dagger} \bigr] \nonumber\\
&=
\mbb{E}_{U_j,U_k} \bigl[ (U_j^{A_r} \otimes U_k^{A_r'}) M^{A_rA_r'BB'} \mbb{F}^{A_rA_r'}(U_j^{A_r} \otimes  U_k^{A_r'})^{\dagger} \bigr] \mbb{F}^{A_rA_r'} \nonumber\\
&=
\mbb{E}_{U_j,U_k} \bigl[ (U_j^{A_r} \otimes U_k^{A_r'}) M^{A_rA_r'BB'} \mbb{F}_{kj}^{A_rA_r'}(U_j^{A_r} \otimes  U_k^{A_r'})^{\dagger} \bigr] \mbb{F}_{jk}^{A_rA_r'} \nonumber\\
&=
\frac{\mbb{F}_{jk}^{A_rA_r'}}{r_jr_k} 
\otimes 
\tr_{A_rA_r'} [\mbb{I}_{jk}^{A_rA_r'} M^{A_rA_r'BB'} \mbb{F}_{kj}^{A_rA_r'} ]\nonumber\\
&=
\frac{\mbb{F}_{jk}^{A_rA_r'}}{r_jr_k} 
\otimes 
\mbb{M}_{\mbb{F},kj}^{BB'}.
\end{align}
Here, we have used relations
\alg{
\mbb{F}_{kj}^{A_rA_r'}
=
(\Pi_k^{A_r}\otm\Pi_j^{A_r'})\mbb{F}^{A_rA_r'}
=
\mbb{F}^{A_rA_r'}(\Pi_j^{A_r}\otm\Pi_k^{A_r'}),
\quad
\mbb{F}_{kj}^{A_rA_r'}\mbb{I}_{jk}^{A_rA_r'}
=
\mbb{F}_{kj}^{A_rA_r'}(\Pi_j^{A_r}\otm\Pi_k^{A_r'})
=
\mbb{F}_{kj}^{A_rA_r'},\nn
}
and used Eq.~(\ref{eq:twtw1}) in the last line.

We finally show Eq.~(\ref{eq:mindthegap}). Consider the operator
$\mbb{E}_{U_j \sim {\sf H}_j} \bigl[ (U_j^{A_r} \otimes U_j^{A_r'}) \outpro{p}{q}^{A_r} \otimes \outpro{s}{t}^{A_r'} (U_j^{A_r} \otimes  U_j^{A_r'})^{\dagger} \bigr]$. 
Since this commutes with $V^{\otimes 2}$ ($\forall V \in \mbb{U}(r_j)$), we obtain from the Schur-Weyl duality~\cite{GR1999} that
\begin{align}
\mbb{E}_{U_j \sim {\sf H}_j} \bigl[ (U_j^{A_r} \otimes U_j^{A_r'}) \outpro{p}{q}^{A_r} \otimes \outpro{s}{t}^{A_r'} (U_j^{A_r} \otimes  U_j^{A_r'})^{\dagger} \bigr]
= \alpha_{pqst} \mbb{I}_{jj}^{A_rA_r'} + \beta_{pqst} \mbb{F}_{jj}^{A_rA_r'}, 
\label{Eq:vfsd0@i}
\end{align}
where $\alpha_{pqst}$ and $\beta_{pqst}$ are determined by
\begin{align}
\delta_{pq} \delta_{st} = \alpha_{pqst} r_j^2 + \beta_{pqst} r_j, 
\quad
\delta_{pt} \delta_{qs} = \alpha_{pqst} r_j + \beta_{pqst} r_j^2. 
\end{align}
Note that the first equation is obtained by taking the trace of Eq.~\eqref{Eq:vfsd0@i}, and the second is by calculating the expectation of $\mbb{F}^{A_rA_r'}$ by both sides in Eq.~\eqref{Eq:vfsd0@i}.
Solving these equalities, we obtain
\begin{align}
&\mbb{E}_{U_j \sim {\sf H}_j} \bigl[ (U_j^{A_r} \otimes U_j^{A_r'}) \outpro{p}{q}^{A_r} \otimes \outpro{s}{t}^{A_r'} (U_j^{A_r} \otimes  U_j^{A_r'})^{\dagger} \bigr] \nonumber\\
&\quad =
\frac{1}{r_j(r_j^2-1)}
\left(
(\delta_{pq} \delta_{st} r_j - \delta_{pt}\delta_{qs} ) \mbb{I}_{jj}^{A_rA_r'} + (\delta_{pt} \delta_{qs} r_j - \delta_{pq}\delta_{st} ) \mbb{F}_{jj}^{A_rA_r'} \right), \label{Eq:zzsletp]q}
\end{align}
from which the equation~\eqref{eq:mindthegap} is obtained after a straightforward calculation. $\hfill \blacksquare$
\end{prf}

\section{Proof of Lemma \ref{lmm:nenchalu}}
\lapp{mother}

We prove Lemma \ref{lmm:nenchalu} based on the twisted twirling (Lemma~\ref{Lemma:av}) and the swap trick, a commonly used method in the context of decoupling given as follows:

\begin{lmm}[Swap trick (see e.g.  \cite{DBWR2010})]\label{Lemma:Swap}
Let $X^A$ and $Y^A$ be linear operators on $\ca{H}^A$, and $\mbb{F}^{AA'}$ be the  swap operator between $\ca{H}^A$ and $\ca{H}^{A'}$ defined by $\sum_{i,j} \outpro{i}{j}^A \otimes \outpro{j}{i}^{A'}$, where $\{ \ket{i}\}$ is any basis of $\ca{H}^A$ and $\ca{H}^{A'} \cong \ca{H}^{A}$. Then, $\tr[X^AY^A] = \tr[(X^A \otimes Y^{A'}) \mbb{F}^{AA'}]$.
\end{lmm}

For simplicity of notations in the proof, we embed a Hilbert space that has the DSP form to the tensor product of three Hilbert spaces. We explain the notation for this embedding in Subsection~\ref{SS:embedding} and then show Lemma \ref{lmm:nenchalu} in Subsection~\ref{SS:PL7}.

\subsection{Embedding of the Hilbert Space} \label{SS:embedding}

Let $A$ be a quantum system described by a finite dimensional Hilbert space ${\ca H}^A$, which is decomposed in the form of
\begin{equation}
{\ca H}^A=\bigoplus_{j=1}^J{\ca H}_j^{A_l}\otimes{\ca H}_j^{A_r}.
\end{equation}
The dimension of each subspace is denoted by $l_j:=\dim{\ca H}_j^{A_l}$, $r_j:=\dim{\ca H}_j^{A_r}$.
Let ${\ca H}^{A_c}$, ${\ca H}^{A_l}$ and ${\ca H}^{A_r}$ be Hilbert spaces such that
\begin{align}
\dim{\ca H}^{A_c}=J,\;\dim{\ca H}^{A_l}=\max_{1\leq j\leq J}l_j,\;{\ca H}^{A_r}=\max_{1\leq j\leq J}r_j, 
\end{align}
and fix linear isometries $W_{j}^{A_l}:{\mathcal H}_j^{A_l} \rightarrow{\mathcal H}^{A_l}$, $W _{j}^{A_r}:{\mathcal H}_j^{A_r}\rightarrow{\mathcal H}^{A_r}$ for each $j$. 
We introduce the following linear isometry, by which 
the Hilbert space $\ca H^A$ is embedded into ${\ca H}^{A_c} \otimes{\ca H}^{A_l} \otimes{\ca H}^{A_r}$: 
\begin{align}
W^{A \rightarrow A_cA_lA_r}:=\sum_{j=1}^J \ket{j}^{A_c} \otimes( W _{j}^{A_l}\otimes W _{j}^{A_r})\Pi_j.
\label{eq:expW}
\end{align}
Here, $\Pi_j$ is the projection onto a subspace ${\mathcal H}_j^{A_l}\otimes{\mathcal H}_j^{A_r}\subset{\mathcal H}^A$, and $\{\ket{j}\}_{j=1}^J$ is a fixed orthonormal basis of ${\ca H}^{A_c}$. The $W$ is indeed an isometry, because
\begin{align}
(W^{A \rightarrow A_cA_lA_r})^{\dagger} W^{A \rightarrow A_cA_lA_r} =I^A.
\end{align}
Noting that ${\ca H}_j^{A_l}={\rm img} W _{j}^{A_l} \subset {\mathcal H}^{A_l}$ and ${\ca H}_j^{A_r}={\rm img} W _{j}^{A_r}\subset {\mathcal H}^{A_r}$,
we have
\begin{align}
{\rm img} (W^{A \rightarrow A_cA_lA_r}) 
=\bigoplus_{j=1}^J{\mathcal H}_j^{A_c}\otimes{\mathcal H}_j^{A_l} \otimes{\mathcal H}_j^{A_r} \subset{\ca H}^{A_c} \otimes{\ca H}^{A_l} \otimes{\ca H}^{A_r},
\end{align}
where ${\mathcal H}_j^{A_c}\subset{\mathcal H}^{A_c}$ is a one-dimensional subspace spanned by $\ket{j}$ for each $j$.
Denoting the projection onto ${\mathcal H}_j^{A_l}\subset{\mathcal H}^{A_l}$ by $\Pi_j^{A_l}\in{\ca L}({\ca H}^{A_l})$ and one onto ${\mathcal H}_j^{A_r}\subset{\mathcal H}^{A_r}$ by $\Pi_j^{A_r}\in{\ca L}({\ca H}^{A_r})$, we also have
\alg{
W _{j}^{A_l}(W _{j}^{A_l})^\dagger=\Pi_j^{A_l},
\quad
W _{j}^{A_r}(W _{j}^{A_r})^\dagger=\Pi_j^{A_r}
}
and thus
\begin{align}
(W^{A \rightarrow A_cA_lA_r})  (W^{A \rightarrow A_cA_lA_r})^\dagger
=\sum_{j=1}^J\outpro{j}{j}^{A_c} \otimes\Pi_j^{A_c} \otimes\Pi_j^{A_r}.\label{eq:GGd}
\end{align}

Let $R$ be another quantum system represented by a finite dimensional Hilbert space $\ca{H}^R$.
Any $X^{A R} \in \ca{L}(\ca{H}^{AR})$ is decomposed by $ W ^{A \rightarrow A_cA_lA_r}$ in the form of 
\begin{align}
 \ca{W}^{A \rightarrow A_cA_lA_r} (X^{AR})=\sum_{j,k\in\ca{J}}\outpro{j}{k}^{A_c}\otimes \tilde{X}_{jk}^{A_l A_r R},
\end{align}
where
\begin{align}
\tilde{X}_{jk}^{A_l A_r R} &:=\bra{j}^{A_c}  \ca{W}^{A \rightarrow A_cA_lA_r}(X^{AR}) \ket{k}^{A_c}
=(W ^{A_l}_j \otimes  W ^{A_r}_j) \Pi_j X^{AR} \Pi_k ( W ^{A_l}_k \otimes  W ^{A_r}_k)^{\dagger}.
\end{align}
Conversely, any $Y^{A_cA_lA_r} \in \ca{L}({\ca H}^{A_c}\otimes{\ca H}^{A_l}\otimes{\ca H}^{A_r})$ such that ${\rm supp}(Y^{A_c A_l A_r}) \subset{\rm img} ( W ^{A \rightarrow A_c A_l A_r})$, is mapped to $(\ca{W}^{A \rightarrow A_c A_l A_r})^\dagger (Y^{A_c A_l A_r} ) \in \ca{L}(\ca{H}^A)$. 
Note that $\tilde{X}_{jk}$ is related to $X_{jk}$ defined by \req{haihai} as $\outpro{j}{k}^{A_c}\otimes \tilde{X}_{jk}^{A_l A_r R}= \ca{W}^{A \rightarrow A_cA_lA_r} (X_{jk}^{AR})$. In the following, we denote $\tilde{X}_{jk}^{A_l A_r R}$ by $X_{jk}^{A_l A_r R}$ for simplicity of notations.

Let $A'$ be a quantum system such that $\ca{H}^A\cong\ca{H}^{A'}$. It is straightforward to verify that the fixed maximally entangled state $|\Phi\rangle$ defined by \req{dfnmaxentDSP} is decomposed by $W$ as
\begin{align}
(W^{A\rightarrow A_cA_lA_r}\otimes W^{A'\rightarrow A_c'A_l'A_r'})|\Phi\rangle^{AA'}
=\sum_{j=1}^J\sqrt{\frac{l_jr_j}{d_A}}\ket{j}^{A_c}\ket{j}^{A_c'}|\Phi_j^l\rangle^{A_lA_l'}|\Phi_j^r\rangle^{A_rA_r'},
\laeq{decompositionmaxent}
\end{align}
where $|\Phi_j^l\rangle\in\ca{H}_j^{A_l}\otimes\ca{H}_j^{A_l'}$ and $|\Phi_j^r\rangle\in\ca{H}_j^{A_r}\otimes\ca{H}_j^{A_r'}$ are fixed maximally entangled states of rank $l_j$ and $r_j$, respectively.

\subsection{Proof of Lemma~\ref{lmm:nenchalu}} \label{SS:PL7}

We now prove Lemma~\ref{lmm:nenchalu}. The statement is given as follows:
for any $\varsigma^{ER} \in \ca{S}_=(\ca{H}^{ER})$ and any $X\in{\rm Her}(\ca{H}^{AR})$ such that
$X_{jj}^{A_lR}=0$, the following inequality holds for any possible permutation $\sigma \in \mbb{P}$:
\begin{align}
\mbb{E}_{U \sim {\sf H}_{\times}} \left[\left\|\ca{T}^{A \rightarrow E} \circ\ca{G}_{\sigma^{-1}}^A  \circ \ca{U}^A ( X^{AR} )\right\|_{2,\varsigma^{ER}}^2\right]
\leq 
\sum_{j,k=1}^J \frac{d_A^2}{r_j r_k}
\left\|   \tr_{A_l}\left[X_{\sigma(j)\sigma(k)}^{A_l^T A_r R}\tau^{A_l \bar{A}_r E}_{jk} \right]  \right\|^2_{2, \varsigma^{ER}} 
.\label{eq:atari}
\end{align}
Here, $A_l^T$ denotes the transposition of $A_l$ with respect to the Schmidt basis of the fixed maximally entangled state used to define the Choi-Jamio\l kowski representation $\tau^{AE}$ of $\mc{T}^{A \rightarrow E}$.

\begin{prf}
 Introducing a notation $\mbb{F}^{RE,R'E'}_{\varsigma}:= ( (\varsigma^{ER})^{\otimes 2})^{-1/4}(\mbb{F}^{RR'}  \otimes \mbb{F}^{EE'}) ( (\varsigma^{ER})^{\otimes 2} )^{-1/4}$, we have
\begin{align}
&
\bigl|\!\bigl| \ca{T}^{A \rightarrow E} \circ \ca{G}_\sigma^{\dagger A} \circ \ca{U}^{\dagger A} (X^{AR} )  \bigr|\!\bigr|_{2,\varsigma^{ER}}^2 
\nn\\
 &=
 \tr \biggl[ \biggl( (\varsigma^{ER})^{-\frac{1}{4}}  \ca{T}^{A \rightarrow E}\! \circ \ca{G}_\sigma^{\dagger A} \!\circ \ca{U}^{\dagger A} ( X^{AR} ) (\varsigma^{ER})^{-\frac{1}{4}} \biggr)^2 \biggr]
 \nonumber\\
&= 
\tr \left[ \biggl( (\varsigma^{ER})^{-\frac{1}{4}}  \ca{T}^{A \rightarrow E} \!\circ \ca{G}_\sigma^{\dagger A} \!\circ \ca{U}^{\dagger A} (X^{AR} ) (\varsigma^{ER})^{-\frac{1}{4}} \biggr)^{\otimes 2} \right.
\bigl( \mbb{F}^{RR'} \otimes \mbb{F}^{EE'} \bigr) \biggr]
\nonumber\\
&= 
\tr \left[ \left(  \ca{T}^{A \rightarrow E} \!\circ \ca{G}_\sigma^{\dagger A} \!\circ \ca{U}^{\dagger A} (X^{AR} ) \right)^{\otimes 2} \mbb{F}^{RE,R'E'}_{\varsigma} \right]
\nonumber\\
&=  
 \tr \bigl[  \bigl(  X^{AR}  \bigr)^{\otimes 2} \bigl[ ( \ca{G}_\sigma^A\circ \ca{U}^A  \circ \ca{T}^{* E \rightarrow A })^{\otimes 2}(\mbb{F}^{RE,R'E'}_{\varsigma})\bigr] \bigr].
\end{align}
Thus, using the fact that $\ca{G}_{\sigma^{-1}}=\ca{G}_{\sigma}^\dagger$ and that $\mbb{E}_{U \sim {\sf H}_{\times}}[f(U)]=\mbb{E}_{U \sim {\sf H}_{\times}}[f(U^\dagger)]$ for any function $f$,
we have
\begin{align}
\mbb{E}_{U } \left[\left\|\ca{T}^{A \rightarrow E} \circ\ca{G}_{\sigma^{-1}}^A  \circ \ca{U}^A ( X^{AR} )\right\|_{2,\varsigma^{ER}}^2\right]
&=\mbb{E}_{U} 
 \biggr[\bigl|\!\bigl| \ca{T}^{A \rightarrow E} \circ \ca{G}_\sigma^{\dagger A} \circ \ca{U}^{\dagger A} (X^{AR} )  \bigr|\!\bigr|_{2,\varsigma^{ER}}^2 \biggl] 
 \nonumber\\
&=  \mbb{E}_{U } \tr \bigl[  \bigl(  X^{AR}  \bigr)^{\otimes 2} \bigl[ ( \ca{G}_\sigma^A\circ \ca{U}^A  \circ \ca{T}^{* E \rightarrow A })^{\otimes 2}(\mbb{F}^{RE,R'E'}_{\varsigma})\bigr] \bigr]\nonumber\\
&=    \tr \bigl[  \bigl(  X^{AR}  \bigr)^{\otimes 2} \mbb{E}_{U }\bigl[ (  \ca{G}_\sigma^A\circ \ca{U}^A  \circ \ca{T}^{* E \rightarrow A })^{\otimes 2}(\mbb{F}^{RE,R'E'}_{\varsigma})\bigr] \bigr]
\nonumber\\
&=\tr [ (X^{AR})^{\otimes 2} \Xi_\sigma^{AA'RR'}],
\label{eq:hoshihoshiii}
\end{align}
where we have defined
$\Xi_\sigma^{AA'RR'}:= \mbb{E}_{U \sim {\sf H}_{\times}} [ ( \ca{G}_\sigma^A\circ \ca{U}^A  \circ \ca{T}^{* E \rightarrow A })^{\otimes 2}(\mbb{F}^{RE,R'E'}_{\varsigma}) ]$.

We first embed the operator $\Xi_\sigma^{AA'RR'}$ into the space $A_c A_l A_r R$ and $A'_c A'_l A'_r R'$.
We introduce the following notations for the embedded map and the embedded operators:
\alg{
&\ca{T}^{A_cA_lA_r \rightarrow E} : =\ca{T}^{A \rightarrow E} \circ (\ca{W}^{A \rightarrow A_cA_lA_r})^{\dagger},
\quad
\tau^{A_cA_lA_rE} : =\ca{W}^{A \rightarrow A_cA_lA_r} (\tau^{AE}),
\\
&\Upsilon_\varsigma
:=
(\ca{T}^{* E\rightarrow A_cA_lA_r})^{\otimes 2}(\mbb{F}^{RE,R'E'}_{\varsigma}),
\quad
\Upsilon_{\varsigma,jkmn}^{A_lA_rR A_l'A_r'R'}
:=(\bra{j}^{A_c}\otimes \bra{k}^{A_c'})\Upsilon_\varsigma (\ket{m}^{A_c}\otimes \ket{n}^{A_c'}).
}
Using these notations, the operator $\Xi_\sigma^{AA'RR'}$ is embedded to be
\begin{align}
&(\ca{W}^{A \rightarrow A_cA_lA_r})^{\otimes 2}(\Xi_\sigma^{AA'RR'})\nonumber\\
&=
\sum_{j,k,m,n =1}^J \bigl[
\outpro{\sigma(j)}{\sigma(m)}^{A_c}\otimes \outpro{\sigma(k)}{\sigma(n)}^{A_c'}\bigr]
\otimes\mbb{E}_{U \sim {\sf H}_{\times}}\bigl[ 
(U_j^{A_r} \otimes U_k^{A_r'})
\Upsilon_{\varsigma,jkmn}^{A_lA_rR A_l'A_r'R'}
(U_m^{\dagger A_r} \otimes U_n^{\dagger A_r'}) 
\bigr].
\nonumber
\end{align}
Due to Lemma \ref{Lemma:av},  
the terms in the summation remain non-zero only in the following three cases: (i) $J\geq2$ and $(j,k)=(m,n)$ ($j \neq k$), (ii) $J\geq2$ and $(j,k)=(n,m)$ ($j \neq k$), and (iii) $j=k=m=n$.
In the following, we assume that $J\geq2$, and separately investigate the three cases using Lemma~\ref{Lemma:av}. 
Our concern is then $\Xi_{\sigma,{\rm (i)}}$, $\Xi_{\sigma,{\rm (ii)}}$ and $\Xi_{\sigma,{\rm (iii)}}$ such that
\begin{align}
&(\ca{W}^{A \rightarrow A_cA_lA_r})^{\otimes 2}(\Xi_{\sigma,{\rm (i)}})
\nn\\
&=
\sum_{j,k =1}^J \bigl[
\outpro{\sigma(j)}{\sigma(j)}^{A_c}\otimes \outpro{\sigma(k)}{\sigma(k)}^{A_c'}\bigr]\otimes
\mbb{E}_{U }\bigl[ 
(U_j^{A_r} \otimes U_k^{A_r'})
\Upsilon_{\varsigma,jkjk}^{A_lA_rR A_l'A_r'R'}
(U_j^{\dagger A_r} \otimes U_k^{\dagger A_r'})
\bigr],\\
&(\ca{W}^{A \rightarrow A_cA_lA_r})^{\otimes 2}(\Xi_{\sigma,{\rm (ii)}})
\nn\\
&=
\sum_{j,k =1}^J \bigl[
\outpro{\sigma(j)}{\sigma(k)}^{A_c}\otimes \outpro{\sigma(k)}{\sigma(j)}^{A_c'}\bigr]\otimes
\mbb{E}_{U }\bigl[ 
( U_j^{A_r} \otimes U_k^{A_r'})
\Upsilon_{\varsigma,jkkj}^{A_lA_rR A_l'A_r'R'}
(U_k^{\dagger A_r} \otimes U_j^{\dagger A_r'})
\bigr],\\
&(\ca{W}^{A \rightarrow A_cA_lA_r})^{\otimes 2}(\Xi_{\sigma,{\rm (iii)}})
\nn\\
&=
\sum_{j =1}^J \bigl[
\outpro{\sigma(j)}{\sigma(j)}^{A_c}\otimes \outpro{\sigma(j)}{\sigma(j)}^{A_c'}\bigr]\otimes
\mbb{E}_{U }\bigl[ 
(U_j^{A_r} \otimes U_j^{A_r'})
\Upsilon_{\varsigma,jjjj}^{A_lA_rR A_l'A_r'R'}
(U_j^{\dagger A_r} \otimes U_j^{\dagger A_r'})
\bigr].
\end{align}
Note that $\Xi_\sigma=\Xi_{\sigma,{\rm (i)}}+\Xi_{\sigma,{\rm (ii)}}+\Xi_{\sigma,{\rm (iii)}}$.

In the case (i), from Lemma \ref{Lemma:av}, we have
\begin{align}
(\ca{W}^{A \rightarrow A_cA_lA_r})^{\otimes 2}(\Xi^{AA'RR'}_{\sigma,{\rm (i)}})
=
\sum_{j\neq k}\frac{1}{r_jr_k}\outpro{\sigma(j)}{\sigma(j)}^{A_c}\otimes \outpro{\sigma(k)}{\sigma(k)}^{A_c'}
\otimes
\mbb{I}_{jk}^{A_rA_r'} \otimes \Xi^{A_lR A_l'R'}_{{\rm(i)},jk}, 
\end{align}
where
$\Xi^{A_lR A_l'R'}_{{\rm(i)},jk} =
\tr_{A_rA_r'}
\bigl[ 
\mbb{I}_{jk}^{A_rA_r'}
\Upsilon_{\varsigma,jkjk}^{A_lA_rR A_l'A_r'R'}
\bigr]$.
It follows that
\begin{align}
\tr
\left[ 
\left( X^{A_lA_rR}_{\sigma(j)\sigma(j)} \otimes  X^{A_l'A'_r R'}_{\sigma(k)\sigma(k)} \right)
\left(\mbb{I}_{jk}^{A_rA_r'} \otimes \Xi^{A_lRA_l'R'}_{{\rm(i)},jk}\right)
\right]
=
\tr
\left[ 
\left( X^{A_lR}_{\sigma(j)\sigma(j)} \otimes  X^{A_l' R'}_{\sigma(k)\sigma(k)} \right)
\Xi^{A_lRA_l'R'}_{{\rm(i)},jk}
\right]
,
\label{eq:8181}
\end{align}
and consequently, from the condition for $X$, i.e. $X_{jj}^{A_lR}=0$, that $\tr[ (X^{AR})^{\otimes 2} \Xi^{AA'RR'}_{\sigma,{\rm (i)}} ] 
=0$.

Let us next consider the case (ii), where $(j,k)=(n,m)$ ($j\neq k)$. This case yields
\begin{align}
(\ca{W}^{A \rightarrow A_cA_lA_r})^{\otimes 2}(\Xi^{AA'RR'}_{\sigma,{\rm (ii)}})
=
\sum_{j\neq k}\frac{1}{r_jr_k}\outpro{\sigma(j)}{\sigma(k)}^{A_c}\otimes \outpro{\sigma(k)}{\sigma(j)}^{A_c'}
\otimes
\mbb{F}_{jk}^{A_r A_r'} \otimes \Xi^{A_lR A_l'R'}_{{\rm(ii)},jk}, 
\end{align}
where
$\Xi^{A_lR A_l'R'}_{{\rm(ii)},jk} =
\tr_{A_rA_r'}
\bigl[ 
\Upsilon_{\varsigma,jkkj}^{A_lA_rR A_l'A_r'R'}
\mbb{F}_{kj}^{A_r A_r'}\bigr]$.
Denoting the $A_r$ part of $\Upsilon$ and $\ca{T}^*$ by $\bar{A}_r$, we have
\begin{align}
&\tr
\left[ 
\left( X^{A_lA_rR}_{\sigma(k)\sigma(j)} \otimes  X^{A_l'A'_r R'}_{\sigma(j)\sigma(k)} \right)
\left(\mbb{F}_{jk}^{A_r A_r'} \otimes \Xi^{A_lRA_l'R'}_{{\rm(ii)},jk}\right)
\right]\nonumber\\
&=
\tr
\left[ 
\left( X^{A_lA_rR}_{\sigma(k)\sigma(j)} \otimes  X^{A_l'A'_r R'}_{\sigma(j)\sigma(k)}\otimes  \mbb{F}_{kj}^{\bar{A}_r\bar{A}_r'} \right)
\left(\mbb{F}_{jk}^{A_r A_r'} \otimes \Upsilon_{\varsigma,jkkj}^{A_l\bar{A}_rR A_l'\bar{A}_r'R'}\right)
\right]\nonumber\\
&=
\tr
\left[ \left(\outpro{k}{j}^{A_c}\!\otimes\! \outpro{j}{k}^{A_c'}\!\otimes\! X^{A_lA_rR}_{\sigma(k)\sigma(j)} \!\otimes\!  X^{A_l'A_r'R'}_{\sigma(j)\sigma(k)} \!\otimes\!  \mbb{F}_{kj}^{\bar{A}_r\bar{A}_r'}\right)
\left(\mbb{F}_{jk}^{A_r A_r'} \otimes(\ca{T}^{* E\rightarrow A_cA_l\bar{A}_r})^{\otimes 2}(\mbb{F}^{RE,R'E'}_{\varsigma})\right)
 \right]\nonumber\\
 &=
 \tr
\left[ \left( (\ca{T}^{A_cA_l\bar{A}_r \rightarrow E} )^{\otimes 2}
\left(\outpro{k}{j}^{A_c} \!\otimes\! \outpro{j}{k}^{A_c'} \otimes  X^{A_lA_rR}_{\sigma(k)\sigma(j)} 
\!\otimes\!  X^{A_l'A_r'R'}_{\sigma(j)\sigma(k)} \!\otimes\!  \mbb{F}_{kj}^{\bar{A}_r\bar{A}_r'}\right)\right)
\left(\mbb{F}_{jk}^{A_r A_r'} \otimes\mbb{F}^{RE,R'E'}_{\varsigma}\right)
 \right]\nonumber\\
 &=d_A^2
\tr
\left[ 
\left(
\left(\outpro{j}{k}^{A_c}\!\otimes\! \outpro{k}{j}^{A_c'}\!\otimes\! X^{A_l^TA_rR}_{\sigma(k)\sigma(j)} \!\otimes \! X^{A_l'^TA_r'R'}_{\sigma(j)\sigma(k)} \!\otimes\!  \mbb{F}_{jk}^{\bar{A}_r\bar{A}_r'} \right)
(\tau^{A_cA_l \bar{A}_r E} )^{\otimes 2}
\right)
\left(\mbb{F}_{jk}^{A_r A_r'} \!\otimes\!\mbb{F}^{RE,R'E'}_{\varsigma}
\right)
 \right]\nonumber\\
&=d_A^2
\tr
\left[ 
\left( X_{\sigma(k)\sigma(j)}^{A_l^T A_r R} \otimes  X_{\sigma(j)\sigma(k)}^{A_l'^T A'_r R'} \right)
\left(\tau_{kj}^{A_l \bar{A}_r E} \otimes \tau_{jk}^{A_l' \bar{A}'_r E'} \right)
\left(
\mbb{F}_{jk}^{A_r A_r'} 
\otimes 
\mbb{F}_{jk}^{\bar{A}_r \bar{A}_r'}
\otimes 
\mbb{F}^{RE,R'E'}_{\varsigma}
\right)  \right]\nonumber\\
 &=d_A^2\tr
\left[ 
\left(\tr_{A_l}
\left[ X_{\sigma(k)\sigma(j)}^{A_l^T A_r R}\tau_{kj}^{A_l \bar{A}_r E}\right]
 \otimes \tr_{A_l'}\left[  X_{\sigma(j)\sigma(k)}^{A_l'^T A'_r R'}\tau_{jk}^{A_l' \bar{A}'_r E'}\right] \right)
\left(
\mbb{F}_{jk}^{A_r A_r'} 
\otimes 
\mbb{F}_{jk}^{\bar{A}_r \bar{A}_r'}
\otimes 
\mbb{F}^{RE,R'E'}_{\varsigma}
\right)
\right]\nonumber\\
&=d_A^2
\left\|
\tr_{A_l}\left[ X^{A_l ^TA_r R}_{\sigma(j)\sigma(k)} \tau^{A_l \bar{A}_r E}_{jk} \right]
\right\|_{2,\varsigma^{ER}}^2,\label{eq:8182}
\end{align}
where the fourth line follows from the Choi-Jamio\l kowski correspondence \req{CJinverse} and the last line from the swap trick (Lemma \ref{Lemma:Swap}).
Hence we obtain
\begin{align}
\tr[ (X^{AR})^{\otimes 2} \Xi^{AA'RR'}_{\sigma,{\rm (ii)}} ] 
=
\sum_{j\neq k}
\frac{d_A^2}{r_j r_k}
\left\|
\tr_{A_l} \left[
X^{A_l^TA_r R}_{\sigma(j)\sigma(k)} \tau^{A_l \bar{A}_r E}_{jk} 
 \right]
\right\|_{2,\varsigma^{ER}}^2.
\label{eq:poincar}
\end{align}

Finally, we investigate the case (iii). Lemma \ref{Lemma:av} leads to
\begin{align}
(\ca{W}^{A \rightarrow A_cA_lA_r})^{\otimes 2}(\Xi^{AA'RR'}_{\sigma,{\rm (iii)}})
=
\sum_{j=1}^J\!\frac{1}{r_j(r_j^2-1)}\outpro{\sigma(j)}{\sigma(j)}^{A_c}\!\otimes\!\outpro{\sigma(j)}{\sigma(j)}^{A_c'}\!\otimes\!\Xi_{{\rm(iii)},jj}^{A_lRA_l'R'}\!
,\laeq{heartheart}
\end{align}
where
\begin{align}
\Xi_{{\rm(iii)},jj}^{A_lRA_l'R'}:=
\left[
r_j\mbb{I}_{jj}^{A_rA_r'}\otimes \Xi_{{\rm(i)},jj}^{A_lRA_l'R'} - \mbb{I}_{jj}^{A_rA_r'}\otimes \Xi^{A_lRA_l'R'}_{{\rm (ii)},jj} 
+r_j\mbb{F}_{jj}^{A_rA_r'}\otimes \Xi_{{\rm(ii)},jj}^{A_lRA_l'R'} - \mbb{F}_{jj}^{A_rA_r'}\otimes \Xi^{A_lRA_l'R'}_{{\rm (i)},jj}
\right].
\laeq{kawashio}
\end{align}
Similarly to (\ref{eq:8181}) and (\ref{eq:8182}), we have
\begin{align}
\tr
\left[ 
\left( X^{A_lA_rR}_{\sigma(j)\sigma(j)} \otimes  X^{A_l'A'_r R'}_{\sigma(j)\sigma(j)} \right)
\left(\mbb{I}_{jj}^{A_r A_r'} \otimes \Xi^{A_lRA_l'R'}_{{\rm(ii)},jj}\right)
\right]
=
0
\end{align}
and
\begin{align}
&\tr
\left[ 
\left(  X^{A_lA_rR}_{\sigma(j)\sigma(j)} \otimes   X^{A_l'A'_r R'}_{\sigma(j)\sigma(j)} \right)
\left(\mbb{F}_{jj}^{A_r A_r'} \otimes \Xi^{A_lRA_l'R'}_{{\rm(i)},jj} \right)\right]\nonumber\\
&=
\tr\left[ \left(\outpro{j}{j}^{A_c} \!\otimes\! \outpro{j}{j}^{A_c'} \!\otimes\! X^{A_lA_rR}_{\sigma(j)\sigma(j)} \!\otimes\!   X^{A_l'A_r'R'}_{\sigma(j)\sigma(j)} \!\otimes\!  \mbb{I}_{jj}^{\bar{A}_r\bar{A}_r'}\right)
\left(\mbb{F}_{jj}^{A_r A_r'} \otimes
(\ca{T}^{* E\rightarrow A_cA_l\bar{A}_r})^{\otimes 2}
\left(\mbb{F}^{RE,R'E'}_{\varsigma}\right)\right)\right]\nonumber\\
 &=
\tr\left[ \left((\ca{T}^{A_cA_l\bar{A}_r \rightarrow E})^{\otimes 2}
\left(\outpro{j}{j}^{A_c} \otimes \outpro{j}{j}^{A_c'} \otimes  X^{A_lA_rR}_{\sigma(j)\sigma(j)} 
\otimes   X^{A_l'A_r'R'}_{\sigma(j)\sigma(j)} \otimes  \mbb{I}_{jj}^{\bar{A}_r\bar{A}_r'}\right)\right)
\left(\mbb{F}_{jj}^{A_r A_r'} \otimes\mbb{F}^{RE,R'E'}_{\varsigma}\right)\right]\nonumber\\
 &=d_A^2
\tr
\left[ 
\left( X_{\sigma(j)\sigma(j)}^{A_l ^TA_r R} \otimes  X_{\sigma(j)\sigma(j)}^{A_l'^T A'_r R'}\right)
\left(\tau_{jj}^{A_l \bar{A}_r E} \otimes \tau_{jj}^{A_l' \bar{A}'_r E'}\right)
\left(
\mbb{F}_{jj}^{A_r A_r'} 
\otimes 
\mbb{I}_{jj}^{\bar{A}_r \bar{A}_r'}
\otimes 
\mbb{F}^{RE,R'E'}_{\varsigma}
\right)  \right]
\nonumber\\
 &=d_A^2\tr
\left[ 
\left(\tr_{A_l}\!\left[X_{\sigma(j)\sigma(j)}^{A_l^T A_r R}\tau_{j}^{A_l  E}\right] \!\otimes\! \tr_{A_l'}\!\left[ X_{\sigma(j)\sigma(j)}^{A_l'^T A'_r R'}\tau_{jj}^{A_l'  E'}\right]\right)
\left(
\mbb{F}_{jj}^{A_r A_r'} 
\otimes 
\mbb{F}^{RE,R'E'}_{\varsigma}
\right)
\right]\nonumber\\
&=d_A^2\left\|
\tr_{A_l}
\left[ X^{A_l^T A_r R}_{\sigma(j)\sigma(j)} \tau^{A_l  E}_{jj} 
\right]
\right\|_{2,\varsigma^{ER}}^2.
\end{align}
Combining this with (\ref{eq:8181}), (\ref{eq:8182}) and \req{kawashio}, we obtain
\begin{align}
\tr\left[ (X_{\sigma(j)\sigma(j)}^{A_lR})^{\otimes 2} \Xi^{A_lA_l'RR'}_{{\rm (iii)},jj} \right]
&
=d_A^2 
r_j \left\|
\tr_{A_l}\left[ X^{A_l^T A_r R}_{\sigma(j)\sigma(j)}\tau^{A_l \bar{A}_r E}_{jj} \right]
\right\|_{2,\varsigma}^2
-d_A^2 \left\|
\tr_{A_l}\left[ X^{A_l^T A_r R}_{\sigma(j)\sigma(j)}\tau^{A_l E}_{jj} \right]
\right\|_{2,\varsigma}^2.
\nonumber
\end{align}
Noting that $\tr_{A_l}[ X^{A_l^T A_r R}_{\sigma(j)\sigma(j)}\tau^{A_l \bar{A}_r E}_{jj}]$ is a Hermitian operator for each $j$, and by using the property of the Hilbert-Schmidt norm (see Lemma~\ref{Lemma:purity}),  the above equality leads to
\begin{align}
\tr\left[ (X_{\sigma(j)\sigma(j)}^{A_lR})^{\otimes 2} \Xi^{A_lA_l'RR'}_{{\rm (iii)},jj} \right]
\leq
d_A^2 
\left(r_j -\frac{1}{r_j}\right)
\left\|
\tr_{A_l} \left[ X^{A_l^T A_r R}_{\sigma(j)\sigma(j)}\tau^{A_l \bar{A}_r E}_{jj}  \right]
\right\|_{2,\varsigma^{ER}}^2.
\end{align}
Combining this with \req{heartheart}, we have
\begin{align}
\tr[ (X^{AR})^{\otimes 2} \Xi^{AA'RR'}_{\sigma,{\rm (iii)}} ]
&=\sum_{j=1}^J \frac{1}{r_j(r_j^2-1)} \tr\left[ (X_{\sigma(j)\sigma(j)}^{A_lR})^{\otimes 2} \Xi^{A_lA_l'RR'}_{{\rm (iii)},jj} \right]
\nonumber\\
&\leq
\sum_{j=1}^J \frac{d_A^2}{r_j^2} 
\left\|
\tr_{A_l} \left[ X^{A_l^T A_r R}_{\sigma(j)\sigma(j)}\tau^{A_l \bar{A}_r E}_{jj}  \right]
\right\|_{2,\varsigma^{ER}}^2
.
\label{eq:poincare}
\end{align}

Since $\Xi_\sigma=\Xi_{\sigma,{\rm (i)}}+\Xi_{\sigma,{\rm (ii)}}+\Xi_{\sigma,{\rm (iii)}}$, we can thus obtain from these evaluations that
\begin{align}
\tr[ (X^{AR})^{\otimes 2} \Xi_\sigma^{AA'RR'}]
\leq 
\sum_{j,k =1}^J \frac{d_A^2}{r_j r_k}
\left\|   \tr_{A_l}\left[X_{\sigma(j)\sigma(k)}^{A_l^T A_r R}\tau^{A_l \bar{A}_r E}_{jk} \right]  \right\|^2_{2, \varsigma^{ER}} 
\end{align}
for any $\varsigma^{ER} \in \ca{S}_=(\ca{H}^{ER})$ and $\sigma\in\mbb{P}$. Combining this with Eq.~(\ref{eq:hoshihoshiii}) concludes the proof.
$\hfill \blacksquare$

\end{prf}

\section{Proof of \rLmm{trianglemother}}
\lapp{trianglemother}

We prove \rLmm{trianglemother}. We start with recalling the statement:
Consider arbitrary unnormalized states $\Psi^{AR},\hat{\Psi}^{AR}\in\ca{P}(\ca{H}^{AR})$ and arbitrary CP maps $\ca{T},\hat{\ca{T}}:A\rightarrow E$.  
Let $\ca{D}_+^{A \rightarrow E}$ and $\ca{D}_-^{A \rightarrow E}$ be arbitrary CP maps such that $\ca{T}-\hat{\ca{T}}=\ca{D}_+-\ca{D}_-$.
Let $\delta_+^{AR}$ and $\delta_-^{AR}$ be linear operators on $\ca{H}^A\otimes\ca{H}^{R}$, such that
\alg{
\delta_+^{AR}\geq0,\quad\delta_-^{AR}\geq0,
\quad
{\rm supp}[\delta_+^{AR}]\perp{\rm supp}[\delta_-^{AR}]
\laeq{dltperp}
}
and that
\alg{
\hat{\Psi}^{AR} -\Psi^{AR}=\delta_+^{AR}-\delta_-^{AR}.
\laeq{pmpdmd}
}
The following inequality holds for any possible permutation $\sigma\in\mbb{P}$ and for both ${\Psi}_*={\Psi}_{\rm av}$ and ${\Psi}_*=\ca{C}^A(\Psi)$:
\begin{multline}
\mbb{E}_{U \sim {\sf H}_{\times}}\left[
\left\|{\ca{T}}^{A \rightarrow E}  \circ \ca{G}_\sigma^A \circ \ca{U}^A ( {\Psi}^{AR}  - {\Psi}_*^{AR} )\right\|_1
\right]\\
\leq
\mbb{E}_{U \sim {\sf H}_{\times}}\left[
\left\|\hat{\ca{T}}^{A \rightarrow E} \circ \ca{G}_\sigma^A \circ \ca{U}^A ( \hat{\Psi}^{AR}  -  \hat{\Psi}_*^{AR} )\right\|_1
\right]+2 \:
\tr[(\ca{D}_+^{A \rightarrow E}+ \ca{D}_-^{A \rightarrow E}) \circ \ca{G}_\sigma^A   ( \Psi_{\rm av}^{AR} )]
\\
+2 \:\mbb{E}_{U \sim {\sf H}_{\times}}\tr[\hat{\ca{T}}^{A \rightarrow E}  \circ \ca{G}_\sigma^A \circ\ca{U}^A (\delta_+^{AR}+\delta_-^{AR})].
\label{eq:boundhathat}
\end{multline}
Here, $\hat{\Psi}_*=\mbb{E}_{U\sim{\sf H}_\times}[\ca{U}^A(\hat{\Psi}^{AR})]$ for ${\Psi}_*={\Psi}_{\rm av}$ and $\hat{\Psi}_*=\ca{C}^A(\hat{\Psi})$ for ${\Psi}_*=\ca{C}^A(\Psi)$.

\begin{prf}
By a recursive application of the triangle inequality, we have
\begin{align}
&\left\|{\ca{T}}^{A \rightarrow E}  \circ \ca{G}_\sigma^A \circ \ca{U}^A ( {\Psi}^{AR}  -  {\Psi}_*^{AR} )\right\|_1\nn\\
&\leq\left\|({\ca{T}}^{A \rightarrow E} - \hat{\ca{T}}^{A \rightarrow E} )\circ \ca{G}_\sigma^A \circ {\ca{U}}^A ( \Psi^{AR} )\right\|_1+\left\|\hat{\ca{T}}^{A \rightarrow E} \circ \ca{G}_\sigma^A \circ \ca{U}^A ( \Psi^{AR}  -  \hat{\Psi}^{AR} )\right\|_1\nonumber\\
&\quad+\left\|\hat{\ca{T}}^{A \rightarrow E} \circ \ca{G}_\sigma^A \circ \ca{U}^A ( \hat{\Psi}^{AR}  - 
 \hat{\Psi}_*^{AR} )\right\|_1
 +\left\|\hat{\ca{T}}^{A \rightarrow E} \circ \ca{G}_\sigma^A \circ \ca{U}^A ( \hat{\Psi}_*^{AR}  -  \Psi_*^{AR} )\right\|_1\nonumber\\
&\quad+\left\|(\hat{\ca{T}}^{A \rightarrow E}  - {\ca{T}}^{A \rightarrow E}) \circ \ca{G}_\sigma^A \circ \ca{U}^A ( {\Psi}_*^{AR} )\right\|_1.
\end{align}
The expectation value of the first term is bounded as
\begin{align}
&\mbb{E}_{U}  \left[
\left\|({\ca{T}}^{A \rightarrow E}  - \hat{\ca{T}}^{A \rightarrow E})\circ \ca{G}_\sigma^A \circ \ca{U}^A ( \Psi^{AR} )
\right\|_1
\right]
 \nonumber\\
&=
\mbb{E}_{U }  \left[
\left\|(\ca{D}_+^{A \rightarrow E} - \ca{D}_-^{A \rightarrow E})\circ \ca{G}_\sigma^A  \circ \ca{U}^A ( \Psi^{AR} ) 
\right\|_1
\right]
\nonumber\\
&\leq
\mbb{E}_{U }
\left[
\left\|\ca{D}_+^{A \rightarrow E} \circ \ca{G}_\sigma^A  \circ \ca{U}^A ( \Psi^{AR} )\right\|_1 
\right]
+\mbb{E}_{U }
\left[
\left\| \ca{D}_-^{A \rightarrow E} \circ \ca{G}_\sigma^A  \circ {\ca{U}}^A ( \Psi^{AR} )\right\|_1
\right]
\nonumber\\
&=
\mbb{E}_{U }
\left[
\tr[\ca{D}_+^{A \rightarrow E} \circ \ca{G}_\sigma^A  \circ {\ca{U}}^A ( \Psi^{AR} )] 
\right]
+
\mbb{E}_{U }
\left[
\tr[ \ca{D}_-^{A \rightarrow E} \circ \ca{G}_\sigma^A  \circ {\ca{U}}^A ( \Psi^{AR} )]
\right]
\nonumber\\
&=
\tr[(\ca{D}_+^{A \rightarrow E}+ \ca{D}_-^{A \rightarrow E}) \circ \ca{G}_\sigma^A  ( \Psi_{\rm av}^{AR} )].
\end{align}
In the same way, the expectation value of the last term is bounded as
\begin{align}
\mbb{E}_{U}  \left[
\left\|(\hat{\ca{T}}^{A \rightarrow E}  - {\ca{T}}^{A \rightarrow E} )\circ \ca{G}_\sigma^A \circ \ca{U}^A ( \Psi_*^{AR} )
\right\|_1
\right]
\leq
\tr[(\ca{D}_+^{A \rightarrow E}+ \ca{D}_-^{A \rightarrow E}) \circ \ca{G}_\sigma^A   ( \Psi_{\rm av}^{AR} )].
\end{align}
For the second term, we have
\begin{align}
&\mbb{E}_{U} \left[
 \left\|\hat{\ca{T}}^{A \rightarrow E} \circ \ca{G}_\sigma^A  \circ {\ca{U}}^A ( {\Psi}^{AR}  
 -  \hat{\Psi}^{AR} )\right\|_1
 \right]\nonumber\\
 &=\mbb{E}_{U } \left[
 \left\|\hat{\ca{T}}^{A \rightarrow E} \circ \ca{G}_\sigma^A \circ \ca{U}^A (\delta_+^{AR}-\delta_-^{AR}) \right\|_1
 \right]\nonumber\\
&\leq
\mbb{E}_{U } \left[
\left\|\hat{\ca{T}}^{A \rightarrow E} \circ \ca{G}_\sigma^A  \circ {\ca{U}}^A (\delta_+^{AR})\right\|_1\right]
+\mbb{E}_{U } \left[ \left\|{\ca{T}}^{A \rightarrow E} \circ \ca{G}_\sigma^A  \circ {\ca{U}}^A (\delta_-^{AR}) \right\|_1
\right]\nonumber\\
&=\mbb{E}_{U } 
\left[
 \tr[\hat{\ca{T}}^{A \rightarrow E}  \circ \ca{G}_\sigma^A \circ {\ca{U}}^A (\delta_+^{AR})]
 \right]
+ \mbb{E}_{U } 
\left[\tr[\hat{\ca{T}}^{A \rightarrow E}  \circ \ca{G}_\sigma^A \circ {\ca{U}}^A (\delta_-^{AR})]
\right]
\nonumber\\
&=\mbb{E}_{U} \left[\tr[\hat{\ca{T}}^{A \rightarrow E}  \circ \ca{G}_\sigma^A \circ \ca{U}^A (\delta_+^{AR}+\delta_-^{AR})]\right]. 
\end{align}
Similarly, the expectation value of the fourth term is bounded as
\begin{align}
\mbb{E}_{U }\left[
\left\|\hat{\ca{T}}^{A \rightarrow E} \circ \ca{G}_\sigma^A \circ \ca{U}^A ( \hat{\Psi}_*^{AR} 
 - {\Psi}_*^{AR} )\right\|_1
 \right]
\leq
\mbb{E}_{U}\left[\tr[\hat{\ca{T}}^{A \rightarrow E}  \circ \ca{G}_\sigma^A \circ \ca{U}^A (\delta_+^{AR}+\delta_-^{AR})]\right].
\end{align}
Combining these all together, we obtain (\ref{eq:boundhathat}).
\hfill$\blacksquare$

\end{prf}

\section{Proof of Lemma \rlmm{dropassump}}
\lapp{prflmmC}

We prove Lemma \rlmm{dropassump}, the statement of which is as follows:
let $\ca{T}^{A\rightarrow E}$ be a   CP map, and introduce a quantum system $E_c$ with dimension $J$. Define an isometry $Y:=\sum_{j}\ket{jj}^{A_cE_c}\bra{j}^{A_c}$, and a linear supermap $\check{\ca{T}}^{A \rightarrow EE_c}$ by $\ca{T}^{A \rightarrow E} \circ \mathcal{Y}^{A_c \rightarrow A_c E_c}$.
Then, $\check{\ca{T}}^{A \rightarrow EE_c}$ is a CP map and, for any $\Psi^{AR}$ that is classically coherent in $A_cR_c$, it holds that
\begin{align}
&\left\|\check{\ca{T}}^{A \rightarrow EE_c}  ( {\Psi}^{AR}  -  \Psi_{\rm av}^{AR} )\right\|_1
=
\left\|\ca{T}^{A \rightarrow E}  ( {\Psi}^{AR}  -  \Psi_{\rm av}^{AR} )\right\|_1,
\laeq{sannnwa1}\\
&\left\|\check{\ca{T}}^{A \rightarrow EE_c}  \circ \ca{G}_\sigma^A \circ \ca{U}^A ( {\Psi}^{AR}  -  \Psi_{\rm av}^{AR} )\right\|_1
=
\left\|\ca{T}^{A \rightarrow E}  \circ \ca{G}_\sigma^A \circ \ca{U}^A ( {\Psi}^{AR}  -  \Psi_{\rm av}^{AR} )\right\|_1.
\laeq{sannnwa2}
\end{align}

\begin{prf}
Define $Z^{R_c\rightarrow R_cE_c}$ by
$Z:=\sum_{j}\ket{jj}^{R_cE_c}\bra{j}^{R_c}$.
Since $\Psi^{AR}$ is classically coherent in $A_cR_c$ and the averaged state is given by
$\Psi_{\rm av}^{AR}=\sum_{j=1}^J\outpro{j}{j}^{A_c}\otimes\pi^{A_r}\otm\Psi_{jj}^{R_r}\otimes\outpro{j}{j}^{R_c}$,
we have
\alg{
\check{\ca{T}}^{A \rightarrow EE_c} ( {\Psi}^{AR}  -  \Psi_{\rm ex}^{AR} )
=
\ca{T}^{A \rightarrow E}  \otm \ca{Z}^{R_c\rightarrow R_cE_c}( {\Psi}^{AR}  -  \Psi_{\rm ex}^{AR} )
}
and 
\alg{
\check{\ca{T}}^{A \rightarrow EE_c}  \circ \ca{G}_\sigma^A \circ \ca{U}^A ( {\Psi}^{AR}  -  \Psi_{\rm ex}^{AR} )
=
\ca{T}^{A \rightarrow E}  \circ \ca{G}_\sigma^A \circ \ca{U}^A \otm \ca{Z}^{R_c\rightarrow R_cE_c}( {\Psi}^{AR}  -  \Psi_{\rm ex}^{AR} ).
}
Therefore, due to the invariance of the the trace distance under linear isometry, we obtain \req{sannnwa1} and \req{sannnwa2}.\QED
\end{prf}

\section{Proof of Lemmas \ref{lmm:propPD2}--\rlmm{projHSN} and \ref{lmm:condmaxCQCQ}--\ref{lmm:inprothree}}
\lapp{prflmma}

\noindent{\bf Proof of \rLmm{propPD2}:}

Property 1 immediately follows from the definition of the purified distance.

To show Property 2, note that for any $\rho,\varsigma\in\ca{S}_\leq(\ca{H})$, we have (see Lemma 6 in  \cite{tomamichel2010duality})
\alg{
\bar{D}(\rho,\varsigma)
\leq
P(\rho,\varsigma)
\leq
\sqrt{2\bar{D}(\rho,\varsigma)},
}
where $\bar{D}$ is the {\it generalized the trace distance} defined by
\alg{
\bar{D}(\rho,\varsigma):=\frac{1}{2}\|\rho-\varsigma\|_1+\frac{1}{2}|{\rm Tr}[\rho]-{\rm Tr}[\varsigma]|.
}
Noting that the second term in the above expression is no greater than the first term, we conclude the proof.

For Property 3, define $\lambda_\phi:=\inpro{\phi}{\phi}$ and consider a normalized pure state $\ket{\phi_{\rm n}}:=\lambda_\phi^{-1/2}\ket{\phi}$. Due to the triangle inequality and the first statement of this lemma, we have
\alg{
P(\psi,\phi)
&
\leq
P(\psi,\phi_{\rm n}) + P(\phi_{\rm n},\phi)
\nn\\
&
=\sqrt{1-|\inpro{\psi}{\phi_{\rm n}}|^2}+\sqrt{1-|\inpro{\phi_{\rm n}}{\phi}|^2}
\nn\\
&
=\sqrt{1-\lambda_\phi^{-1}|\inpro{\psi}{\phi}|^2}+\sqrt{1-\lambda_\phi^{-1}|\inpro{\phi}{\phi}|^2}\nn\\
&
\leq
\sqrt{1-|\inpro{\psi}{\phi}|^2}+\sqrt{1-\lambda_\phi},
}
which completes the proof.
\QED\\

\noindent{\bf Proof of Lemma \ref{lmm:kazuhito}:}
Since $\rho^{ABK}$ and $\rho_k^{AB}$ are normalized, the purified distances are given by
\alg{
P(\rho^{ABK},\hat{\rho}^{ABK})=\sqrt{1-\|\sqrt{\rho^{ABK}}\sqrt{\hat{\rho}^{ABK}}\|_1^2},
\quad
\delta_k:=P(\rho_k,\hat{\rho}_k)=\sqrt{1-\|\sqrt{\rho_k}\sqrt{\hat{\rho}_k}\|_1^2}.
\laeq{penganai}
}
The latter equality leads to
\alg{
\sum_kp_k\|\sqrt{\rho_k}\sqrt{\hat{\rho}_k}\|_1=\sum_kp_k\sqrt{1-\delta_k^2}\geq\sum_kp_k(1-\delta_k)=1-\sum_kp_k\delta_k.
}
In addition, a simple calculation yields $\|\sqrt{\rho^{ABK}}\sqrt{\hat{\rho}^{ABK}}\|_1=\sum_kp_k\|\sqrt{\rho_k}\sqrt{\hat{\rho}_k}\|_1$.
Combining these relations with the first one in \req{penganai}, and by using $\sqrt{1-(1-x)^2}\leq\sqrt{2x}$, we obtain the desired result. \QED \\

\noindent{\bf Proof of Lemma \ref{lmm:kazufusa}:}
Define $\varsigma'^{AK}:=\sum_kp_k\varsigma_k^{A}\otm\proj{k}^K$. By the triangle inequality, we have
\alg{
\left\|\rho^{AK}-\varsigma^{AK}\right\|_1
\leq
\left\|\rho^{AK}-\varsigma'^{AK}\right\|_1
+
\left\|\varsigma'^{AK}-\varsigma^{AK}\right\|_1
=
\sum_kp_k\left\|\rho_k-\varsigma_k\right\|_1
+
\sum_k|p_k-q_k|.
}
We also have
\alg{
\sum_kp_k\left\|\rho_k-\varsigma_k\right\|_1
&
=
\sum_k\left\|p_k\rho_k-p_k\varsigma_k\right\|_1
\nn\\
&
\leq
\sum_k\left\|p_k\rho_k-q_k\varsigma_k\right\|_1
+
\sum_k\left\|q_k\varsigma_k-p_k\varsigma_k\right\|_1
\nn\\
&
=
\left\|\rho^{AK}-\varsigma^{AK}\right\|_1
+
\sum_k|p_k-q_k|,
}
which implies the first inequality in \req{somodey}. The second inequality simply follows from the monotonicity of the trace distance under discarding of system $A$.\QED\\

\noindent{\bf Proof of Lemma \ref{lmm:triDSP}:}
Consider arbitrary finite dimensional quantum system $C$ and any subnormalized state $\xi$ on $AC$ such that the reduced state on $A$ takes the form of $\xi^A=\bigoplus_{j=1}^Jq_j \varpi_j^{A_l}\otimes \pi_j^{A_r}$. Due to the triangle inequality for the trace norm, it holds that
\alg{
\|\ca{E}^{A\rightarrow B}(\xi^{AC})+\ca{F}^{A\rightarrow B}(\xi^{AC})\|_1
\leq
\|\ca{E}^{A\rightarrow B}(\xi^{AC})\|_1+\|\ca{F}^{A\rightarrow B}(\xi^{AC})\|_1
\leq
\|\ca{E}^{A\rightarrow B}\|_{\rm DSP}+\|\ca{F}^{A\rightarrow B}\|_{\rm DSP}.\nn
}
By taking the supremum over all $C$ and $\xi$ in the first line, we obtain Lemma \ref{lmm:triDSP}.
\QED\\

\noindent{\bf Proof of \rLmm{projHSN}:}
Due to the completeness of the set of projectors, it holds that
$
\varrho=\sum_{j,k}\Pi_j\varrho \Pi_k.
$.
This yields
$
{\rm Tr}[\varrho^\dagger\varrho]
=
\sum_{j,j',k}{\rm Tr}[\Pi_j\varrho \Pi_k \Pi_k \varrho \Pi_{j'}]
=
\sum_{j,k}{\rm Tr}[\Pi_j\varrho \Pi_k \Pi_k \varrho \Pi_{j}]
\nn
$
and completes the proof.\QED\\

\noindent{\bf Proof of \rLmm{condmaxCQCQ}:}
Let $\ket{\varphi_k}^{ABC}$ be a purification of $\rho_k^{AB}$ for each $k$. A purification of $\rho^{ABK_1K_2}$ is given by $\ket{\varphi}^{ABCK_1K_2K_3}:=\sum_k\sqrt{p_k}\ket{\varphi_k}^{ABC}\ket{k}^{K_1}\ket{k}^{K_2}\ket{k}^{K_3}$. Due to the duality of the conditional entropies (\rLmm{duality}), \rLmm{condminCQCQ} and isometric invariance (\rLmm{invCEmaxiso}), we have
\alg{
H_{\rm max}^\epsilon(AK_1|BK_2)_\rho
&
=
-H_{\rm min}^\epsilon(AK_1|CK_3)_\varphi
=
-H_{\rm min}^\epsilon(A|CK_3)_\varphi
\nn\\
&
=
H_{\rm max}^\epsilon(A|BK_1K_2)_\rho
=
H_{\rm max}^\epsilon(A|BK_2)_\rho,
}
which completes the proof. \QED\\

\noindent{\bf Proof of \rLmm{clchtwo}:}
Consider $\rho'\in\ca{B}^\epsilon(\rho)$ such that $H_{\rm max}^\epsilon(K_1A|K_2B)_\rho=H_{\rm max}(K_1A|K_2B)_{\rho'}$. Introduce a projector $\Pi^{K_1K_2}:=\sum_k\proj{k}^{K_1}\otm\proj{k}^{K_2}$, and define $\hat{\rho}^{K_1K_2AB}:=\Pi^{K_1K_2}\rho'^{K_1K_2AB}\Pi^{K_1K_2}$. Using the monotonicity of purified distance under trace non-increasing CP map (Property 2 in \rLmm{propPD}), and noting that $\rho^{K_1K_2AB}=\Pi^{K_1K_2}\rho^{K_1K_2AB}\Pi^{K_1K_2}$ by assumption, we have $P(\hat{\rho}^{K_1K_2AB},\rho^{K_1K_2AB})\leq P(\rho'^{K_1K_2AB},\rho^{K_1K_2AB})$, which yields $\hat{\rho}\in\ca{B}^\epsilon(\rho)$. Due to the operator monotonicity of the square root function (see e.g.~\cite{hiai2010matrix}) and $\rho'^{K_1K_2AB}\geq\hat{\rho}^{K_1K_2AB}$, we have, for any $\varsigma\in\ca{S}(\ca{H}^{K_2B})$,
\alg{
\left\|\sqrt{\rho'^{K_1AK_2B}}\sqrt{\varsigma^{K_2B}}\right\|_1
&
=
{\rm Tr}\left[\sqrt{\sqrt{\varsigma^{K_2B}}\rho'^{K_1AK_2B}\sqrt{\varsigma^{K_2B}}}\right]
\nn\\
&
\geq
{\rm Tr}\left[\sqrt{\sqrt{\varsigma^{K_2B}}\hat{\rho}^{K_1AK_2B}\sqrt{\varsigma^{K_2B}}}\right]
\nn\\
&
=\left\|\sqrt{\hat{\rho}^{K_1AK_2B}}\sqrt{\varsigma^{K_2B}}\right\|_1.
}
Recalling the definition of the conditional max entropy \req{condmaxs}, \req{condmax} and \req{condmaxsm}, this implies
\alg{
H_{\rm max}(K_1A|K_2B)_{\rho'}\geq H_{\rm max}(K_1A|K_2B)_{\hat{\rho}}\geq H_{\rm max}^\epsilon(K_1A|K_2B)_\rho,
}
and consequently, $H_{\rm max}^\epsilon(K_1A|K_2B)_\rho=H_{\rm max}(K_1A|K_2B)_{\hat{\rho}}$. 
If $\rho$ is also diagonal in $K_1K_2$, we may, without loss of generality, assume that $\rho'$ is diagonal in $K_1K_2$ (see Proposition 5.8 in \cite{TomamichelThesis}), which completes the proof.
\QED
\\

\noindent{\bf Proof of \rLmm{minentave}:}
Let $\hat{\rho}_k^{AB}\in\ca{B}^{\epsilon_k}(\rho_k^{AB})$ be such that $H_{\rm min}^{\epsilon_k}(A|B)_{\rho_k}=H_{\rm min}(A|B)_{\hat{\rho}_k}$ for each $k$, and define a subnormalized state $\hat{\rho}^{ABK}:=\sum_kp_k\hat{\rho}_k^{AB}\otm\proj{k}^K$. From \rLmm{condminCQ}, we have $H_{\rm min}(A|BK)_{\hat\rho}=-\log(\sum_kp_k\cdot2^{-H_{\rm min}(A|B)_{\hat{\rho}_k}})$. Due to the property of the purified distance (\rLmm{kazuhito}), we also have $\hat{\rho}^{ABK}\in\ca{B}^{\sqrt{2\varepsilon}}(\rho^{ABK})$, where $\varepsilon=\sum_kp_k\epsilon_k$. This completes the proof. 
\QED\\

\noindent{\bf Proof of Lemma \ref{Lemma:PTtrace}:}
Let $\{\ket{i}\}_{i=1}^{d_A}$ and $\{\ket{j}\}_{j=1}^{d_B}$ be the Schmidt bases of $\ket{\Phi}^{AA'}$ and $\ket{\Phi}^{BB'}$, respectively, and suppose that $X=\sum_{i,j}x_{ij}\outpro{j}{i}$ and $Y=\sum_{i,j}y_{ij}\outpro{j}{i}$. The statement follows by noting that ${\rm Tr}[X^TY]=\sum_{i,j}x_{ij}y_{ij}$.
\hfill$\blacksquare$\\

\noindent{\bf Proof of \rLmm{clchsq}:}
Suppose that $\varrho^2$ is classically coherent. For any $x\neq y$, 
it holds that
\alg{
0
&=
\bra{x}^X\bra{y}^Y\varrho^2\ket{x}^X\ket{y}^Y
\nn\\
&
=
\sum_{x',y'}\bra{x}^X\bra{y}^Y\varrho\ket{x'}^X\ket{y'}^Y\cdot\bra{x'}^X\bra{y'}^Y\varrho\ket{x}^X\ket{y}^Y
\nn\\
&
\geq
(\bra{x}^X\bra{y}^Y\varrho\ket{x}^X\ket{y}^Y)^2,
}
which implies
$\bra{x}^X\bra{y}^Y\varrho\ket{x}^X\ket{y}^Y=0$
and completes the proof. \QED \\

\noindent{\bf Proof of \rLmm{OPvar}:}
The first inequality is proved as
\alg{
\left\|\rho^{AR}-\pi^A\otm\rho^{R}\right\|_2^2
&
=
{\rm Tr}[(\rho^{AR}-\pi^A\otm\rho^{R})^2]
\nn\\
&
=
{\rm Tr}[(\rho^{AR})^2-\rho^{AR}(\pi^A\otm\rho^{R})
-(\pi^A\otm\rho^{R})\rho^{AR}+(\pi^A\otm\rho^{R})^2]
\nn\\
&
=
{\rm Tr}[(\rho^{AR})^2]
-
\frac{1}{d_A}
{\rm Tr}[(\rho^{R})^2]
\nn\\
&
\leq
{\rm Tr}[(\rho^{AR})^2]
=\left\|\rho^{AR}\right\|_2^2.
}
Similarly, we obtain the second one as
\alg{
&
\left\|\rho^{AR}-\ca{C}^A(\rho^{AR})\right\|_2^2
\nn\\
&=
{\rm Tr}[(\rho^{AR}-\ca{C}^A(\rho^{AR}))^2]
\nn\\
&
=
{\rm Tr}[(\rho^{AR})^2]+{\rm Tr}[(\ca{C}^A(\rho^{AR}))^2]
-
2{\rm Tr}[
\rho^{AR}\ca{C}^A(\rho^{AR})]
\nn\\
&
=
{\rm Tr}[(\rho^{AR})^2]+
\sum_{i,j}{\rm Tr}[(\proj{i}^A\otm\rho_{ii}^{R})(\proj{j}^A\otm\rho_{jj}^{R})]
-
2\sum_j{\rm Tr}[
\rho^{AR}(\proj{j}^A\otm\rho_{jj}^{R})]
\nn\\
&
=
{\rm Tr}[(\rho^{AR})^2]
-
\sum_j
{\rm Tr}[(\rho_{jj}^{R})^2]
\nn\\
&
\leq
{\rm Tr}[(\rho^{AR})^2]
=\left\|\rho^{AR}\right\|_2^2,
}
which concludes the proof.
\QED\\

\noindent{\bf Proof of Lemma \ref{lmm:inprothree}:}
There exist normalized state vectors $\ket{\psi'},\ket{\phi'}\in\ca{H}$ such that
\alg{
\ket{\psi}=\inpro{e}{\psi}\ket{e}+\alpha\ket{\psi'},
\quad
\ket{\phi}=\inpro{e}{\phi}\ket{e}+\beta\ket{\phi'},
\quad
\inpro{e}{\psi'}=\inpro{e}{\phi'}=0,
}
where the coefficients $\alpha$ and $\beta$ are given by
\alg{
\alpha=\sqrt{1-\inpro{e}{\psi}^2},
\quad
\beta=\sqrt{1-\inpro{e}{\phi}^2}.
}
Since $\inpro{e}{\psi}\geq c$, and $\inpro{e}{\phi}\geq c$, we have
$\alpha,\beta
\leq
\sqrt{1-c^2}$, which implies
\alg{
|\inpro{\psi}{\phi}|
=
|\inpro{\psi}{e}\inpro{e}{\phi}+\alpha\beta\inpro{\psi'}{\phi'}|
\geq
|\inpro{\psi}{e}\inpro{e}{\phi}|-|\alpha\beta\inpro{\psi'}{\phi'}|
\geq
c^2-(1-c^2).
}
This completes the proof.
\QED

\section{Proof of Ineq.~\req{laaa}}
\lsec{Evdelta}

We prove Ineq.~\req{laaa}, i.e.
\alg{
P(\theta_X^{BER},\Theta^{BER})
\leq
2\sqrt{\iota+4\sqrt{20\upsilon+2\delta}}
+\sqrt{2\sqrt{20\upsilon+2\delta}}+2\sqrt{2\delta}
+2\sqrt{20\upsilon+2\delta}
+3\upsilon,
\laeq{laaatt}
}
under the following conditions that are presented in \rSec{converse}:
\benum
\renewcommand{\theenumi}{(\roman{enumi})}
\item The $\delta$-partial decoupling condition is satisfied, that is, there exists a state
\alg{
\Omega^{ER}:=\sum_{j=1}^J\varsigma_j^E\otm\Psi_{jj}^{R_r}\otm\proj{j}^{R_c},
\laeq{dfnOME}
}
where $\{\varsigma_j\}_{j=1}^J$ are normalized states on $E$, such that
\alg{
\left\|
\ca{T}^{A \rightarrow E} ( \Psi^{AR} ) -\Omega^{ER}
\right\|_1
\leq
\delta.
\laeq{keshikeshi}
}

\item
The operator $X\in\ca{P}(\ca{H}^{ER})$ satisfies
\alg{
[(X^{ER})^{-\frac{1}{2}},\omega^{ER}]=0
\laeq{mewotoji2}
} 
and 
\alg{
(\theta^E)^{-\frac{1}{2}}(X^{ER})^{-\frac{1}{2}}\omega^{ER}(X^{ER})^{-\frac{1}{2}}(\theta^E)^{-\frac{1}{2}}
=\sum_{k:q_k>0}\proj{k}^{E_c}\otm I_k^{E_r}\otm I_k^{R_r}\otm\proj{k}^{R_c},\laeq{moisttt2}
}
where $\omega$ is a subnormalized state defined by
\alg{
\omega^{ER}&:=\sum_{k:q_k>0}q_k\proj{k}^{E_c}\otm\theta_k^{E_r}\otm\theta_k^{R_r}\otm\proj{k}^{R_c}.
\laeq{dfnbas3}
}
and
\alg{
q_k :=\|\bra{k}^{R_c}\ket{\theta}\|_1^2,
\quad
\ket{\theta_k}^{E_rR_r}:=q_k^{-1/2}\bra{k}^{E_c}\bra{k}^{R_c}\ket{\theta}.
\laeq{dfnbas4}
}
\ennum
To this end, we evaluate the distances between purifications of $\Omega^{ER}\in\ca{S}_=(\ca{H}^{ER})$ and $\omega^{ER}\in\ca{S}_\leq(\ca{H}^{ER})$, in addition to a normalized pure state
$\ket{\Theta}$ and subnormalized pure states $\ket{\theta}$, $\ket{\theta_X}$ and $\ket{\omega_X}$ on $BERD$. Recall that $\ket{\Theta}$ and $\ket{\theta}$ are defined as follows:
\bitem

\item $\ket{\Theta}:=V\ket{\Psi}$, where $\ket{\Psi}^{ARD}$ is a purification of $\Psi^{AR}$ and $V^{A\rightarrow BE}$ is a Stinespring dilation of $\ca{T}^{A\rightarrow E}$.

\item $\ket{\theta}$ : A subnormalized pure state such that
\alg{
H_{\rm max}(RD|E)_\theta=H_{\rm max}^{\upsilon}(RD|E)_{\Theta},
\quad
P(\theta^{BERD},\Theta^{BERD})\leq\upsilon,
\laeq{sigmatildeup2}
}
which is classically coherent in $E_cR_c$.

\entem
With $\Gamma_X^{ER}$ being a linear operator
\alg{
\Gamma_X^{ER}&:=
\sqrt{1-\iota}\cdot (X^{ER})^{\frac{1}{2}}((1-\iota)\cdot X^{ER}+\iota\cdot Y^{ER})^{-\frac{1}{2}}, \laeq{erever}
}
the subnormalized pure states $\ket{\theta_X}$ and $\ket{\omega_X}$ are define by
\alg{
\ket{\theta_X}:=\Gamma_X^{ER}\ket{\theta}, \quad \ket{\omega_X}:=\Gamma_X^{ER}\ket{\omega}.
\laeq{shinjutsu}
} 
Due to the operator monotonicity of the inverse function (see e.g.~\cite{hiai2010matrix}),
we have
\alg{
\Gamma_X\Gamma_X^\dagger
&=
(1-\iota)\cdot (X^{ER})^{\frac{1}{2}}((1-\iota)\cdot X^{ER}+\iota\cdot Y^{ER})^{-1}(X^{ER})^{\frac{1}{2}}
\nn\\
&
\leq
(1-\iota)\cdot (X^{ER})^{\frac{1}{2}}((1-\iota)\cdot X^{ER})^{-1}(X^{ER})^{\frac{1}{2}}
=
I^{ER}.
}
Consequently, $\Gamma_X^{ER}$ is contractive, and thus  $\ket{\theta_X}$ and $\ket{\omega_X}$ are indeed subnormalized states.
Relations among these states are depicted in Figure \ref{fig:A}.

\begin{figure}[t]
\begin{center}
\includegraphics[bb={0 0 673 389}, scale=0.45]{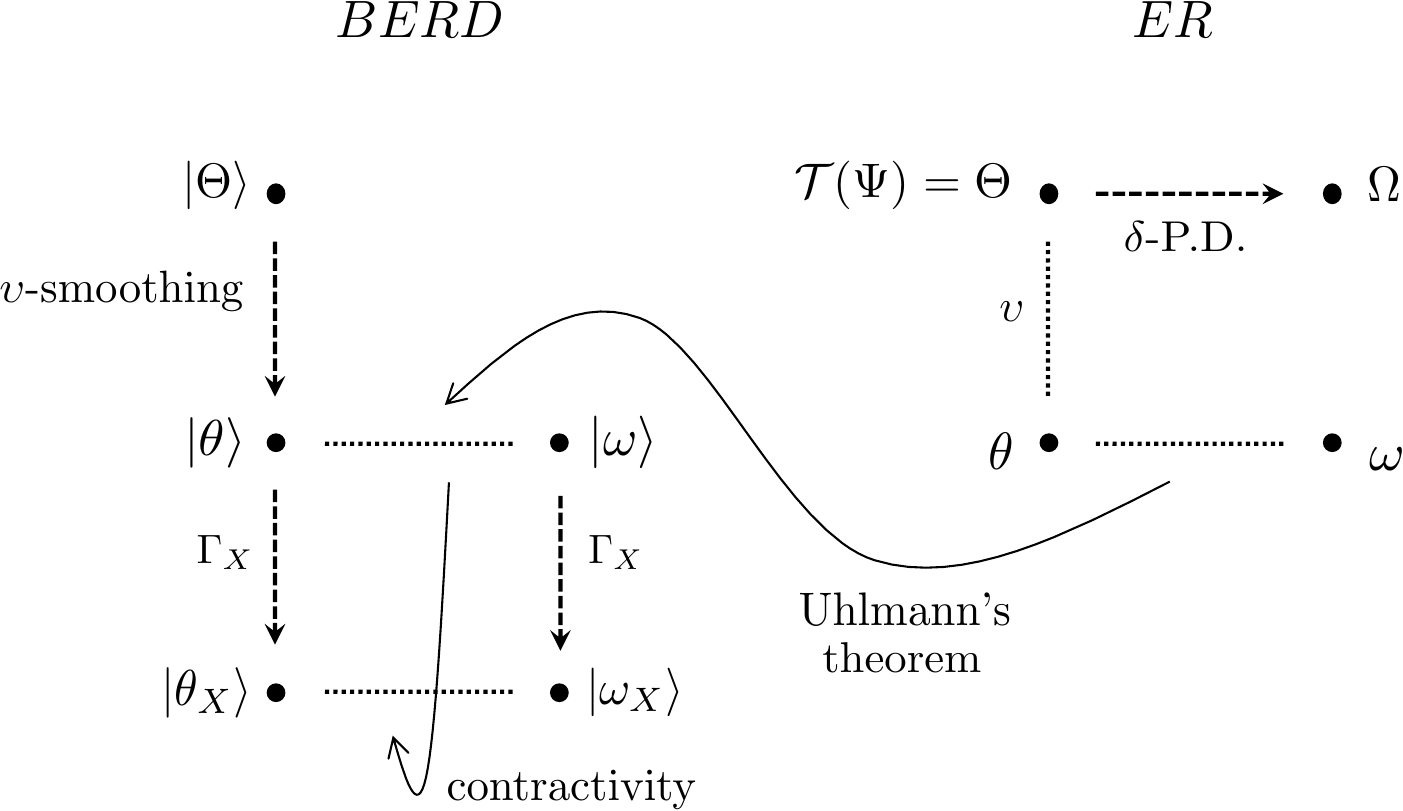}
\end{center}
\caption{
Relations among subnormalized states $\Theta$, $\theta$, $\theta_X$, $\Omega$, $\omega$ and $\omega_X$ are depicted.  Dashed arrows represent how the states are defined, and the dotted lines represent the distances between the states. The goal of the proof is to evaluate the distance between $\ket{\Theta}$ and $\ket{\theta_X}$, by expressing it in terms of distances between other states on the whole system $BERD$ as depicted in the left. To evaluate the distance between $\ket{\theta}$ and $\ket{\omega}$, we also consider those states on subsystem $ER$, as depicted in the right. 
}
\label{fig:A}
\end{figure}

\subsection{Application of triangle inequality}

Consider a subnormalized state $\omega^{ER}$ defined by \req{dfnbas3}.
Due to Uhlmann's theorem (\rLmm{propPD}), there exists a purification $|\omega\rangle^{BERD}$ of $\omega^{ER}$ such that
\alg{
P(\theta^{BERD},\omega^{BERD})=P(
\theta^{ER},\omega^{ER}
).
\laeq{kashkash}
}
By the triangle inequality for the the purified distance, it holds that
\alg{
&
P(\theta^{BER}_X,\Theta^{BER})
\nn\\
&
\leq
P(\theta^{BER}_X,\omega_X^{BER})
+
P(\omega_X^{BER},\omega^{BER})
+
P(\omega^{BER},\theta^{BER})
+
P(\theta^{BER},\Theta^{BER})
\nn\\
&
\leq
P(\theta^{BERD}_X,\omega_X^{BERD})
+
P(\omega_X^{BERD},\omega^{BERD})
+
P(\omega^{BERD},\theta^{BERD})
+
P(\theta^{BERD},\Theta^{BERD})
\nn\\
&\leq
2P(\omega^{BERD},\theta^{BERD})
+
P(\omega_X^{BERD},\omega^{BERD})+\upsilon
\nn\\
&=
2P(\omega^{ER},\theta^{ER})
+
P(\omega_X^{BERD},\omega^{BERD})+\upsilon
\nn\\
&\leq
2P(\omega^{ER},\Omega^{ER})
+
2P(\Omega^{ER},\Theta^{ER})
+
2P(\Theta^{ER},\theta^{ER})
+
P(\omega_X^{BERD},\omega^{BERD})+\upsilon
\nn\\
&\leq
2P(\omega^{ER},\Omega^{ER})
+
2P(\Omega^{ER},\Theta^{ER})
+
P(\omega_X^{BERD},\omega^{BERD})+3\upsilon.
\laeq{zinccopper}
}
Here, the third line follows from the monotonicity of the purified distance under partial trace (see \rLmm{propPD}); the fourth line from the monotonicity of the purified distance under the trace-nonincreasing CP map $\Gamma_X^{ER}$ and from the condition for $\theta$ given by \req{sigmatildeup2}; the fifth line due to Eq.~\req{kashkash}; and the last line again from \req{sigmatildeup2}. 
Noting that we have $\Theta^{ER}=\ca{T}^{A\rightarrow E}(\Psi^{AR})$ from the definition of $\Theta$, and by using the partial decoupling condition \req{keshikeshi} as well as the relation between the purified distance and the the trace distance (\rLmm{propPD2}), we have
\alg{
P(\Omega^{ER},\Theta^{ER})
\leq
\sqrt{2\left\|\Omega^{ER}-\Theta^{ER}\right\|_1}
\leq\sqrt{2\delta}
}
for the second term in \req{zinccopper}.
In the following, we prove that the first and the third term in \req{zinccopper} are bounded as
\alg{
&
P(\omega^{ER},\Omega^{ER})
\leq
\sqrt{20\upsilon+2\delta},
\laeq{gogogo}
\\
&
P(\omega_X^{BERD},\omega^{BERD})
\leq
2\sqrt{\iota+4P(\omega^{ER},\Omega^{ER})}
+
\sqrt{2P(\omega^{ER},\Omega^{ER})},
\laeq{tingogo}
}
respectively.
Combining these all together, we arrive at \req{laaatt}.

\subsection{Evaluation of $P(\Omega^{ER},\omega^{ER})$}

We first evaluate $P(\Omega^{ER},\omega^{ER})$ by using the partial decoupling condition \req{keshikeshi}.
From the normalized state $\ket{\Theta}$,
define
\alg{
p_k:=\|\bra{k}^{R_c}\ket{\Theta}\|_1^2,
\quad
\ket{\Theta_k}:=p_k^{-1/2}\bra{k}^{E_c}\bra{k}^{R_c}\ket{\Theta}.
}
From the condition that $\Psi^{AR}$ is classically coherent in $A_cR_c$ and $\ca{T}^{A\rightarrow E}$ is trace-preserving, it follows that
\alg{
p_k\Theta_k^{R_r}
&
=
{\rm Tr}_{BED}[\bra{k}^{R_c}\proj{\Theta}^{BER_cR_rD}\ket{k}^{R_c}]
\nn\\
&
=
{\rm Tr}_{AD}[\bra{k}^{R_c}\proj{\Psi}^{AR_cR_rD}\ket{k}^{R_c}]
\nn\\
&
=\bra{k}^{R_c}\Psi^{R_cR_r}\ket{k}^{R_c}=\Psi_{kk}^{R_c}.
}
Consequently, the state $\Omega^{ER}$ defined by \req{dfnOME}
is represented as
\alg{
\Omega^{ER}=\sum_{k=1}^Jp_k\varsigma_k^{E_cE_r}\otm\Theta_k^{R_r}\otm\proj{k}^{R_c}.
\laeq{dfnbas2}
}
Thus, from the definition of $\omega$ given by \req{dfnbas3} and \req{dfnbas4},  and by using the property of the trace distance (\rLmm{kazufusa}),we have
\alg{
&
\left\|\Omega^{ER}-\omega^{ER}\right\|_1
\nn\\
&\leq
\sum_{k=1}^J|p_k-q_k|
+
\sum_{k=1}^J p_k\left\|\varsigma_k^{E_cE_r}\otm\Theta_k^{R_r}-\proj{k}^{E_c}\otm\theta_k^{E_r}\otm\theta_k^{R_r}\right\|_1
\nn\\
&\leq
\sum_{k=1}^J|p_k-q_k|
+
\sum_{k=1}^J p_k\left\|\varsigma_k^{E_cE_r}\otm\Theta_k^{R_r}-\proj{k}^{E_c}\otm\Theta_k^{E_r}\otm\Theta_k^{R_r}\right\|_1
\nn\\
&\quad
+
\sum_{k=1}^J p_k\left\|\Theta_k^{E_r}\otm\Theta_k^{R_r}-\Theta_k^{E_r}\otm\theta_k^{R_r}\right\|_1
+
\sum_{k=1}^J p_k\left\|\Theta_k^{E_r}\otm\theta_k^{R_r}-\theta_k^{E_r}\otm\theta_k^{R_r}\right\|_1
\nn\\
&=
\sum_{k=1}^J|p_k-q_k|
+
\sum_{k=1}^J p_k\left\|\varsigma_k^{E_cE_r}-\proj{k}^{E_c}\otm\Theta_k^{E_r}\right\|_1
\nn\\
&\quad
+
\sum_{k=1}^J p_k\left\|\Theta_k^{R_r}-\theta_k^{R_r}\right\|_1
+
\sum_{k=1}^J p_k\left\|\Theta_k^{E_r}-\theta_k^{E_r}\right\|_1.
\laeq{tsuyogari}
}
Noting that $\Theta^{E_c}$ and $\theta^{E_c}$ are both diagonal in $\{\ket{k}\}_k$, the first term is equal to $\|\Theta^{E_c}-\theta^{E_c}\|_1$. 
By using \rLmm{kazufusa} again, the third and the fourth terms are bounded as $\sum_{k=1}^J p_k\|\Theta_k^{R_r}-\theta_k^{R_r}\|_1\leq2\|\Theta^{R_cR_r}-\theta^{R_cR_r}\|_1$ and $\sum_{k=1}^J p_k\|\Theta_k^{E_r}-\theta_k^{E_r}\|_1\leq2\|\Theta^{E_cE_r}-\theta^{E_cE_r}\|_1$, respectively.
In addition, denoting by $\ca{C}^{R_c}$ the completely dephasing operation on $R_c$ with respect to the basis $\{\ket{k}\}_k$, the second term is bounded as
\alg{
&
\sum_{k=1}^J p_k\left\|\varsigma_k^{E_cE_r}-\proj{k}^{E_c}\otm\Theta_k^{E_r}\right\|_1
\nn\\
&
=
\left\|\sum_{k=1}^J p_k\varsigma_k^{E_cE_r}\otm\proj{k}^{R_c}-\sum_{k=1}^J p_k\proj{k}^{E_c}\otm\Theta_k^{E_r}\otm\proj{k}^{R_c}\right\|_1
\nn\\
&=
\left\|\Omega^{E_cE_rR_c}-\ca{C}^{R_c}(\Theta^{E_cE_rR_c})\right\|_1
\nn\\
&\leq
\left\|\Omega^{E_cE_rR_c}-\Theta^{E_cE_rR_c}\right\|_1,
}
where we used $\Omega^{E_cE_rR_c}=\ca{C}^{R_c}(\Omega^{E_cE_rR_c})$ in the last line.
Substituting all these inequalities to \req{tsuyogari}, we arrive at
\alg{
&
\left\|\Omega^{ER}-\omega^{ER}\right\|_1
\nn\\
&
\leq
\left\|\Theta^{E_c}-\theta^{E_c}\right\|_1
+
2\left\|\Theta^{R_cR_r}-\theta^{R_cR_r}\right\|_1
+
2\left\|\Theta^{E_cE_r}-\theta^{E_cE_r}\right\|_1
+
\left\|\Omega^{E_cE_rR_c}-\Theta^{E_cE_rR_c}\right\|_1
\nn\\
&\leq
5\left\|\Theta^{ER}-\theta^{ER}\right\|_1
+
\left\|\Omega^{ER}-\Theta^{ER}\right\|_1
\nn\\
&
\leq
5\left\|\Theta^{ER}-\theta^{ER}\right\|_1
+
\delta,
\laeq{speeee}
}
where the last line follows from the partial decoupling condition \req{keshikeshi} and $\Theta^{ER}=\ca{T}^{A\rightarrow E}(\Psi^{AR})$.
From the relation between the trace distance and the purified distance (see \rLmm{propPD}), and from the definition of $\theta$, the first term is bounded as
\alg{
\left\|\Theta^{ER}-\theta^{ER}\right\|_1
\leq
2P(\Theta^{ER},\theta^{ER})
\leq
2\upsilon.
}
Substituting this to \req{speeee}, and again using \rLmm{propPD}, it follows that
\alg{
P(\Omega^{ER},\omega^{ER})
\leq
\sqrt{2
\left\|\Omega^{ER}-\omega^{ER}\right\|_1}
\leq
\sqrt{
20\upsilon+2\delta
},
}
which implies \req{gogogo}.

\subsection{Evaluation of $P(\omega_X^{BERD},\omega^{BERD})$}

Due to the property of the purified distance for subnormalized pure states (Property 3 in \rLmm{propPD2}), we have
\alg{
P(\omega_X^{BERD},\omega^{BERD})
\leq
\sqrt{1-|\inpro{\omega_X}{\omega}|^2}
+
\sqrt{\chi_\omega}
=
\sqrt{1-|\bra{\omega}(\Gamma_X^{ER})^\dagger\ket{\omega}|^2}
+
\sqrt{\chi_\omega},
\laeq{dejavoo}
}
where $\chi_\omega:=1-\inpro{\omega}{\omega}$ and the last line follows from the definition of $\omega_X$ given by \req{shinjutsu}. 
To bound the first term, define
\alg{
\ket{\tilde\omega}^{BERD}:=\sqrt{\frac{1-\iota}{\alpha}}\cdot(\Gamma_X^{ER})^{-1}\ket{\omega}^{BERD},
\laeq{moistttaa}
} 
where
\alg{
\alpha:=(1-\iota)\cdot\langle\omega|\omega\rangle+\iota.
\laeq{dfnalphasaga}
}
Note that $\alpha\leq1$ due to the condition $\iota\leq1$.
As we prove below, $\ket{\tilde\omega}$ is a normalized pure state. 
In addition, since $\Gamma_X^\dagger$ is a contraction, $(\Gamma_X^{ER})^\dagger|\omega\rangle$ is a subnormalized pure state. Hence, we can apply \rLmm{inprothree} for subnormalized pure states $\ket{\omega},(\Gamma_X^{ER})^\dagger|\omega\rangle$ and a normalized pure state $\ket{\tilde\omega}$ to bound the first term in \req{dejavoo}.

Due to the definition of $\tilde{\omega}$ in \req{moistttaa} and $\alpha\leq1$,
we have
\alg{
\langle\tilde\omega|(\Gamma_X^{ER})^\dagger|\omega\rangle
=
\sqrt{\frac{1-\iota}{\alpha}}\cdot
\langle\omega|\omega\rangle
\geq
\sqrt{1-\iota}\cdot
(1-\chi_\omega).
}
In addition, we have
\alg{
\langle\omega|\tilde\omega\rangle
&
=
\sqrt{\frac{1-\iota}{\alpha}}\cdot 
\langle\omega|
(\Gamma_X^{ER})^{-1}
|\omega\rangle
\nn\\
&=
\alpha^{-1/2}\cdot
{\rm Tr}[
\omega^{ER}
((1-\iota)\cdot X^{ER}+\iota\cdot Y^{ER})^{\frac{1}{2}}
(X^{ER})^{-\frac{1}{2}}]
\nn\\
&=
\alpha^{-1/2}\cdot
{\rm Tr}[
(X^{ER})^{-\frac{1}{4}}
\omega^{ER}
(X^{ER})^{-\frac{1}{4}}
((1-\iota)\cdot X^{ER}+\iota\cdot Y^{ER})^{\frac{1}{2}}
]
\nn\\
&\geq
\sqrt{1-\iota}
\cdot
{\rm Tr}[
(X^{ER})^{-\frac{1}{4}}
\omega^{ER}
(X^{ER})^{-\frac{1}{4}}
\cdot
(X^{ER})^{\frac{1}{2}}
]
\nn\\
&=
\sqrt{1-\iota}
\cdot
{\rm Tr}[
\omega^{ER}
]
=
\sqrt{1-\iota}
\cdot
(1-\chi_\omega).
\laeq{nagaseru}
}
Here, the second line follows from the definition of $\Gamma_X$ by \req{erever}, the third line from the commutativity of $(X^{ER})^{-1/2}$ and $\omega^{ER}$, given by \req{mewotoji2}, and the fourth line due to $\alpha\leq1$ and the matrix monotonicity of the square root function. Thus, \rLmm{inprothree} yields
\alg{
|\bra{\omega}(\Gamma_X^{ER})^\dagger\ket{\omega}|
\geq
2(1-\iota)
\cdot
(1-\chi_\omega)^2-1
\geq
1-2(\iota+2
\chi_\omega).
}
Combining this with \req{dejavoo}, and by using $\sqrt{1-(1-x)^2}\leq\sqrt{2x}$, we obtain
\alg{
\!\!
P(\omega_X^{BERD},\omega^{BERD})
\leq
2
\sqrt{\iota+2\chi_\omega}
+
\sqrt{\chi_\omega}.
\laeq{dejavoo2}
}
Noting that $\Omega^{ER}$ is a normalized state,
the triangle inequality for the trace norm and the relation between the trace distance and the purified distance (\rLmm{propPD2}) lead to
 \alg{
\chi_\omega
={\rm Tr}[
\Omega^{ER}
]-{\rm Tr}[
\omega^{ER}
]
=\|
\Omega^{ER}
\|_1
-\|
\omega^{ER}
\|_1
\leq
\|\omega^{ER}-\Omega^{ER}\|_1
\leq
2P(\omega^{ER},\Omega^{ER}).
 }
Substituting this to \req{dejavoo2}, we arrive at \req{tingogo}.

To prove that $\ket{\tilde\omega}$ is a normalized pure state, we observe, from the definition of $\Gamma_X$ in \req{erever} and that of $\tilde{\omega}$ in \req{moistttaa}, that
\alg{
\alpha\cdot\langle\tilde\omega|\tilde{\omega}\rangle
&
=
(1-\iota)\cdot\bra{\omega}(\Gamma_X^{ER})^{-1\dagger}(\Gamma_X^{ER})^{-1}\ket{\omega}
\nn\\
&=
\bra{\omega}(X^{ER})^{-\frac{1}{2}}((1-\iota)\cdot X^{ER}+\iota\cdot Y^{ER})(X^{ER})^{-\frac{1}{2}}\ket{\omega}
\nn\\
&=
(1-\iota)\cdot\langle\omega|\omega\rangle
+
\iota\cdot 
\bra{\omega}(X^{ER})^{-\frac{1}{2}}Y^{ER}(X^{ER})^{-\frac{1}{2}}\ket{\omega}
\nn\\
&=
(1-\iota)\cdot\langle\omega|\omega\rangle
+
\iota\cdot 
{\rm Tr}[
(X^{ER})^{-\frac{1}{2}}\omega^{ER}(X^{ER})^{-\frac{1}{2}}Y^{ER}
].
\laeq{moistttt}
}
Noting that 
$Y^{ER}$
is classically coherent in $E_cR_c$ due to \rLmm{clchsq}, we obtain from the property \req{moisttt2} of $X^{ER}$ that
\alg{
(X^{ER})^{-\frac{1}{2}}\omega^{ER}(X^{ER})^{-\frac{1}{2}}Y^{ER}
&=
(\theta^E)^{\frac{1}{2}}\cdot
(\theta^E)^{-\frac{1}{2}}(X^{ER})^{-\frac{1}{2}}\omega^{ER}(X^{ER})^{-\frac{1}{2}}(\theta^E)^{-\frac{1}{2}}
\cdot(\theta^E)^{\frac{1}{2}}Y^{ER}
\nn\\
&
=
\left(\sum_{k:q_k>0}\proj{k}^{E_c}\otm \theta_k^{E_r}\otm I_k^{R_r}\otm\proj{k}^{R_c}\right)Y^{ER}
\nn\\
&=
(\theta^E\otm I^R)Y^{ER}.
\laeq{yokogaga}
}
Substituting this to \req{moistttt}, we obtain 
\alg{
\alpha\cdot\langle\tilde{\omega}|\tilde{\omega}\rangle=
(1-\iota)\cdot\langle\omega|\omega\rangle
+
\iota\cdot 
{\rm Tr}[
\theta^EY^{ER}
].
}
Note that we have
$
{\rm Tr}[\theta^EY^{ER}]={\rm Tr}[\theta^EY^{ERD}]=1
$
from the definition of the conditional max-entropy and the definition of $Y^{ERD}$. Thus, using the definition of $\alpha$ in \req{dfnalphasaga}, we arrive at
$
\langle\tilde{\omega}|\tilde{\omega}\rangle=1
$.

\QED

\section{Proof of Ineq.~\req{tsuuwamo}}
\lapp{tsuuwamoprf}

We prove Ineq.~\req{tsuuwamo}, that is,
\alg{
\bar{\lambda}:=\sum_kr_k\lambda_k\leq
\lambda\left(\iota,\sqrt{2}\sqrt[4]{24\upsilon+2\delta}\right)+\lambda(\iota,4)\cdot\sqrt{2}\sqrt[4]{24\upsilon+2\delta},
\laeq{tsuuwamo2}
}
under the partial decoupling condition \req{keshikeshiit}. 
Recall that $\lambda(\iota,x)$ is defined by $\lambda(\iota,x):=2\sqrt{\iota+2x}+\sqrt{x}+2x$, and that $r_k$ and $\lambda_k$ are given by
\alg{
&
r_k:=\|\bra{k}^{R_c}\ket{\hat{\Psi}}\|_1^2,
\quad
\lambda_k:=2\sqrt{\iota+4\sqrt{2\delta_k}}
+\sqrt{2\sqrt{2\delta_k}}+4\sqrt{2\delta_k},
\laeq{tsubokatte}
}
where
\alg{
\delta_k:=
\left\|
\hat{\ca{T}}_{\ca{C},k}^{A_r\rightarrow E}(\hat{\Psi}_{k}^{A_rR_r})
-
\varsigma_k^E\otm\hat{\Psi}_{k}^{R_r}
\right\|_1
}
and
\alg{
|\hat{\Psi}_{k}\rangle^{A_rR_rD}:=r_k^{-1/2}\bra{k}^{E_c}\bra{k}^{R_c}\ket{\hat{\Psi}},
\quad
\hat{\ca{T}}_{\ca{C},k}^{A_r\rightarrow E}(\tau)
=
\proj{k}^{E_c}\hat{\ca{T}}_{\ca{C}}^{A\rightarrow E}(\proj{k}^{A_c}\otm\tau^{A_r})\proj{k}^{E_c}.
}
We introduce similar notations for $\ket{\Psi}$ and $\ca{T}_{\ca{C}}^{A\rightarrow E}:=\ca{T}^{A\rightarrow E}\circ\ca{C}$ as follows:
\alg{
&
p_k:=\|\bra{k}^{R_c}\ket{\Psi}\|_1^2,
\quad
|\Psi_{k}\rangle^{A_rR_rD}:=p_k^{-1/2}\bra{k}^{E_c}\bra{k}^{R_c}\ket{\Psi},
\nn\\
&
\ca{T}_{\ca{C},k}^{A_r\rightarrow E}(\tau)
=
\proj{k}^{E_c}\ca{T}_{\ca{C}}^{A\rightarrow E}(\proj{k}^{A_c}\otm\tau^{A_r})\proj{k}^{E_c}.
}
Note that $\Psi_{kk}^{R_r}=p_k\Psi_{k}^{R_r}$.
It is straightforward to verify that the states $\hat{\ca{T}}_\ca{C}^{A \rightarrow E}( \hat{\Psi}^{AR} )$ and $\ca{T}_\ca{C}^{A \rightarrow E} ( \Psi^{AR} )$ are represented by
\alg{
&
\hat{\ca{T}}_\ca{C}^{A \rightarrow E} ( \hat{\Psi}^{AR} )
=
\sum_kr_k\proj{k}^{E_c}\otm\proj{k}^{R_c}\otm\hat{\ca{T}}_{\ca{C},k}^{A_r\rightarrow E}(\hat{\Psi}_{k}^{A_rR_r}),
\laeq{ondosa1}\\
&
\ca{T}_\ca{C}^{A \rightarrow E} ( \Psi^{AR} )
=
\sum_kp_k\proj{k}^{E_c}\otm\proj{k}^{R_c}\otm\ca{T}_{\ca{C},k}^{A_r\rightarrow E}(\Psi_{k}^{A_rR_r}).
\laeq{ondosa2}
}
Since $\Psi^{AR}$ is assumed to be classically coherent in $A_cR_c$ (Converse Condition 2), the partial decoupling condition \req{keshikeshiit} implies that there exists $\{ \varsigma_j^E \}$ ($\varsigma_j^E \in \mathcal{S}_={\mathcal{H}^E} $) satisfying
\alg{
\|
\ca{T}^{A \rightarrow E}\circ\ca{C}^A ( \Psi^{AR} ) -\Omega^{ER}
\|_1
\leq
\delta,
\laeq{tiluanin}
}
where $\Omega^{ER}:=\sum_{j=1}^J\varsigma_j^E\otm\Psi_{jj}^{R_r}\otm\proj{j}^{R_c}=\sum_{j=1}^Jp_j\varsigma_j^E\otm\Psi_{j}^{R_r}\otm\proj{j}^{R_c}$.

From \req{tsubokatte} and the definition of $\lambda(\iota,x)$, we have
 $\lambda_k=\lambda(\iota,2\sqrt{2\delta_k})$.
Noting that $\delta_k\leq2$ by the definition of the trace distance, and that $\sum_kr_k\cdot2\sqrt{2\delta_k}\leq2\sqrt{2\bar{\delta}}$ by Jensen's inequality, where $\bar{\delta}:=\sum_kr_k\delta_k$, we can apply \rLmm{yoshiki} for $f(x)=\lambda(\iota,x)$, $c=4$ and $\epsilon_k=2\sqrt{2\delta_k}$ to obtain
\alg{
\bar{\lambda}=\sum_kr_k\lambda(\iota,2\sqrt{2\delta_k})
\leq
\lambda\left(\iota,\sqrt{2\sqrt{2\bar{\delta}}}\right)+\lambda(\iota,4)\cdot\sqrt{2\sqrt{2\bar{\delta}}}.
\laeq{arriveat}
}
The $\bar\delta$ can further be calculated as follows.
By the triangle inequality, we have
\alg{
\bar{\delta}
&=\sum_kr_k\left\|
\hat{\ca{T}}_{\ca{C},k}^{A_r\rightarrow E}(\hat{\Psi}_{k}^{A_rR_r})
-
\varsigma_k^E\otm\hat{\Psi}_{k}^{R_r}
\right\|_1
\nn\\
&\leq\sum_kr_k\left\|
\hat{\ca{T}}_{\ca{C},k}^{A_r\rightarrow E}(\hat{\Psi}_{k}^{A_rR_r})
-
\ca{T}_{\ca{C},k}^{A_r\rightarrow E}(\Psi_{k}^{A_rR_r})
\right\|_1
+
\sum_kr_k\left\|
\ca{T}_{\ca{C},k}^{A_r\rightarrow E}(\Psi_{k}^{A_rR_r})
-
\varsigma_k^E\otm\Psi_{k}^{R_r}
\right\|_1
\nn\\
&
\quad
+
\sum_kr_k\left\|
\Psi_{k}^{R_r}
-
\hat{\Psi}_{k}^{R_r}
\right\|_1
\nn\\
&\leq
2\sum_kr_k\left\|
\hat{\ca{T}}_{\ca{C},k}^{A_r\rightarrow E}(\hat{\Psi}_{k}^{A_rR_r})
-
\ca{T}_{\ca{C},k}^{A_r\rightarrow E}(\Psi_{k}^{A_rR_r})
\right\|_1
+
\sum_kr_k\left\|
\ca{T}_{\ca{C},k}^{A_r\rightarrow E}(\Psi_{k}^{A_rR_r})
-
\varsigma_k^E\otm\Psi_{k}^{R_r}
\right\|_1,
\laeq{kagenonai}
}
where the last line follows from the monotonicity of the trace distance under partial trace.

Using the property of the trace distance (\rLmm{kazufusa} and \rlmm{propPD2}), and Eqs.~\req{ondosa1} and \req{ondosa2}, the first term in \req{kagenonai}  is bounded as
\alg{
&
\sum_kr_k\left\|
\hat{\ca{T}}_{\ca{C},k}^{A_r\rightarrow E}(\hat{\Psi}_{k}^{A_rR_r})
-
\ca{T}_{\ca{C},k}^{A_r\rightarrow E}(\Psi_{k}^{A_rR_r})
\right\|_1
\nn\\
&
\leq
2\left\|
\hat{\ca{T}}_{\ca{C}}^{A\rightarrow E}(\hat{\Psi}^{AR})
-
{\ca{T}}_{\ca{C}}^{A\rightarrow E}(\Psi^{AR})
\right\|_1
\nn\\
&\leq
4P(
\hat{\ca{T}}_{\ca{C}}^{A\rightarrow E}(\hat{\Psi}^{AR})
,
{\ca{T}}_{\ca{C}}^{A\rightarrow E}(\Psi^{AR})
).
}
Noting that $\hat{\ca{T}}_{\ca{C}}^{A\rightarrow E}(\hat{\Psi}^{AR})=\hat{\theta}_{\ca{C}}^{ER}$ and $\ca{T}_{\ca{C}}^{A\rightarrow E}(\Psi^{AR})=\Theta_{\ca{C}}^{ER}$ from the definitions of $\hat{\ca{T}}_{\ca{C}}$ and $\Theta_{\ca{C}}$, and recalling Ineq.~\req{tsukawanai},
we have
$
P(
\hat{\ca{T}}_{\ca{C}}^{A\rightarrow E}(\hat{\Psi}^{AR})
,
{\ca{T}}_{\ca{C}}^{A\rightarrow E}(\Psi^{AR})
)
\leq
P(
\hat{\theta}_{\ca{C}}^{ERD}
,
\Theta_{\ca{C}}^{ERD}
)
\leq
\upsilon
$.
Whereas, noting that the total variation distance is no greater than 2, the second term is calculated to be
\alg{
&
\sum_kr_k\left\|
\ca{T}_{\ca{C},k}^{A_r\rightarrow E}(\Psi_{k}^{A_rR_r})
-
\varsigma_k^E\otm\Psi_{k}^{R_r}
\right\|_1
\nn\\
&
\leq
2\sum_k|p_k-r_k|
+
\sum_kp_k\left\|
\ca{T}_{\ca{C},k}^{A_r\rightarrow E}(\Psi_{k}^{A_rR_r})
-
\varsigma_k^E\otm\Psi_{k}^{R_r}
\right\|_1
\nn\\
&
\leq
2\left\|
\hat{\ca{T}}_{\ca{C}}^{A\rightarrow E}(\hat{\Psi}^{AR})
-
{\ca{T}}_{\ca{C}}^{A\rightarrow E}(\Psi^{AR})
\right\|_1
+
\sum_kp_k\left\|
\ca{T}_{\ca{C},k}^{A_r\rightarrow E}(\Psi_{k}^{A_rR_r})
-
\varsigma_k^E\otm\Psi_{k}^{R_r}
\right\|_1
\nn\\
&
\leq
4\upsilon
+
\sum_kp_k\left\|
\ca{T}_{\ca{C},k}^{A_r\rightarrow E}(\Psi_{k}^{A_rR_r})
-
\varsigma_k^E\otm\Psi_{k}^{R_r}
\right\|_1
\nn\\
&=
4\upsilon
+
\left\|
\sum_kp_k
\ca{T}_{\ca{C},k}^{A_r\rightarrow E}(\Psi_{k}^{A_rR_r})
\otm\proj{k}^{R_c}
-
\sum_kp_k
\varsigma_k^{E}\otm\Psi_{k}^{R_r}
\otm\proj{k}^{R_c}
\right\|_1
\nn\\
&=
4\upsilon
+
\left\|
\ca{T}^{A\rightarrow E}\circ\ca{C}^A(\Psi^{AR})
-
\Omega^{ER}
\right\|_1
\leq
4\upsilon
+
\delta,
}
where the fourth line follows from the similar argument to show the bound of the first term, and the last line follows from the partial decoupling condition \req{tiluanin}.

Combining these all together, we obtain
$
\bar{\delta}
\leq
12\upsilon+\delta
$.
Substituting this to \req{arriveat}, we arrive at
\alg{
\bar{\lambda}
\leq
\lambda\left(\iota,\sqrt{2}\sqrt[4]{24\upsilon+2\delta}\right)+\lambda(\iota,4)\cdot\sqrt{2}\sqrt[4]{24\upsilon+2\delta}.
}
\QED

\section{List of Notations}
\lapp{listofnotations}

The followings are the lists of notations used in the proofs of the main theorems.

\begin{table}[t]
\renewcommand{\arraystretch}{1.6}
  \begin{center}
    \begin{tabular}{l|l} 
\multicolumn{2}{l}{General notation}\\ \hline\hline 
  $\ca{L}(\ca{H})$        & The set of linear operators on $\ca{H}$ \\ \hline
    $\ca{L}(\ca{H}^A,\ca{H}^B)$        & The set of linear operators from $\ca{H}^A$ to $\ca{H}^B$ \\ \hline
  ${\rm Her}(\ca{H})$        & $\{\rho \in \ca{L}(\ca{H}) : \rho = \rho^\dagger \}$ \\ \hline
  $\ca{P}(\ca{H})$        & $\{\rho \in {\rm Her}(\ca{H}) : \rho \geq 0 \}$ \\ \hline
    $\ca{S}_\leq(\ca{H})$        & $\{\rho \in \ca{P}(\ca{H}) : \tr [\rho] \leq 1 \}$ \\ \hline
      $\ca{S}_=(\ca{H})$        & $\{\rho \in \ca{P}(\ca{H}) : \tr [\rho]=1 \}$ \\ \hline
        $\ca{C}\ca{P}(A\rightarrow B)$        & The set of CP maps from $A$ to $B$ \\ \hline
        $\ca{C}\ca{P}_\leq(A\rightarrow B)$        & The set of trace non-increasing CP maps from $A$ to $B$ \\ \hline
        $\ca{C}\ca{P}_=(A\rightarrow B)$        & The set of trace preserving CP maps from $A$ to $B$ \\ \hline
  $\Psi^{AR}$       & A subnormalized (resp. normalized) state on $AR$ in Theorem \rthm{SmoothMarkov} and \rthm{SmoothExMarkov} (resp. Theorem \rthm{converse}) \\ \hline
  $\ca{T}^{A\rightarrow E}$       & A completely-positive superoperator from $\ca{L}(\ca{H}^A)$ to $\ca{L}(\ca{H}^B)$ (trace-preserving in \rThm{converse}) \\ \hline
$\ca{T}^{A\rightarrow B}$       & A complementary superoperator of   $\ca{T}^{A\rightarrow E}$ \\ \hline
$\Phi^{AA'}$  & Maximally entangled state between $A$ and $A'$ ($\ca{H}^A\cong\ca{H}^{A'}$) \\ \hline
\multirow{2}{*}{$\tau^{AE}$,  $\tau^{AB}$}
      & The Choi-Jamio\l kowski state of $\ca{T}^{A\rightarrow E}$ and $\ca{T}^{A\rightarrow B}$:\\
      &  $\tau^{AE}=\ca{T}^{A'\rightarrow E}(\Phi^{AA'})$, $\tau^{AB}=\ca{T}^{A'\rightarrow B}(\Phi^{AA'})$\\ \hline
 $\mbb{U}(d)$ & Unitary group of degree $d$ \\ \hline
            \multicolumn{2}{l}{}\\ 
 
  \multicolumn{2}{l}{Norms and distances}\\ \hline\hline 
  $\|X\|_1$       & The trace norm of a linear operator $X$: $\|X\|_1={\rm Tr}[\sqrt{XX^\dagger}]$ \\ \hline
$\|X\|_2$       & The Hilbert-Schmidt norm of a linear operator $X$: $\|X\|_2=\sqrt{{\rm Tr}[XX^\dagger]}$ \\ \hline
$|\!| X^{VW} |\!|_{2,\varsigma^W}$ & $|\!| (\varsigma^W)^{-1/4} X^{VW} (\varsigma^W)^{-1/4} |\!|_{2}$ for $\varsigma\in\ca{S}_=(\ca{H}^W)$ \\ \hline
\multirow{2}{*}{$\bar{F}(\rho,\rho')$}    
  & Generalized fidelity between subnormalized states $\rho,\rho'\in\ca{S}_\leq(\ca{H})$:  \\
  & $\bar{F}(\rho,\rho')=\|\sqrt{\rho}\sqrt{\rho'}\|_1+\sqrt{(1-{\rm Tr}[\rho])(1-{\rm Tr}[\rho'])}$\\ \hline
    $P(\rho,\rho')$      & Purified distance between subnormalized states $\rho,\rho'\in\ca{S}_\leq(\ca{H})$:  $P(\rho,\rho')=\sqrt{1-\bar{F}(\rho,\rho')^2}$\\ \hline
   $\ca{B}^\epsilon(\rho)$	&	The $\epsilon$-ball of a subnormalized state $\rho$: $\ca{B}^\epsilon(\rho)=\{\rho'\in\ca{S}_\leq(\ca{H})|\:P(\rho,\rho')\leq\epsilon\}$ \\ \hline
            \multicolumn{2}{l}{}\\

     \end{tabular}
  \end{center}
  \label{tb:values4}
\end{table}

    \begin{table}[h]
\renewcommand{\arraystretch}{1.6}
  \begin{center}
    \begin{tabular}{l|l} 
               
              \multicolumn{2}{l}{Conditional Entropies for $\rho\in\ca{P}(\ca{H}^{AB})$ and $\varsigma\in\ca{S}_=(\ca{H}^B)$}\\ \hline\hline 
  $H_{\rm min}(A|B)_{\rho|\varsigma} $       & $\sup \{ \lambda \in \mathbb{R}| 2^{-\lambda} I^A \otimes \varsigma^B \geq \rho^{AB} \}$ \\ \hline
$H_{\rm max}(A|B)_{\rho|\varsigma}$       & $\log{\|\sqrt{\rho^{AB}}\sqrt{I^A\otm\varsigma^B}\|_1^2}$ \\ \hline
$H_2(A|B)_{\rho|\varsigma} $ & $- \log \tr \bigl[ \bigl(  (\varsigma^B)^{-1/4} \rho^{AB} (\varsigma^B)^{-1/4} \bigr)^2 \bigr]$ \\ \hline 
$H_{\rm min}(A|B)_{\rho}$  & $\sup_{\varsigma^B \in \ca{S}_=(\ca{H}^B)}H_{\rm min}(A|B)_{\rho|\varsigma}$  \\ \hline
$H_{\rm max}(A|B)_{\rho}$  & $\sup_{\varsigma^B \in \ca{S}_=(\ca{H}^B)}H_{\rm max}(A|B)_{\rho|\varsigma}$\\ \hline
    $H_2(A|B)_{\rho}$      & $\sup_{\varsigma^B \in \ca{S}_=(\ca{H}^B)}H_2(A|B)_{\rho|\varsigma}$\\ \hline
   $H_{\rm min}^\epsilon(A|B)_{\rho}$	&	$\sup_{\hat{\rho}^{AB} \in \ca{B}^\epsilon(\rho)}H_{\rm min}(A|B)_{\hat\rho}$ for $\rho\in\ca{S}_\leq(\ca{H}^{AB})$ \\ \hline
$H_{\rm max}^\epsilon(A|B)_{\rho}$ & $\inf_{\hat{\rho}^{AB} \in \ca{B}^\epsilon(\rho)}H_{\rm max}(A|B)_{\hat\rho}$ for $\rho\in\ca{S}_\leq(\ca{H}^{AB})$  \\ \hline   
            \multicolumn{2}{l}{}\\ 
    
  \multicolumn{2}{l}{Notations when a Hilbert space $\ca{H}^A$ is decomposed into $\bigoplus_{j=1}^J\ca{H}_j^{A_l}\otm\ca{H}_j^{A_r}$ (\rThm{SmoothMarkov})}\\ \hline\hline 
  $l_j$ and $r_j$ & $\dim{\ca{H}_j^{A_l}}$ and $\dim{\ca{H}_j^{A_r}}$, respectively \\ \hline
  $\Pi_j^A\in\ca{P}(\ca{H}^A)$ & The projection onto $\ca{H}_j^{A_l}\otm\ca{H}_j^{A_r}$  \\ \hline
  $\Phi_j^l$, $\Phi_j^r$ & Maximally entangled states on $\ca{H}_j^{A_l}\otm\ca{H}_j^{\bar{A}_l}$ and $\ca{H}_j^{A_r}\otm\ca{H}_j^{\bar{A}_r}$ ($\ca{H}_j^{A_l}\cong\ca{H}_j^{\bar{A}_l}$, $\ca{H}_j^{A_r}\cong\ca{H}_j^{\bar{A}_r}$)\\ \hline
 \multirow{2}{*}{$\Phi^{AA'}$} & Maximally entangled state between $A$ and $A'$: \\
  & $\ket{\Phi}^{AA'}=\sum_{j=1}^J\sqrt{l_jr_j/d_A}|\Phi_j^l\rangle^{A_lA_l'}|\Phi_j^r\rangle^{A_rA_r'}$\\ \hline
 \multirow{2}{*} {$A^*$}	&  A quantum system represented by a Hilbert space \\ 
 &${\ca H}^{A^*}:=\bigoplus_{j=1}^J{\ca H}_j^{A_r}\otimes{\ca H}_j^{\bar{A}_r}$ (${\ca H}_j^{A_r}\cong{\ca H}_j^{\bar{A}_r}$)\\ \hline
 \multirow{2}{*}{$F^{A\bar{A}\rightarrow A^*}$} & A linear operator from $\ca{H}^A\otm\ca{H}^{\bar A}$ to $\ca{H}^{A^*}$: \\ 
  &$F^{A\bar{A}\rightarrow A^*}=
\bigoplus_{j=1}^J \sqrt{d_Al_j/r_j} \langle\Phi_j^l|^{A_l\bar{A}_l}(\Pi_j^{A} \otimes \Pi_j^{\bar{A}})$ \\ \hline
${\Lambda}(\Psi,\ca{T})$ & An unnormalized state on $A^*RE$: ${\Lambda}(\Psi,\ca{T})=F(\Psi^{AR}\otimes\tau^{\bar{A}E})F^\dagger\in\ca{P}(\ca{H}^{A^*RE})$\\ \hline
$\Psi_{jk}^{A_lA_rR}$ & $\Pi_j^A\Psi^{AR}\Pi_k\in\ca{L}(\ca{H}_k^{A_l}\otm\ca{H}_k^{A_r}\otm\ca{H}^R,\ca{H}_j^{A_l}\otm\ca{H}_j^{A_r}\otm\ca{H}^R)$  \\ \hline 
$\tau_{jk}^{A_lA_rE}$ & $\Pi_j^A\tau^{AE}\Pi_k\in\ca{L}(\ca{H}_k^{A_l}\otm\ca{H}_k^{A_r}\otm\ca{H}^E,\ca{H}_j^{A_l}\otm\ca{H}_j^{A_r}\otm\ca{H}^E)$  \\ \hline 
$\pi_j^{A_r}\in\ca{S}(\ca{H}_j^{A_r})$ & The maximally mixed state on $\ca{H}_j^{A_r}$ \\ \hline
${\sf H}_j$ & The Haar measure on $\mbb{U}(r_j)$ \\ \hline
${\sf H}_\times$ & A product measure ${\sf H}_1\times\cdots\times{\sf H}_J$ on $\mbb{U}(r_1)\times\cdots\times\mbb{U}(r_J)$ \\ \hline
 $\Psi_{\rm av}^{AR}$ & A subnormalized state on $AR$: $\Psi_{\rm av}^{AR}=\mbb{E}_{U\sim{\sf H}_\times}[\ca{U}^A(\Psi^{AR})]$ \\ \hline
 \multirow{2}{*}{ $\| \ca{E}^{A \rightarrow B} \|_{\rm DSP}$ }
       &  The DSP-diamond norm of a supermap $\ca{E}$ from $\ca{L}(\ca{H}^A)$ to $\ca{L}(\ca{H}^B)$: \\
       &  $\| \ca{E}^{A \rightarrow B} \|_{\rm DSP}=\sup_{C,\:\xi} \{\|  \ca{E}^{A \rightarrow B}(\xi^{AC}) \|_1:\xi\in\ca{S}_\leq(\ca{H}^{AC}),\:\xi^A=\bigoplus_{j=1}^Jq_j \varpi_j^{A_l}\otimes \pi_j^{A_r}\}$ \\ \hline
  $\ca{B}_{\rm DSP}^\epsilon(\ca{E})$      &  $\{\ca{E}'\in\ca{C}\ca{P}_=(A\rightarrow B)\:|\:\|\ca{E}'-\ca{E}\|_{\rm DSP}\leq\epsilon\}$ \\ \hline
$ H_{\rm min}^{\epsilon,\mu}(A^*|RE)_{{\Lambda}(\Psi,\ca{T})}$ & 
$\sup_{\Psi'\in\ca{B}^\epsilon(\Psi)}\sup_{\ca{T}'\in\ca{B}_{\rm DSP}^{\mu}(\ca{T})}
H_{\rm min}(A^*|RE)_{{\Lambda}(\Psi',\ca{T}')}$ \\ \hline
 
       \multicolumn{2}{l}{}\\ 
       
        \end{tabular}
  \end{center}
  \label{tb:values14}
\end{table}

    \begin{table}[h]
\renewcommand{\arraystretch}{1.6}
  \begin{center}
    \begin{tabular}{l|l}

  \multicolumn{2}{l}{Notations when $l_j=1$ and $r_j=r$ for $1\leq j\leq J$ (\rThm{SmoothExMarkov} and \rthm{converse})}  \\ \hline\hline 
    $\alpha(J)$ & A function that is equal to $0$ when $J=1$ and to $1/(J-1)$  if $J\geq2$ \\ \hline
  $\mbb{P}$       & The permutation group on $[1,\cdots,J]$ \\ \hline
  $\sf P$      & The uniform distribution on $\mbb{P}$ \\ \hline
  $G_\sigma$ & A unitary in $\ca{H}^A$: $G_\sigma=\sum_{j=1}^J\outpro{\sigma(j)}{j}^{A_c}  \otimes I^{A_r}$ for any $\sigma\in{\mbb{P}}$ \\ \hline
  $\ca{C}$ & The completely dephasing operation on $A_c$ with respect to the basis $\{\ket{j}\}_{j=1}^J$ \\ \hline
  $\Psi_{\rm dp}^{AR}$ & A normalized state on $AR$: $\Psi_{\rm dp}^{AR}=\ca{C}(\Psi^{AR})$ \\ \hline
  $\pi^{A_r}$ & The maximally mixed state on $\ca{H}^{A_r}$ \\ \hline
  
    \end{tabular}
  \end{center}
  \label{tb:values2}
\end{table}

\end{document}